\documentclass[a4paper, 12pt,letterpaper, headsepline,singlespacing, oneside, french]{report}
\usepackage{pifont}
\usepackage{footnote}
\usepackage{bm}
\usepackage{wrapfig}
\usepackage{enumitem}
\usepackage[utf8]{inputenc} 
\usepackage{blindtext}
\usepackage{setspace}
\usepackage[french]{babel}
\usepackage{geometry}
\usepackage[toc,page]{appendix} 
\usepackage{geometry}
\usepackage[export]{adjustbox}
\usepackage{wrapfig}
\AddThinSpaceBeforeFootnotes 
\FrenchFootnotes 
\usepackage{float}
\addtocounter{tocdepth}{3}
\usepackage{wrapfig}
\usepackage[T1]{fontenc}
\usepackage{hyperref}
\usepackage{caption}
\usepackage{indentfirst}
\usepackage[autolanguage]{numprint}
\usepackage[table]{xcolor}
\usepackage{graphicx}
\usepackage{lmodern}
\usepackage{amsmath}
\usepackage{amssymb}
\usepackage{mathrsfs}
\usepackage{lipsum}
\usepackage{float}
\usepackage{multicol}
\usepackage{booktabs} 
\usepackage[ampersand]{easylist} 
\usepackage{algorithm}
\usepackage{algorithmic}
\usepackage { varwidth }
\usepackage{fancyhdr} 
\usepackage{titlesec}
\usepackage{amsmath}
\usepackage{array,multirow,makecell}
\newcolumntype{R}[1]{>{\raggedleft\arraybackslash }b{#1}}
\newcolumntype{L}[1]{>{\raggedright\arraybackslash }b{#1}}
\newcolumntype{C}[1]{>{\centering\arraybackslash }b{#1}}
\newcommand*{\savedfootnotes}{}
\newcommand*{\resetsavedfootnotes}{\global\let\savedfootnotes\empty}
\usepackage{titlesec}
\usepackage{amssymb}

\newcommand{\bi}{\begin{itemize}}
\newcommand{\ei}{\end{itemize}}

\newcommand{\be}{ \begin{equation} }
\newcommand{\ee}{ \end{equation} }

\makeatletter
\usepackage[french]{minitoc}
\usepackage[nottoc]{tocbibind}
\usepackage[french]{nomencl}
\makenomenclature
\begin{document}
\begin{titlepage}
\center
\textbf{RÉPUBLIQUE TUNISIENNE  } \\
\vspace{0.1em}
\textbf{MINISTÈRE DE L'ENSEIGNEMENT SUPÉRIEUR,} \\
\vspace{0.1em}
\textbf{DE LA RECHERCHE SCIENTIFIQUE}\\ 
\vspace{0.5em}
 \textbf{UNIVERSITÉ DE TUNIS EL MANAR} \\
\vspace{1em}
\begin{figure}[H]
\centering
\includegraphics[width=0.29\textwidth]{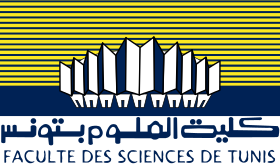}
\end{figure}
\vspace{0.1em}
\textbf{FACULTÉ DES SCIENCES DE TUNIS} \\
\vspace{0.1em}
\textbf{DÉPARTEMENT DES SCIENCES DE L'INFORMATIQUE} \\
\vspace{2em} 
{\bfseries \Large Mémoire de Mastère } \\
\vspace{0.1em}
{\slshape \normalsize Présenté en vue de l'obtention du diplôme de } \\
\vspace{1em}
{\textrm {\large \textbf{MASTÈRE DE RECHERCHE EN INFORMATIQUE}}} \\

{\slshape \normalsize Par } \\
\vspace{0.1em}
{\bfseries \large 	Halima NEFZI } \\
\vspace{2em}
 \textbf{ \large{PROPOSITION D'UNE NOUVELLE APPROCHE DE RECOMMANDATION CONTEXTUELLE EN SE BASANT SUR LA MÉTHODE D'ANALYSE HIÉRARCHIQUE DES PROCÉDÉS (AHP)}}\\
\vspace{2em}
\textbf{Soutenue le 17/02/2018 devant le jury composé de :} \\ 
\begin{flushleft}

\textbf{\footnotesize  Mme. Amel GRISSA TOUZI} \hspace{10ex}\textbf{\textit{{\footnotesize Professeur à l'ENIT}}} \hspace{11ex}\textbf{\textit{{\footnotesize Président de jury}}} \\
\textbf{{\footnotesize  Mme. Narjes DOGGAZ}}\hspace{16ex}\textbf{\textit{{\footnotesize Maître-Assistante à la FST}}} \hspace{4ex}\textbf{\textit{{\footnotesize Rapporteur}}} \\
\textbf{{\footnotesize  M. Sadok BEN YAHIA}}\hspace{16ex}\textbf{\textit{{\footnotesize  Professeur à la FST}}} \hspace{11ex}\textbf{\textit{{\footnotesize  Directeur de mémoire}}}
\end{flushleft}
\vspace{2mm}
\textbf{ Au sein du laboratoire LIPAH : FST }
\pagenumbering{gobble}
\end{titlepage}


\chapter*{Dédicace}

\begin{center}
\textrm{\textbf{À ma chère mère}}
\end{center}

Pour son sacrifice, sa tendresse, son amour infini et le soutien qu'elle m'a apporté durant toute ma vie.

\begin{center}
\textrm{\textbf{À mon père }} 

\end{center}
\begin{center}
Pour son soutien, réconfort et dévouement continus.
\end{center}

\begin{center}
\textrm{\textbf{À mes chers frères}}
\end{center}

À qui je souhaite toute la réussite et le bonheur.
À toute personne qui m'a soutenue durant la période de ce projet.
À tous ceux qui m'ont soutenu de près ou de loin à réaliser ce travail.
Ainsi que tous ceux dont je n'ai pas indiqué le nom mais qui ne sont pas moins chers.

\chapter*{Remerciements}



\emph{Je remercie le bon DIEU, qui m'a donné la force, la volonté et le courage pour terminer ce modeste travail.}
  
\emph{Mes remerciements vont également à Monsieur \textbf{SADOK BEN YAHIA}, Professeur à la Faculté des Sciences de Tunis, mon directeur de mémoire, son encadrement, son écoute, ses élucidations, ses conseils, ses directives et encouragements qu'il m'a afflué. Merci pour tout!}

\emph{Je tiens aussi à remercier Mlle \textbf{IMEN BEN SASSI} pour l'aide qu'elle m'avait apportée tout le long de mon travail.}

\emph{Je tiens à exprimer ma profonde gratitude à Madame \textbf{AMEL GRISSA TOUZI}, Professeur à l'École Nationale d'Ingénieurs de Tunis, pour l'honneur qu'elle m'a fait en acceptant de présider le jury de soutenance.}

\emph{Mes remerciements s'adressent aussi à Madame \textbf{NARJES DOGGAZ}, Maître-Assistante à la Faculté des Sciences de Tunis, d'avoir accepté de rapporter mon travail avec patience et pertinence.}

\emph{J'exprime mes reconnaissances et mes respects envers tous les enseignants du Département des Sciences de l'informatique de la Faculté des Sciences de Tunis. Je tiens à remercier tous ceux, non cité ici, qui m'ont soutenu tout le long de ce travail.}


\pagenumbering{roman}
\tableofcontents
\addstarredchapter{Liste des symboles} 
\nomenclature{\textbf{CF}}{Collaborative Filtering (Filtrage collaboratif)}
\nomenclature{\textbf{CBF}}{Content-Based Filtering (Filtrage basé sur le contenu)} 
\nomenclature{\textbf{ACP ou PCA}}{Analyse en composantes principales (Principal component analysis)}
\nomenclature{\textbf{SVD}}{Décomposition en valeurs singulières (singular value decomposition)}
\nomenclature{\textbf{WUM}}{A Web Utilization Miner}
\nomenclature{\textbf{CARS}}{Context Aware Recommender System (systèmes de recommandation sensibles au contexte )}
\nomenclature{\textbf{MCDM}}{Multiple-criteria decision analysis (Méthodes de prise de décision multicritères)}
\nomenclature{\textbf{AHP}}{The Analytic Hierarchy Process (méthode d'Analyse Hiérarchique des Procédés)}
\nomenclature{\textbf{TF-IDF}}{(Term Frequency-Inverse Document Frequency)}
\printnomenclature
\listoffigures 
\listoftables
\chapter*{Introduction Générale}
\pagenumbering{arabic}
\addstarredchapter{Introduction générale} 
\markboth{INTRODUCTION GÉNÉRALE}{}
\section*{Contexte et problématique :}
La masse des données échangées aujourd'hui sur Internet constitue un atout sans précédent pour l'accès de tous à l'information. De même, l'explosion des services de recommandation de nos jours ont propulsé la recherche d'information (RI). En effet, la surabondance de l'information a engendré la dégradation de la qualité des résultats retournés par un système de recommandation et a apporté de nouveaux problèmes au domaine de recherche d'information. L'arrivée des systèmes de recommandation permet de résoudre le problème de la surcharge d'information auxquels ils sont confrontés, aujourd'hui, avec l'avènement d'Internet, en leur fournissant des recommandations. Ce sont alors des systèmes qui présentent aux utilisateurs les contenus les plus pertinents, en utilisant certaines informations concernant leurs historiques préférences. 
\newline 
\hspace*{3ex} Depuis les années 90, les systèmes de recommandation sont apparus comme un domaine de recherche indépendant. Les chercheurs ont commencé à se concentrer sur les problèmes de recommandation en s’appuyant sur la notion de classement « rating » \textbf{(R : USER x ITEM -> RATING)} pour exprimer les préférences des utilisateurs. Comme les gens disposent de leurs appareils mobiles à tout moment et partout dans le monde, l'utilisation des capacités de ces dispositifs intelligents offre une occasion importante pour améliorer la qualité des items recommandés aux utilisateurs \textbf{(Rcontext : USER x ITEM x CONTEXT -> RATING)} \cite{adomavicius2005incorporating}. De ce fait, afin de choisir le film à regarder, les gens choisissent un film ou plusieurs à partir d'un grand nombre de films stockés dans une base de données (exemple : \textbf{MovieLens}). Toutefois, le choix de l'utilisateur est fortement lié à son contexte représenté par : sa localisation, le temps, son activité, et son humeur. Par exemple, un utilisateur peut regarder un film de genre "Romance" durant un jour de pluie et choisir un autre film de genre "Comedy" lorsqu'il est de bonne humeur.
\\
Dans ce contexte, citons le travail de Imen et al. (2017) \cite{sassi2017context}, membre de notre laboratoire LIPAH, qui a exploité le contexte dans les systèmes de recommandation afin de suggérer des éléments qui aident les utilisateurs à prendre des décisions parmi un grand nombre d'actions possibles, telles que le lieu à visiter, le film à regarder ou l'ami à ajouter à un réseau social.
\\
	Alors, l'idée de proposer un système de recommandation, pouvant limiter les actions d'un ami proche ou d'un expert en adaptant les recommandations non seulement aux préférences de l'utilisateur, mais aussi à son contexte, devient de plus en plus intéressante.

\section*{Contributions}
En recommandation, il peut être intéressant de placer les utilisateurs (ou les items) dans un contexte précis afin d'en extraire plus d'informations et ainsi obtenir une meilleure prédiction des préférences. L'objectif du présent mémoire est de proposer une nouvelle approche permettant l'introduction du contexte dans le processus de recommandation tout en se basant sur l'une des méthodes de prise de décision à critères multiples (MCDM) : c'est la méthode d'Analyse Hiérarchique des Procédés (AHP) qui permet de faciliter le choix de décision pour un utilisateur et aider à choisir la meilleure solution.

\section*{Plan du mémoire :}
Les travaux de recherche sont synthétisés dans ce mémoire composé de quatre chapitres : 
\begin{itemize}
\item Le premier chapitre récapitule quelques concepts de base, à savoir les systèmes de recommandation classiques et sensibles au contexte, leurs définitions et leurs techniques et leurs architectures.
\item Le deuxième chapitre est consacré à la représentation des diverses approches des recommandations basées sur le contexte dans le domaine des films. 
\item Dans le troisième chapitre, nous allons présenter les méthodes de prise de décision à critères multiples (MCDM) et plus précisément la méthode  d'Analyse Hiérarchique des Procédés (AHP). Ensuite, nous allons  proposer une nouvelle approche de recommandation contextuelle en utilisant la méthode AHP.
\item Le quatrième chapitre exposera une étude expérimentale de notre nouvelle approche. Nous clôturons 
ce chapitre avec une synthèse de l'ensemble de nos travaux ainsi que quelques perspectives futures de recherche. 
\end{itemize}

\chapter{\textbf{Notion de base}}
\section{Introduction} 
Dans ce chapitre, nous présentons les notions de bases importantes que nous utilisons tout le long de ce mémoire. Nous commençons par les notions de base des systèmes de recommandations classiques en détaillant les techniques utilisées et les limites de ces systèmes. Ensuite, nous allons définir quelques notions de base des systèmes de recommandation contextuels et nous allons terminer par quelques définitions des façons d'intégration du contexte.
\section{Systèmes de recommandation classiques}
\subsection{Définition}  
Les systèmes de recommandation peuvent être définis de plusieurs façons, vue la diversité des classifications proposées pour ces systèmes. La définition que nous utiliserons dans ce mémoire est une définition générale de J. Bobadilla :
\vspace*{3mm}
\newline
les systèmes de recommandation peuvent être définis comme des programmes qui essayent de recommander les articles les plus appropriés (des produits ou des services) aux utilisateurs particuliers (des individus ou des affaires (activités)) en prévoyant l'intérêt d'un utilisateur dans un article basé sur des informations liées sur les articles, les utilisateurs et les interactions entre des articles et des utilisateurs \cite{bobadilla2013recommender}.
\vspace*{3mm}
\newline 
\textbf{Définition de Burke et Robin} " Des systèmes capables de fournir des recommandations personnalisées permettant de guider l'utilisateur vers des ressources intéressantes et utiles au sein d'un espace de données important " \cite{burke2002hybrid}.
\vspace*{3mm}

Voici les travaux qui ont menés en collaboration avec notre laboratoire LIPAH concernant les systèmes de recommandation personnalisés : 
Jelassi et al. (2013 \cite{jelassi2013personalized}, 2014 \cite{jelassi2014vers}, 2015 \cite{jelassi2015towards}, 2016 \cite{jelassi2016etude}).

\vspace*{3mm}
D'après Chris Anderson dans " The Long Tail ", les bouleversements qu'a subi le Web et la masse de données qui constituent Internet font que " nous quittons progressivement l'âge de l'information pour rentrer dans l'âge de la recommandation " \cite{anderson2006long}.
\vspace*{3mm}
\newline 
La tâche d'un système de recommandation consiste à accomplir un filtrage d'information afin de suggérer à un utilisateur des articles à acheter (ex. e-commerce) ou bien d'autres utilisateurs avec qui interagir/se connecter (ex. réseaux sociaux) \cite{ricci2011introduction}. Ces recommandations peuvent concerner un article à lire, un livre à commander, un film à regarder, un restaurant à choisir, une musique à écouter, etc. Ainsi, l'utilisation des systèmes de recommandation est arrivée pour résoudre le problème de surcharge et de profusion d'informations disponibles notamment à travers le Web ou les e-services.

\subsection{Les données}
Les deux entités de base qui apparaissent dans tous les systèmes de recommandations sont l'utilisateur et l'article (Item en Anglais). En effet, l'utilisateur est la personne qui utilise un système de recommandation, donne son opinion sur divers articles et reçoit les nouvelles recommandations du système et l'item est le terme général utilisé pour désigner ce que le système recommande aux utilisateurs. Les données d'entrée pour un système de recommandation dépendent du type de l'algorithme de filtrage employé. Généralement, elles appartiennent à l'une des catégories suivantes :
\begin{itemize}
\item \textbf{Des notes} aux items consultés indiquant le degré d'appréciation d'un item par cet utilisateur. Les notes sont souvent numériques et limitées par une échelle de valeurs.
\newline
Une note élevée signifie que l'utilisateur accorde un grand intérêt à l'item et qu'il correspond bien à ses goûts. Par contre, une note faible signifie que l'utilisateur ne s'intéresse pas à l'item. Dans d'autres cas, les notes peuvent être exprimées sous une forme binaire telle que "Aime" ou "Aime pas".
\vspace*{3mm}
\newline 
Les notes permettent de faciliter l'apprentissage des appréciations vu que les notes sont faciles à traiter par le système de recommandation. Les utilisateurs n'ayant pas les mêmes façons de noter, les notes peuvent ne pas être fiables. En effet, certains utilisateurs attribuent des notes élevées et d'autres non. Par exemple, sur une échelle $\mathopen{[}1 - 5\mathclose{]}$, une note qui vaut 3 peut être négative pour un utilisateur et plutôt neutre pour un autre. 
\newline  
\textbf{Exemple 1 (Site d'Amazon)}
Considérons l'exemple ci-dessous des notes dans le site "Amazon" : 
\begin{figure}[H]
   \centering
    \includegraphics[width=14cm]{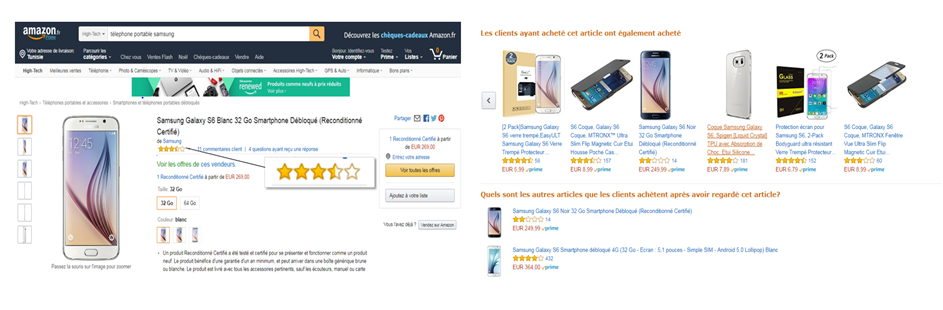}
    \caption{Exemple de notes : Site d'Amazon}
    \label{fig1:my_label}
\end{figure}
\hspace*{3ex} Le tableau \ref{tab1} donne les échelles des notes avec leurs descriptions : 
\begin{center}
\begin{tabular}{|c||c|}
\hline 
\textbf{Type d'échelle}& \textbf{Description}\\
\hline \hline
Unaire &  "Aime" ou "Je sais pas" \\
\hline 
Binaire & "Aime" ou "Aime pas" \\
\hline
Entier & $\mathopen{[}1 - 5\mathclose{]}$,$\mathopen{[}1 - 7\mathclose{]}$ ou $\mathopen{[}1 - 10\mathclose{]}$\\
\hline 
\end{tabular}
\end{center}
\captionof{table}{Les échelles des notes les plus connues}
\label{tab1}
\item \textbf{Des commentaires, des mots-clés ou des tags sur des items} ces tags ont eu l'intention d'inclure chaque artiste disponible sur LastFM indépendamment du genre et promouvoir un sens communautaire. Quand complété (achevé), toute la radio d'étiquette devrait fournir tout le monde la capacité d'avoir accès à une sélection largement éclectique de musique. Ceci permettra à l'utilisateur de découvrir et développer des intérêts dans des styles de musique auxquels ils ne seraient pas autrement normalement exposés.

\vspace*{3mm}
\textbf{Exemple 2 (LastFM)}
La figure \ref{fig2:my_label} montre un exemple d'ajout de tags sur le site de recommandation de musique "LastFM".

\begin{figure}[H]
   \centering
    \includegraphics[width=8cm]{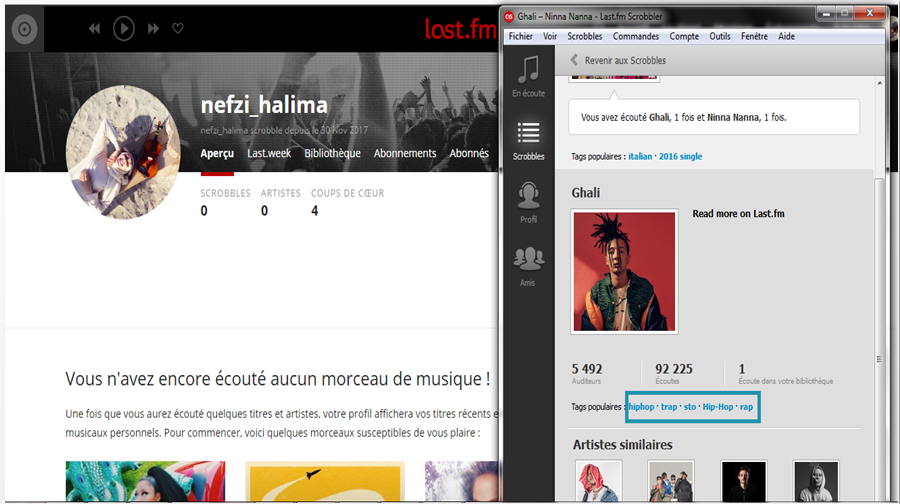}
    \caption{LastFM: Exemple de tags}
    \label{fig2:my_label}
\end{figure}
\item \textbf{Des attributs démographiques} ces attributs concernant l'utilisateur, tels que : l'âge, le sexe, la catégorie socio-professionnelle, le niveau d'étude, la localité géographique, le statut personnel, etc. Ils ne fournissent pas d'informations sur les appréciations, mais ils permettent notamment d'affiner le profil utilisateur afin d'y adapter les recommandations.
\end{itemize}
\subsection{La notion de profil dans les systèmes de recommandation}
\begin{figure}[H]
   \centering
    \includegraphics[width=10cm]{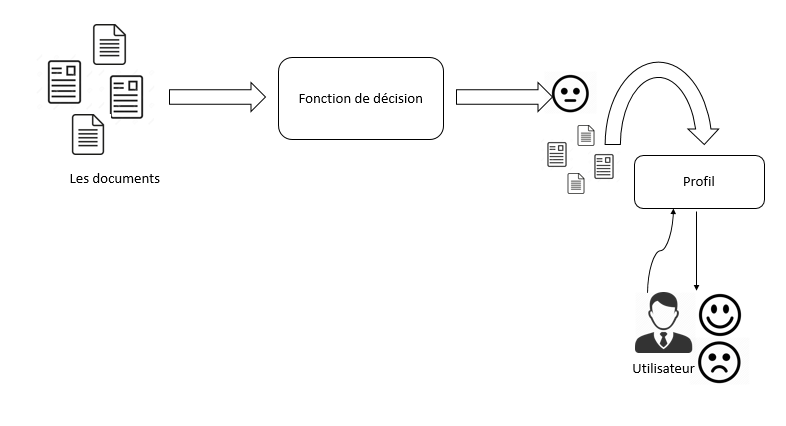}
    \caption{Système général de filtrage d'information}
    \label{fig3:my_label}
\end{figure}

Les profils pour réaliser le filtrage, les systèmes de recommandation utilisent \textbf{\textit{les profils}} représentant des préférences relativement stables des utilisateurs pour calculer des recommandations. Ce calcul se fait par la prédiction des scores qu'un utilisateur est susceptible d'attribuer aux contenus.
\newline 
Les systèmes de recommandation adaptent ce profil au cours du temps en exploitant au mieux le retour de pertinence que les utilisateurs fournissent sur les documents reçus. Par exemple, dans la figure \ref{fig3:my_label}, la fonction de décision du système traite les flux entrant de documents pour suggérer à l'utilisateur, en consultant son profil, les documents qu'il préfère. Par la suite, l'utilisateur doit évaluer fréquemment les recommandations pour que le système comprenne mieux ses besoins en informations et lui fournisse par conséquent de meilleures nouvelles recommandations.
\subsection{Classification des systèmes de recommandation}
\hspace*{3ex} Il existe plusieurs classifications des systèmes de recommandations: 
\begin{itemize}
\item \textbf{La classification classique} cette classification de Adomavicius et Tuzhilin \cite{adomavicius2005toward} est reconnue par trois types de filtrage : un filtrage collaboratif (CF), un filtrage basé sur le contenu (CBF) et le filtrage hybride.
\item \textbf{La classification de Su et al.\cite{su2009survey}} elle est utilisée dans les systèmes de collaboration. Ils proposent une sous classification qui comprend les techniques hybrides et les classer dans les méthodes de collaboration. Su et al. \cite{su2009survey} classent le filtrage collaboratif en trois catégories :
\begin{itemize}
\item Approches CF à base de mémoire : pour K-plus proches voisins.
\item Approches CF basé sur un modèle englobant une variété de techniques telles que : Clustering, les réseaux bayésiens, factorisation de matrices, les processus de décision de Markov.
\item CF hybride qui combine une technique de  recommandation CF avec une ou plusieurs autres méthodes.
\end{itemize}
\item \textbf{La classification de Rao and Talwar \cite{rao2008application}} c'est une classification en fonction de la source d'information utilisée.
\end{itemize} 
\hspace*{3ex} Pour tous les systèmes de recommandation développés jusqu'à nos jours, la collecte de données relatives aux utilisateurs et/ou aux items, représente une phase clé dans le processus de personnalisation. La sous-section qui suit décrit les techniques utilisées par les systèmes de recommandation.
\subsection{Les techniques de recommandation}
\hspace*{3ex} Il existe une large variété de techniques de recommandation. À travers les travaux de recherche, différentes tentatives de classification des approches ou des techniques ont été réalisées. La classification de ces approches dépend notamment du type de données exploitées et de la méthode d'apprentissage utilisée par le système de recommandation.
Dans cette partie, nous présentons les principales techniques de recommandation avec leurs apports et leurs limites.
\newline 
\hspace*{3ex} Plusieurs facteurs entrent en considération afin de catégoriser les systèmes de recommandation:
\begin{itemize}
\item La connaissance de l'utilisateur c'est à dire le profil de l'utilisateur en fonction de ses préférences.
\item La notion de classes ou réseaux d'utilisateurs : le positionnement d'un utilisateur par rapport aux autres.
\item La connaissance des items à recommander.
\item La connaissance des différentes classes d'items à recommander.
\end{itemize}
\hspace*{3ex} À partir de ces facteurs, divers types de recommandations ont été produits. Parmi les techniques les plus utilisées dans la littérature sont le filtrage basé sur le contenu, le filtrage collaboratif et les méthodes hybrides.
\subsubsection{Filtrage basé sur le contenu}
\hspace*{3ex} Dans cette partie, nous définissons dans un premier lieu le filtrage basé sur le contenu, puis nous donnons l'architecture générale de ce filtrage ainsi que les deux types de recommandation basée sur le contenu.
\begin{enumerate}[label=\alph*)]
\item \textbf{\underline {Définition du filtrage basé sur le contenu}}
\vspace*{3mm}
\newline
Le filtrage basé sur le contenu est une évolution générale des études sur le filtrage d'information s'appuie sur des évaluations effectuées par un utilisateur sur un ensemble des documents ou items \cite{celma2009music}.
\vspace*{3mm}
\newline
Naak \cite{naak2009papyres}  a défini le filtrage basé sur le contenu comme ceci : \textbf{\textit{" les méthodes basées sur le contenu, comme leur nom l'indique, se basent sur la compréhension de pourquoi l'usager, à qui la recommandation est destinée, a donné une haute valeur à certains items qu'il a évalués dans le passé ? Une fois cette question résolue, le système cherche parmi les nouveaux items ceux qui maximisent ces caractéristiques pour les lui recommander "}}.
\vspace*{3mm}
\newline 
La technique de recommandation basée sur le contenu peut être appliquée à la recommandation de pages Web, de films, d'articles actualités, de restaurants, etc. Cette technique a pour avantage de pouvoir générer des recommandations en dépit d'une situation de démarrage à froid.

\item \textbf{\underline{Architecture générale}}
\vspace*{3mm}
\newline
Cette technique de recommandation basée sur le contenu s'articule autour de trois modules principaux :
\begin{itemize}
  \item \textbf{L'analyseur de contenu}: selon la nature des données à recommander (texte, éléments multimédia, pages Web, produits commerciaux, etc.), une étape de pré-traitement est nécessaire afin de décrire les objets à recommander et d'en extraire les caractéristiques. Le module d'analyse de contenu est responsable de produire une description structurée de ces objets. Cette description va servir d'élément d'entrée aux autres modules.
  
  \item   \textbf{Le module d'apprentissage de profils} : ce module est responsable de l'analyse des interactions passées de l'utilisateur sur les objets du système. En utilisant des méthodes empruntées au monde de l'apprentissage, ce module construit une description des préférences des utilisateurs.

\item \textbf{Le module de filtrage} : à partir des profils utilisateurs et des descriptions des objets à recommander, ce module construit des listes de suggestions à présenter aux utilisateurs.
\end{itemize}
\begin{figure}[H]
   \centering
    \includegraphics[width=12cm]{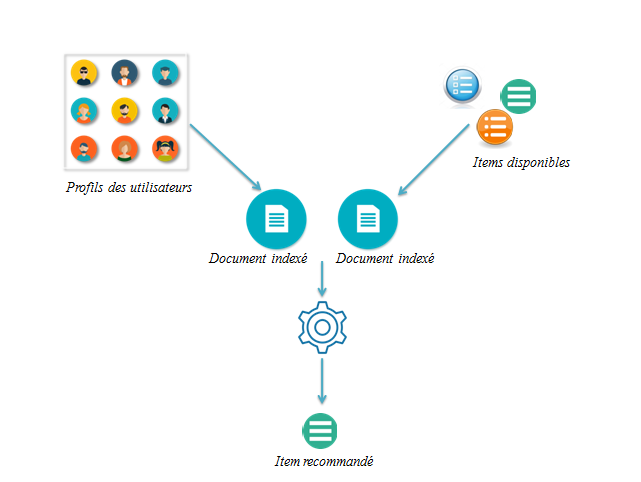}
    \caption{Recommandation basée sur le contenu}
    \label{fig4:my_label}
\end{figure}
\item \textbf{\underline {Les deux types de recommandation basée sur le contenu}}                                                                                                                                                                                                                                                                                                                                                                                                                                                                                                                                                                                                                                                                                                                                                                                               
\vspace*{3mm}
\newline 
On distingue deux types de recommandation basée sur le contenu : la recommandation basée sur les mots clefs ainsi que la recommandation basée sur la sémantique.
\begin{itemize}
\item \textbf{Recommandation basée sur les mots clefs} 
lorsqu'un utilisateur a tendance à consulter souvent des articles portant sur le domaine de la génétique, le système lui proposera des recommandations liées à la génétique. En effet, ces articles disposent de mots-clés communs tels que : "ADN", "gène" ou "protéine".
\newline
Lorsque des caractéristiques plus complexes sont nécessaires, les approches à base de mots clefs montrent leurs limites. Si l'utilisateur, par exemple, aime "l'impressionnisme Français", les approches à base de mots-clefs chercheront seulement des documents dans lesquels les mots "Français" et "impressionnisme" apparaissent. Des documents concernant Claude Monet ou Renoir n'apparaîtront pas dans l'ensemble des recommandations, même s'ils sont susceptibles d'être pertinents pour l'utilisateur.

\item \textbf{Recommandation basée sur la sémantique}
les systèmes de recommandation basés sur la sémantique évoluent au rythme des méthodes et outils proposés dans le domaine du Web sémantique. La sémantique a été introduite par plusieurs méthodes dans le processus de recommandation. Ces méthodes sont abordées en tenant compte de plusieurs critères : 
\end{itemize}
\begin{itemize}
\item Le type de source de connaissance impliquée (lexique, ontologie, etc.).
\item Les techniques adoptées pour l'annotation ou la représentation d'items.
\item Le type de contenu inclus dans le profil utilisateur.
\item La stratégie de correspondance entre items et profil.
\end{itemize}
\end{enumerate}
\hspace*{3ex} Il existe différents méthodes basées sur le contenu qui ont donné des résultats pertinents et plus précis comparés aux méthodes traditionnelles basées sur le contenu.
\newline
\textbf{SiteIF} c'est le premier système à adopter une représentation basée sur le sens des documents pour construire un modèle des intérêts de l'utilisateur. SiteIF est un agent personnel pour un site Web de nouvelles multilingues \cite{magnini2001improving}.
\newline 
\textbf{ITR : ITerm Recommender} un système capable de fournir des recommandations d'items dans plusieurs domaines (films, musique, livres), à condition que les descriptions d'articles soient disponibles sous forme de documents texte \cite{degemmis2007content}.
\newline 
\textbf{SEWeP : Semantic Enhancement for Web Personalization} est un système de personnalisation Web, qui utilise à la fois les logs d'utilisation et la sémantique du contenu du site Web dans le but de le personnaliser.
\newline 
\textbf{Quickstep} est un système de recommandation d'articles de recherche académique. Le système adopte une ontologie d'articles de recherche basée sur la classification scientifique du projet DMOZ open directory \cite{middleton2002exploiting}.
\vspace*{3mm}
\newline
\textbf{Informed Recommender} ce système utilise les avis des utilisateurs sur les produits pour faire des recommandations. Le système convertit les opinions des clients dans une forme structurée en utilisant une ontologie de traduction, qui est exploitée pour la représentation et le partage de connaissance \cite{aciar2007informed}.
\subsubsection{Filtrage collaboratif}
\begin{enumerate}[label=\alph*)]
\item \textbf{\underline{Définition du filtrage collaboratif}}
\newline 
Lorsque les systèmes de recommandation basés sur le contenu trouvent des propriétés semblables, le filtrage collaboratif (FC) trouve des évaluations semblables. Ces systèmes comparent des utilisateurs et des articles seulement par le comportement d'utilisateur passé (c'est-à-dire les rangées et les colonnes de la matrice utilitaire), sans regarder leurs propriétés. 
\newline
Cette deuxième grande famille de systèmes de recommandation est basée sur l’hypothèse que les utilisateurs qui ont aimé des articles similaires par le passé ont un goût similaire et vont donc apprécier les mêmes articles dans le futur. Un des exemples les plus connus d’un tel système a été popularisé par le site de commerce en ligne "Amazon.com" et son algorithme de Item-to-item Collaborative Filtering qui se traduit sur le site par la fonctionnalité "Les gens qui ont acheté le produit x ont aussi acheté le produit y" \cite{linden2003amazon}. La figure \ref{fig5:my_label} décrit la technique de filtrage collaboratif.
\begin{figure}[H]
\centering
    \includegraphics[width=10cm]{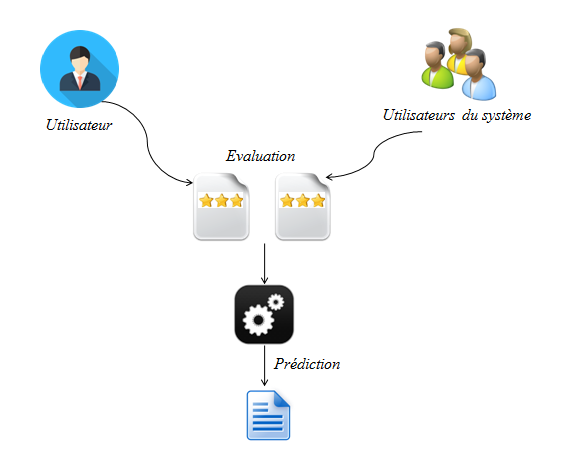}
    \caption{Recommandation basée sur le filtrage collaboratif}
    \label{fig5:my_label}
\end{figure}
Dans un système de filtrage collaboratif, il faut que les utilisateurs fournissent des évaluations des items qu'ils ont déjà utilisés, sous forme des notes, pour constituer leurs profils. Il n'y a aucune analyse du sujet ou du contenu des objets à recommander. Ce type de système est très efficace en cas où le contenu des objets est complexe, il est compliqué ou impossible de l'analyser, l'utilisateur peut apercevoir divers domaines intéressants, car le principe du filtrage collaboratif ne se fonde absolument pas sur la dimension thématique des profils, et n'est pas soumis à l'effet « entonnoir ».
\newline
L'avantage principal de cette approche est qu'elle ne nécessite pas de description précise des objets à recommander. Les recommandations étant basées sur l'ensemble des interactions des utilisateurs avec les objets/produits, cette méthode permet de recommander des objets complexes sans avoir à les analyser. La plupart des services de recommandation de musique en ligne fonctionnent sur ce mode (ex. \textbf{LastFM}) car les fichiers multimédia sont difficiles à analyser.
\newline
\textbf{Exemple 3 (un groupe d'amis sur MovieLens)} la figure \ref{fig6:my_label} représente un tableau de films : sur un axe les utilisateurs d'un même système et sur un autre les films. Chaque cellule de la matrice contient l'avis donné par un utilisateur pour un film, la cellule vide signifie qu'il n'a pas d'avis particulier sur ce film. Afin de prédire si \textbf{Mourad} apprécierait le film \textbf{"Harry Potter"} et probablement lui recommander ce film, on compare les votes de \textbf{Mourad} à ceux des autres utilisateurs choisis. On peut alors voir que \textbf{Mourad} et \textbf{Rahma} ont des votes identiques, et que \textbf{Rahma} n'a pas aimé le film \textbf{"Harry Potter"} et \textbf{"les Reliques de la Mort"}, on pourrait alors prédire que \textbf{Mourad} n'aimera pas aussi ce film et de ne lui pas faire cette suggestion.
\begin{figure}[H]
\centering
    \includegraphics[width=7cm]{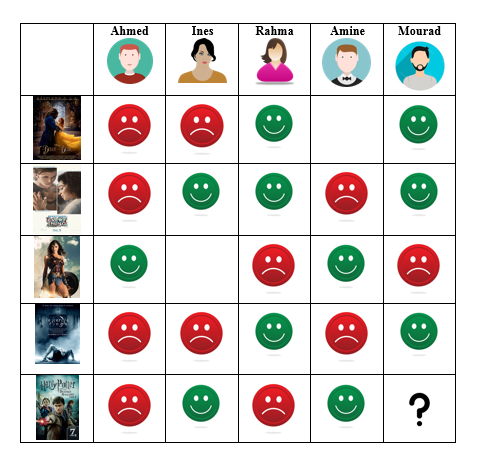}
    \caption{Exemple de recommandation basée sur le filtrage collaboratif}
    \label{fig6:my_label}
\end{figure}
\item \textbf{\underline {Processus du filtrage collaboratif }}
\vspace*{3mm}
\newline
\textbf{Évaluation des recommandations : }
pour pouvoir fonctionner, le système a besoin de collecter des données sur les utilisateurs et leurs préférences, cette collecte peut se faire de deux façons : Collecte explicite et Collecte implicite
\begin{itemize}
\item \textbf{Collecte explicite :} dans ce cas, les utilisateurs sont sollicités pour émettre leurs avis sur des produits/objets. Ils peuvent le faire via un système de notation (ex. une grille de 5 étoiles, un questionnaire de satisfaction), ou bien en publiant leurs avis sur un élément donné (ex. La fonctionnalité "J'aime" sur le réseau social Facebook permet aux utilisateurs d'exprimer leur intérêt pour un élément donné). 
\item \textbf{Collecte implicite :} s'intéresse aux interactions des utilisateurs sur le système. Les exemples de cette collecte incluent la surveillance du nombre de visites sur une page, le nombre de vues sur une vidéo, le temps passé sur une section donnée ou de l'historique des achats sur une plateforme de e-commerce.  
\end{itemize}
\textbf{Production des recommandations : }
une fois la communauté de l'utilisateur est crée, le système prédit l'intérêt qu'un document particulier peut présenter pour l'utilisateur en s'appuyant sur les évaluations que les membres de la communauté ont faites sur ce même document. Lorsque l'intérêt prédit dépasse un certain seuil, le système recommande le document à l'utilisateur.
\vspace*{3mm}
\newline
\textbf{Profil de l'utilisateur : }
le profil de l'utilisateur est composé d'un ensemble des prédicats pondérés. Ce profil s'enrichit progressivement au fur et à mesure que l'utilisateur évalue des documents reçus. Outre les informations d'identification de base, le profil de l'utilisateur peut regrouper des informations très diverses selon les besoins.
\begin{center}
   \begin{tabular}{|c||p{7cm}|}
        \hline           
 \textbf{ Informations  d'identification }  & \textbf{ \centering  Description} \\
         \hline  \hline
        \textbf{Les caractéristiques personnelles} & Ces caractéristiques peuvent influencer fortement l'interaction (âge, sexe, etc.).  \\
     \hline
  \textbf{Sécurité} & Cette dimension est le niveau de confidentialité concernant tous les autres critères.\\
\hline
 \textbf{Livraison} & On peut citer la modalité de livraison des informations comme le format, le standard,  le volume, le mode de visualisation et le délai.\\

        \hline 
\textbf{Historique des interactions avec le service} & Cet historique permet de modéliser les habitudes comportementales.\\
        \hline 
\end{tabular} 
\captionof{table}{Description des informations d'identification}
\label{tab2}
\end{center}
\textbf{Communautés : }
la notion des communautés est définie comme le regroupement des utilisateurs en fonction de l'historique de leurs évaluations, afin que le système calcule des recommandations. Selon cette optique, les profils sont un facteur interactif alors  que les communautés sont considérées comme un facteur interne du système.

\textbf{Calcul de la prédiction pour un système de filtrage collaboratif : } 
l'exploitation des données disponibles dans un système de filtrage peut se faire de plusieurs manières. Ces méthodes sont classées en deux principales familles : les algorithmes basés sur la mémoire et les algorithmes basés sur le modèle.
\newline 
\textbf{Algorithmes basés sur la mémoire :} ils utilisent l'ensemble de la base de données des évaluations des utilisateurs pour faire les prédictions : les évaluations de l'utilisateur actif sont prédites à partir d'informations partielles concernant l'utilisateur actif, et un ensemble de poids calculés à partir de la base de données des évaluations des utilisateurs.
\begin{itemize}
\item \textbf{\textit{Filtrage collaboratif basé sur la mémoire (utilisateurs) :}}
les systèmes basés sur le voisinage utilisateur, évaluent l'intérêt d'un utilisateur pour un item en utilisant les notes de cet item. Ces notes sont données par d'autres utilisateurs, appelés voisins, qui ont des habitudes de notation similaires.
En se basant sur le profil d'un utilisateur $ u_{i} $, le système recherche les utilisateurs $ u_{j} $ ($j$ diffère de $i$) qui lui sont les plus similaires. Alors, Les deux mesures de similarité qui sont très utilisées sont : la corrélation de Pearson et la similarité vectorielle.
\newline 
\textbf{La corrélation de Pearson : } la corrélation de Pearson est une méthode issue des statistiques. Elle est aussi très utilisée dans le domaine des systèmes de recommandation pour mesurer la similarité entre deux utilisateurs. La formule suivante, nous donne cette valeur pour deux utilisateurs A et B :
\vspace{-3em}
\begin{center}
\begin{equation} 
Sim(A,B) = \frac {\sum_{j}(v_{A,j} - v_{\bar{A},j})(v_{B,j} - v_{\bar{B},j})}{\sum_{j}(v_{A,j} - v_{\bar{A},j})^2(v_{B,j} - v_{\bar{B},j})^2}
\end{equation}
\end{center}
\vspace{-1em}
\begin{table}[h!]
\begin{center}
\begin{tabular}{|c||c|}
\hline 
\textbf{Notation} &  \textbf{Signification} \\
\hline \hline 
\textbf{$j$} & Indice d'objets ayant été voté à la fois par A et B \\
\hline
\textbf{$ v_{A,j}$} & Vote de A pour l'item $j$ \\
\hline
\textbf{$ v_{\overline{A},j}$} & Moyenne des votes de A  \\ 
\hline
\end{tabular}
\end{center}
\caption{Notations utilisées dans la méthode " Corrélation de Pearson "}
\label{tab3}

\end{table}
\textbf{Cosinus des vecteurs : } dans cette méthode, les utilisateurs A et B sont considérés comme deux vecteurs de même origine dans un espace de $m$ dimensions, $m$ est égal au nombre d'items évalués par les deux utilisateurs. 
\newline
Empiriquement, la similarité entre ces deux utilisateurs est calculée par la formule du Cosinus suivante :
\vspace{-3em}
\begin{center}
\begin{equation}
Sim(A,B) = \sum_{j=1}^{n} \frac{ v_{A,j} } {\sqrt{ \sum_{j=1}^{n}  v ^2_{A,j}}}  \times \frac{ v_{B,j} } {\sqrt{ \sum_{j=1}^{n}  v ^2_{B,j}}}
\end{equation}
\end{center}
 \vspace{-1em}
 \begin{table}[h!]
\begin{center}
\begin{tabular}{|c||c|}
\hline 
\textbf{Notation} &  \textbf{Signification} \\
\hline \hline 
\textbf{$n$} & Nombre d'items communs entre A et B votés par $v$\\
\hline
\textbf{$ v_{A,j}$} & Vote de A pour l'item $j$ \\
\hline
\textbf{$ v_{B,j}$} & Vote de B pour l'item $j$  \\ 
\hline
\end{tabular}
\end{center}
\caption{Notations utilisées dans la méthode " Cosinus des vecteurs "}
\label{tab4}
\end{table}
\textbf{La distance de Spearman : } est équivalente à la distance de Pearson, mais au lieu d'utiliser les évaluations comme critère pour la distance, elle utilise le classement des préférences. Si un utilisateur a évalué 20 items, l'item  préféré a une note de 20 et l'item le moins préféré a une note de 1. 
\newline  
Une fois que toutes les similarités de l'utilisateur cible A par rapport aux autres utilisateurs sont calculées et que les n utilisateurs les plus similaires qui constituent le voisinage de cet utilisateur cible sont définis, la prédiction de la valeur d'un item j évaluée par l'utilisateur A ($P_{A,j}$) est calculée à l'aide de la somme pondérée des estimations des voisins les plus proches qui ont déjà estimé l'item j :
\vspace{-3em}
\begin{center}
\begin{equation}
P_{A,j} = \sum_{i=1}^{n} \frac{Sim(A,i) \times  v_{i,j} } {\sum_{i=1}^{n} Sim(A,i)} 
\end{equation}
\end{center}
\vspace{-1em}
\begin{table}[h!]
\begin{center}
\begin{tabular}{|c||c|}
\hline 
\textbf{Notation} &  \textbf{Signification} \\
\hline \hline 
\textbf{$n$} & Nombre d'utilisateurs présents dans le voisinage de A, \\
& ayant déjà voté sur l'item $j$\\
\hline
\textbf{$ v_{i,j}$} & Vote de l'utilisateur $i$ pour l'objet $j$ \\
\hline
\end{tabular}
\end{center}
\caption{Notations utilisées de " La distance de Spearman "}
\end{table}
\label{tab5}
\item \textbf{\textit{Filtrage collaboratif basé sur la mémoire (items):}} alors que les méthodes basées sur le voisinage utilisateur s'appuient sur l'avis d'utilisateurs partageant les mêmes idées pour prédire une note, les approches basées sur les items prédisent la note d'un utilisateur $u$ pour un item $i$ en se basant sur les notes de $u$ pour des items similaires à $i$.
Les choix possibles pour calculer la similarité $|sim (i, j)|$ entre les items $i$ et $j$ sont aussi la corrélation Pearson et la similarité vectorielle \cite{arnautu2013mures}.
\newline 
\textbf{Similarité vectorielle} la similarité vectorielle se sert de l'estimation moyenne d'utilisateur de chaque paire évaluée, et fait face à la limitation de la similarité vectorielle. Empiriquement, la similarité entre deux items est calculée par la formule du Cosinus suivante : 
\vspace{-3em}
\begin{center}
\begin{equation}
Sim(i, j) = \frac{(v_{A,i} - v_{\bar{A}})}{\sqrt{ \sum_{A=1}^{m} (v_{A,i} -v_{\bar{A}})^2}}  \times \frac{(v_{A,j} - v_{\bar{A}})}{\sqrt{ \sum_{A=1}^{m} (v_{A,j} -v_{\bar{A}})^2}} 
\end{equation}
\end{center}
\vspace{-1em}
\begin{table}[h!]
\begin{center}
\begin{tabular}{|c||c|}
\hline 
\textbf{Notation} &  \textbf{Signification} \\
\hline \hline 
\textbf{$m$} & Nombre d'utilisateurs qui ont votés pour les deux items \\
\hline
\textbf{$ v_{A,i}$} & Vote de A pour l'item $i$\\
\hline 
\textbf{$ v_{A,j}$} & Vote de A pour l'item $j$\\
\hline 
\textbf{$v_{\bar{A}}$} & Moyenne des votes de l'utilisateur A\\
\hline 
\end{tabular}
\end{center}
\caption{Notations utilisées de la méthode " La similarité vectorielle "}
\label{tab6}
\end{table}
\newpage
Une fois que la similarité parmi les items ait été calculée, la prochaine étape est de prévoir pour l'utilisateur cible A, une valeur pour l'item actif i. Une manière commune est de capturer comment l'utilisateur a évalué les items similaires \cite{arnautu2013mures}. La valeur prévue est basée sur la somme pondérée des estimations de l'utilisateur ainsi que les déviations des estimations moyennes et peut être calculée à l'aide de la formule suivante :
\vspace{-3em}
\begin{center}
\begin{equation}
P_{A,i} = \bar{v}_{i} + \frac{\sum_{j=1}^{m} Sim(A, B) \times (v_{A,j} - \bar{v}_{j})}{\sum_{j=1}^{m}|Sim(i,j)|}
\end{equation}
\end{center}
\vspace{-1em}
\begin{table}[h!]
\begin{center}
\begin{tabular}{|c||c|}
\hline 
\textbf{Notation} &  \textbf{Signification} \\
\hline \hline 
\textbf{$m$} &  Nombre d'items présents dans le voisinage de item i,\\
 & ayant déjà été voté par l'utilisateur A\\
 \hline 
\textbf{$ v_{A,j}$} & Vote de l'utilisateur A pour l'objet j\\
\hline 
\textbf{$v_{\bar{j}}$} & Moyenne des votes pour l'item j\\
\hline 
\textbf{$|Sim(i, j)| $} & Similarité moyenne\\
\hline
\end{tabular}
\end{center}
\caption{Notations utilisées de " La similarité moyenne "}
\label{tab7}
\end{table}
\textbf{Algorithmes basés sur le modèle :} 
ce type d'algorithme est comme le nom l'indique est basé sur des modèles, supposés réduire la complexité. Ces modèles utilisent la base de données des évaluations des utilisateurs pour estimer ou apprendre un modèle qui est alors utilisé pour les prédictions. Ils peuvent être basés sur des classificateurs permettant de créer des classes pour réduire la complexité.
Exemple d'algorithmes utilisés : (Modèle de Clustering, K-Means, RecTree.)
\end{itemize}
\item \textbf{\underline{Algorithme général d'un système de filtrage collaboratif}}
\vspace*{3mm}
\newline
L'algorithme général d'un système de filtrage collaboratif suit les étapes suivantes (\cite{adomavicius2005toward}, \cite{berrut2003filtrage}):
\begin{itemize}
\item Collecter les appréciations de l'utilisateur sur les documents qu'il consulte.
\item Intégrer ces informations dans le profil de l'utilisateur.
\item Utiliser ce profil pour aider l'utilisateur dans ces prochaines recherches d'information.
\end{itemize}
\item \textbf{\underline{Architecture générale d'un système de filtrage collaboratif}}
\vspace*{3mm}
\newline
L'architecture générale d'un système de filtrage collaboratif s'articule autour de deux fonctionnalités centrales :
\begin{itemize}
\item Le calcul de la proximité entre les utilisateurs.
\item Le calcul de la prédiction de l'évaluation qu'un utilisateur fera d'un document. 
\end{itemize}
S'ajoute la fonctionnalité de mise à jour perpétuelle des profils d'utilisateurs, au fur et à mesure de la collecte de leurs évaluations.
\begin{figure}[H]
   \centering
   \includegraphics[width=11cm]{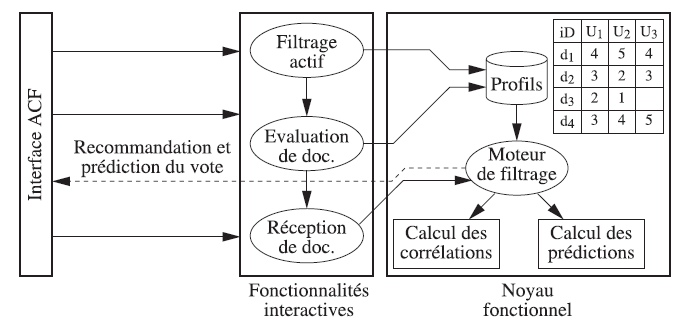}
    \caption{Architecture générale d'un système de filtrage collaboratif \cite{berrut2003filtrage}}
       \label{fig7:my_label}
\end{figure}
\end{enumerate}
\subsubsection{Étude comparative des méthodes collaboratives et des méthodes basées sur le contenu}
\hspace*{3ex} Les différentes techniques exploitées par les systèmes de recommandation ont chacune leurs apports mais aussi leurs limites. Le tableau \ref{tab8} présente une synthèse comparant les avantages et les inconvénients des techniques de recommandation qui ont été présentées dans cette section.
\begin{center}
\begin{tabular}{|p{3cm}||p{4cm}|p{4cm}|p{5cm}|}
 \hline 
\textbf{Catégorie }& \textbf{Exemples d'algorithmes utilisés} & \textbf{Avantages} & \textbf{Inconvénients} \\
 \hline  \hline 
\textbf{Technique basée sur le contenu} &  
\begin{itemize}
\item Analyse de similarité de contenu (TF/IDF)
\item Clustering
\item Arbres de décision
\end{itemize} &
\begin{itemize}
\item Amélioration de la qualité des recommandations
\item Réduction du problème de manque de données 
\end{itemize} & 
\begin{itemize}
\item Manque de diversité des recommandations
\item Nécessité d'indexation de contenus (extraction d'attributs représentatifs)
\item Problème d'indexation de documents multimédia 
\end{itemize} \\
\hline 
\textbf{FC basé sur la mémoire} &
\begin{itemize}
\item FC exploitant l'approche kNN (basée sur l'utilisateur ou sur l'item)  
\item Utilisation des mesures Pearson ou cosinus
\end{itemize} &
\begin{itemize}
\item Implémentation simple 
\item Intégration facile de nouvelles données
\item Précision des recommandations 
\end{itemize}
&
\begin{itemize}
\item Dépendance aux données de notes
\item Détérioration de la qualité de recommandations à cause du manque de données
\item Problème de passage à l'échelle
\end{itemize} \\ 
\hline 
\end{tabular}
 \end{center}
 \newpage 
\begin{center}
\begin{tabular}{|p{3cm}||p{4cm}|p{4cm}|p{5cm}|}
 \hline
 
\textbf{FC basé sur un modèle} &
\begin{itemize}
\item Clustering
\item Approches probabilistes (réseaux bayésiens)
\item Méthodes de réduction de dimensionnalité (SVD, PCA)
WUM (règles d'association, motifs séquentiels, modèles de Markov)
\end{itemize}
& 
\begin{itemize}
\item Amélioration de la qualité des recommandations
\item Réduction du problème de manque de données
\item Prédiction des futurs comportements de navigation
\end{itemize}
& 
\begin{itemize}
\item Construction coûteuse de modèles 
\item Risque de perte d'information pertinente dû à la réduction de dimensionnalité 
\item Problème de calcul des règles ou de motifs quand le système manque de données 
\item Pas de considération du profil utilisateur (pour les modèles du WUM)
\end{itemize}
  \\
  \hline 
\end{tabular}
 \end{center} 
\captionof{table}{Synthèse comparative des techniques de recommandation}
\label{tab8}

\subsubsection{Les méthodes hybrides}
\hspace*{3ex} Constatant les avantages et inconvénients de chacune des deux approches ci-dessus, on comprend que de nombreux systèmes reposent sur leur combinaison, ce qui en fait des systèmes de filtrage dits \textbf{" hybrides "}.
\newline
Plus généralement, les systèmes hybrides gèrent des profils d'utilisateurs orientés contenu, et la comparaison entre ces profils donne lieu à la formation de communautés d'utilisateurs permettant le filtrage collaboratif. En général, l'hybridation s'effectue en deux phases : 
\begin{itemize}
\item Appliquer séparément le filtrage collaboratif et autres techniques de filtrage pour générer des recommandations.
\item Combiner ces ensembles de recommandations préliminaires selon certaines méthodes telles que la pondération, la cascade, la commutation, etc., afin de produire les recommandations finales pour les utilisateurs \cite{nguyen2006cocofil2}.
\end{itemize}
\section{Limites des systèmes de recommandation}
\hspace*{3ex} Malgré leur popularité croissante, les systèmes de recommandation ont des apports mais ont aussi des limites.
\begin{itemize}
\item \textbf{\underline{Démarrage à froid :}} souvent, on se trouve confronté au problème qu'un utilisateur ne soit comparable avec aucun autre. Ce problème est du au fait que peu ou pas d'utilisateurs ont évalué un article donné, ou qu'un utilisateur donné a évalué très peu ou pas d'articles. Généralement, ce problème survient quand un nouvel utilisateur ou une nouvelle ressource est ajoutée à la base de la recommandation.
\item \textbf{\underline{Masse critique :}} afin de former de meilleures communautés, le système exige un nombre suffisant des évaluations en commun entre les utilisateurs pour les comparer entre eux.
Malgré  la taille énorme de l'ensemble des documents dans les systèmes, le nombre des évaluations en commun entre les utilisateurs risque d'être faible.
\item \textbf{\underline{Principe d'induction :}}
les systèmes de recommandation se basent sur le principe qu'un utilisateur qui a offert un comportement dans le passé tendra à offrir un comportement semblable dans le futur. Cependant, ce principe n'est pas nécessairement vrai dans le contexte réel. Par exemple, un utilisateur peut changer complètement d'intérêt ou en avoir plusieurs. 
\item \textbf{\underline{Centralisé ou distribué :}} les systèmes de recommandation tels qu'Amazon, sont basés sur une architecture centralisée. En effet, le moteur de recommandation centralisé permet de sauvegarder le profil de l'utilisateur et le calcul des recommandations dans un serveur central, par contre, les systèmes de recommandation centralisés souffrent de plusieurs problèmes tels que : le coût, la robustesse, la sécurité, la portabilité, etc. Une des solutions à ces problèmes est de répartir le système.
Alors un système de recommandation centralisé pourrait être conçu pour tirer profit de la puissance de calcul disponible sur les ordinateurs des utilisateurs.
\item \textbf{\underline{Sécurité ou crédibilité :}}
les systèmes de recommandation ne peuvent pas empêcher les actes de tromperie. Il est difficile de contrôler l'identité des utilisateurs et de pénaliser le comportement malveillant. Par conséquent, Il est indispensable d'avoir des moyens permettant à chaque utilisateur de décider en quels utilisateurs et en quels contenus avoir confiance.
\item \textbf{\underline{Protection de la vie privé :}}
la protection des informations sensibles constituant le profil de l'utilisateur est considérée comme un autre problème qui touche les systèmes de recommandation. Vu la nature de l'information, ces systèmes doivent assurer une telle protection. Ainsi, des moyens de préserver l'anonymat des utilisateurs et chiffrer les données transmises sont nécessaires. 
\end{itemize}
\section{Les systèmes de recommandation sensibles au contexte}
\hspace*{3ex} Les systèmes de recommandation jouent un rôle important dans la manipulation de grandes quantités d'information. Souvent, le contenu et les objets qui pourraient intéresser une personne, dépendant de sa situation spécifique : l'emplacement actuel, la saison, le rôle de l'utilisateur, l'heure, lieu, la compagnie d'autres personnes par exemple (pour regarder des films ou aller au restaurant), etc. Les systèmes de recommandation sensibles aux contextes tentent d'exploiter l'utilisation du contexte afin d'améliorer le processus de génération des recommandations.
\subsection{Définition}
Gorgoglione et al.\cite{gorgoglione2011effect} ont conclu que l'utilisation du contexte dans les systèmes de recommandation avait donné plus de confiance dans les recommandations. Cette augmentation du confiance conduit à son tour à des clients prêts à payer des prix plus élevés pour les produits, ce qui améliore les ventes. 
\newline
\hspace*{3ex} Avant de discuter du rôle et de l'apport des informations contextuelles dans le système de recommandation, nous commençons par discuter la notion générale de contexte. Puis, nous nous concentrons sur les systèmes de recommandation et expliquons comment le contexte est spécifié et modélisé ici.

\hspace*{3ex} Voici quelques travaux des systèmes de recommandation sensibles au contexte réalisés au sein  de notre laboratoire LIPAH concernant les  : Imen et al. (2012) \cite{sassi2012situation}; (2013) \cite{sassi2013recherche}; (2017) \cite{sassi2017fuzzy}; Hazem et al. (2017) \cite{souid2017hypergraph}.
\subsection{Contexte}
Le contexte est un concept à multiples facettes qui a été étudié dans différentes disciplines de recherche, y compris l'informatique (principalement dans l'intelligence artificielle et l'informatique ubiquitaire), les sciences cognitives, la linguistique, la philosophie, la psychologie et les sciences organisationnelles.
\newline
Abowed, Day et al \cite{abowd1999towards}  
\textit{\textbf{"Le contexte est n'importe quelle information qui peut être utilisé pour caractériser une situation d'une entité. Une entité peut être une personne, un endroit ou un objet que l'on considère comme étant pertinent à l'interaction entre un utilisateur et une application y compris ces deux derniers"}}
\newline
Dey el al,.\cite{dey2001understanding} ont défini le contexte : 
\begin{itemize}
\item Comme étant toute information qui peut être utilisée pour caractériser la situation d'une entité. Une entité est une personne, un lieu ou un objet considéré comme pertinent pour l'interaction entre un utilisateur et une application, y compris l'utilisateur et les applications elles-mêmes.
\item Un système est dit sensible au contexte s'il utilise un contexte pour fournir des informations et/ou des services pertinents à l'utilisateur, lorsque la pertinence dépend de la tâche de l'utilisateur.
\end{itemize} 
\subsubsection{Dimensions du contexte}
\hspace*{3ex} M.Daoud \cite{daoud2007recherche} a proposé deux dimensions principales du contexte de recherche: 
\begin{itemize}
\item \textbf{Le contexte lié à l'utilisateur} : c'est l'ensemble d'éléments qui peuvent être donnés explicitement par l'utilisateur ou implicitement déduits par le système. Il existe des préférences qui sont  données explicitement et comportent les deux aspects suivants : 
\begin{itemize}
\item \textbf{\textit{Le fraîcheur}} : l'utilisateur peut s'intéresser à l'information la plus récente ou bien à des documents liés à des dates  bien précises.
\item \textbf{\textit{La granularité}} : l'utilisateur peut s'intéresser à un certain niveau de détail de la réponse attendue, cela peut déterminer la structure de la réponse retournée.
\end{itemize}
\item \textbf{Le contexte lié à la requête} : On peut définir ce contexte par différents paramètres descriptifs et mesurables à partir de la requête ou bien à partir du profil de n top documents retournés par la requête.
\begin{itemize}
\item \textbf{\textit{Clarté de la requête}} : ce paramètre quantifie le degré d'ambiguïté de la requête et permet de mieux cibler la recherche.
\item \textbf{\textit{Distribution de la datation des documents}} : c'est la distribution de n meilleurs documents selon le paramètre date, et qui permet de mettre en évidence le fraîcheur de l'information.
\item \textbf{\textit{Degré de couverture de la requête par les services web}} : ce paramètre permet d'orienter la réponse retournée vers un document ou service.  
\end{itemize}
\end{itemize}
\subsubsection{Pertinence de l'information contextuelle}
\hspace*{3ex} Les informations contextuelles ne sont pas toutes pertinentes pour la formulation des recommandations
\cite{adomavicius2011context} par exemple quelle est l'information contextuelle pertinente au moment de recommander un livre.
\newline 
\hspace*{3ex} La détermination de la pertinence des informations contextuelles se fait selon les deux manières suivantes:
\begin{itemize}
\item Manuellement : se fait à l'aide de la connaissance du domaine du concepteur du système de recommandation.
\item Automatiquement : se fait en utilisant des procédures de sélection de l'apprentissage automatique \cite{koller1996toward}, Fouille de données \cite{liu2012feature} et des statistiques \cite{chatterjee2015regression}. 
\end{itemize}
\subsection{Modélisation des informations contextuelles dans les systèmes de recommandation}
\hspace*{3ex} Dans cette partie, nous allons nous intéresser aux systèmes traditionnels, ainsi qu'aux systèmes de recommandation contextuels. 
\subsubsection{Les systèmes traditionnels ou bidimensionnels 2D}
\hspace*{3ex} Le processus de recommandation commence généralement par la spécification de l'ensemble initial d'évaluations qui est explicitement fourni par les utilisateurs ou implicite dans le système. 
\newline 
Une fois ces évaluations initiales ont été précisées, un système de recommandation tente d'estimer la fonction des évaluations \textbf{R} :
\vspace{-3em}
\begin{center}
\begin{equation}
\bm{ R : User \times Item \to Rating }
\end{equation}
\end{center} 
\vspace{-1em}
\hspace*{3ex} Pour les paires (User, Item) qui n'ont pas encore été évaluées par les utilisateurs. Une fois que la fonction \textbf{R} est estimée pour l'ensemble de l'espace Utilisateur \textbf{x}, un système de recommandation peut recommander l'article le mieux noté pour chaque utilisateur. Nous appelons de tels systèmes traditionnels ou bidimensionnels (2D) puisqu'ils ne prennent en compte que les dimensions User et Item dans le processus de recommandation.
\newline 
\hspace*{3ex} 
Les systèmes de recommandation fournissent aux utilisateurs des suggestions personnalisées de produits ou de services. Ils jouent un rôle important dans le succès du commerce électronique et sont utilisés dans la plupart des sites Web de vidéos tels que YouTube et Hulu. De nombreux réseaux sociaux permettent aux utilisateurs de noter des vidéos ou des films. En effet, les contributions de notre laboratoire LIPAH ont atteint un niveau avancé dans ce domaine, ce qui va énormément nous aider. Citons en quelques travaux intéressants : 
(Chiraz et al. (2011) \cite{trabelsi2011folksonomy}, \cite{trabelsi2011auto}; (2012) \cite{trabelsi2012hmm, trabelsi2012scalable, trabelsi2012bgrt}; (2013) \cite{trabelsi2013integrated}; (2016) \cite{trabelsi2016harnessing}).
\subsubsection{Les systèmes de recommandation contextuels }
\hspace*{3ex} La plupart des recherches effectuées se concentrent sur la recommandation d'items aux utilisateurs ou items aux utilisateurs et ne prennent pas en considération d'autres informations contextuelles telles que l'heure, le lieu et la compagnie d'autres personnes (par exemple pour regarder des films).
\newline 
\hspace*{3ex} Alors nous explorons le domaine du système de recommandation contextuelle avec la modélisation et la prédiction des goûts et des préférences des utilisateurs en incorporant les informations contextuelles disponibles dans le processus de recommandation en tant que des catégories explicites de données supplémentaires. Ces préférences et goûts à long terme sont généralement exprimés sous la forme d'évaluations et sont modélisés en fonction non seulement des items et des utilisateurs, mais aussi du contexte.
\newline 
\hspace*{3ex} Contrairement au modèle traditionnel, les systèmes de recommandation contextuels tentent d'incorporer ou d'utiliser des preuves supplémentaires (au-delà des informations sur les utilisateurs et les items) pour estimer les préférences des utilisateurs sur des éléments non vus. 
\vspace{-5em}
\begin{center}
\begin{equation}
\bm{ R : Users \times Items\times Contexts\to Rating }
\end{equation}
\end{center}  
\vspace{-1em}
\hspace*{3ex} L'intégration de ce contexte dans les systèmes de recommandation peut se faire de trois façons: (le pré-filtrage contextuel, le post-filtrage contextuel et la modélisation contextuelle).
\begin{enumerate}[label=\alph*)]
\item \textbf{\underline{Le pré-filtrage contextuel}}
dans l'approche de pré-filtrage contextuel, l'information contextuelle est utilisée comme une étiquette permettant de filtrer les évaluations qui ne correspondent pas à l'information contextuelle spécifiée. Cela se fait avant que la méthode principale de recommandation soit lancée sur le reste de données sélectionnées.
\newline 
Si un contexte d'intérêt particulier est \textbf{$C$}, alors cette méthode sélectionne à partir de la série initiale toutes les évaluations relatives au contexte spécifié \textbf{$C$}, et elle génère la matrice \textbf{"User} x \textbf{Item"} ne contenant que les données relatives au contexte \textbf{$C$}. Puis, la méthode des systèmes de recommandation, comme le filtrage collaboratif, est lancée sur la base de donnée réduite afin d'obtenir les recommandations liées au contexte \textbf{$C$} (Voir la figure \ref{fig9:my_label}).
\begin{figure}[H]
\centering
\includegraphics[width=8cm]{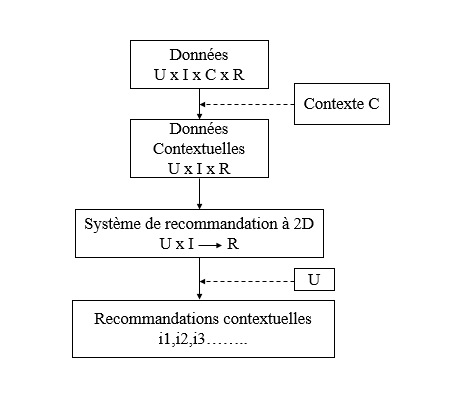}
\caption{Pré-filtrage contextuel \cite{adomavicius2015context}}
\label{fig9:my_label}
\end{figure}
\newpage
\item \textbf{\underline{Le post-filtrage contextuel}}
l'information contextuelle est utilisée après le lancement de la méthode principale de recommandation à deux dimensions (2D). Une fois les évaluations inconnues sont estimées et les recommandations sont produites, le système analyse les données pour un utilisateur donnée dans un contexte précis pour trouver les modèles  d'utilisation des articles spécifiques, et utilise ces modèles pour contextualiser les recommandations obtenues à partir de la méthode classique de recommandation (2D), comme le filtrage collaboratif (Voir la figure \ref{fig10:my_label}).
\begin{figure}[H]
\centering
\includegraphics[width=8cm]{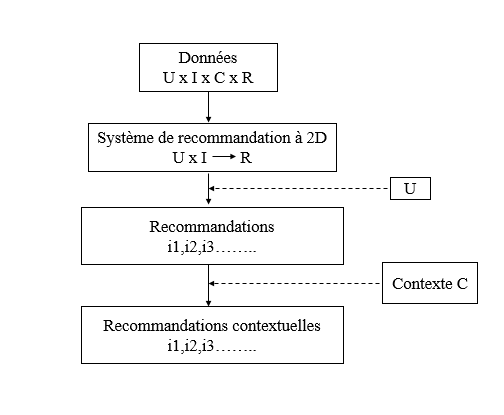}
\caption{Post-filtrage contextuel \cite{adomavicius2015context}}
\label{fig10:my_label}
\end{figure}
\item \textbf{\underline{La modélisation contextuelle}}
est une méthode dans laquelle l'information contextuelle est utilisée directement à l'intérieur des algorithmes de génération de recommandation.
La figure \ref{fig11:mylabel} montre l'intégration du contexte dans le processus de recommandation.
\begin{figure}[H]
\centering
\includegraphics[width=8cm]{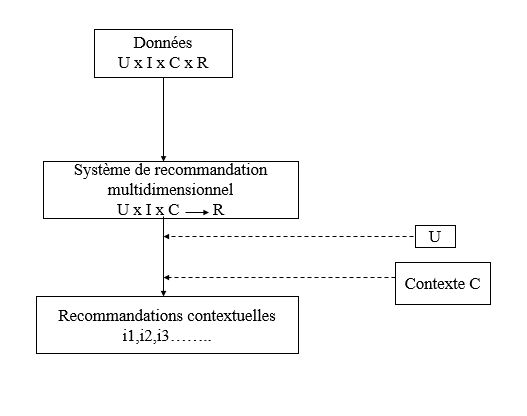}
\caption{Modélisation contextuelle \cite{adomavicius2015context}}
\label{fig11:mylabel}
\end{figure}
\end{enumerate}

\section{Conclusion}
\hspace*{3ex}Dans ce chapitre, nous avons d'abord présenté dans la première partie la notion des systèmes de recommandation en détaillant les différentes techniques utilisées par ces systèmes. Ensuite, nous avons défini la notion de profil utilisateur avec ses deux types de collecte des données (explicite et implicite). Puis, nous avons terminé par citer quelques problèmes rencontrés par les systèmes de recommandation classiques. Parmi ces problèmes, le fait que ces systèmes ne tiennent en compte du contexte dans lequel se trouve l'utilisateur lorsqu'il décide de faire une recommandation. Ainsi dans ce chapitre, nous avons mis l'accent sur les systèmes sensibles au contexte, qui apportent une solution pour le problème de contexte.

\chapter{État de l'art}
\begin{quote}
\begin{center}
\textit{“There’s too many men  \\ 
Too many people \\
Making too many problems\\
And not much love to go round \\
Can’t you see\\
This is \textbf{a land of confusion}”\\
— Genesis, Land of confusion, 1986}

\end{center}
\end{quote}
\section{Introduction}
\hspace*{3ex} Dans ce chapitre, nous allons présenter les principales approches proposées des systèmes de recommandation contextuels dans le domaine des films et nous allons terminer par une étude comparative de ces approches. 

\section{Approche de  Ostuni et al}
\hspace*{3ex} Dans cette section, nous nous intéressons à l'approche de Ostuni et al. qui a été introduite dans le contexte d'utilisation des systèmes de recommandation sensibles au contexte. Cependant, ces auteurs arrivent à développer une application appelée \textbf{" Cinemappy "}, basée sur la localisation qui calcule les recommandations contextuelles de films.
\subsection{Contexte}
\hspace*{3ex} Grâce aux grands progrès technologiques réalisés ces dernières années, en particulier dans l'informatique ubiquitaire, 
les utilisateurs peuvent exécuter presque n'importe quelle sorte de demande (d'application) et exécuter presque n'importe quelle tâche sur des petits dispositifs mobiles. Les smartphones et les tablettes deviennent une plate-forme principale (primaire) pour l'accès à l'information \cite{ricci2010mobile}. 
\subsection{Objectif}
\hspace*{3ex} " Cinemappy " est une application qui met en œuvre un moteur de recommandation contextuelle. Cette application affine les résultats de recommandation d'un système de recommandation basé sur le contenu en exploitant des informations contextuelles liées à la position spatiale et temporelle actuelle de l'utilisateur. 
\subsection{Les techniques utilisées}
\hspace*{3ex} L'application de " Cinemappy " utilise certaines techniques pour suggérer des films à l'utilisateur.
\begin{itemize}
\item \textbf{DBPEDIA} : Le système choisit le graphique localisé dans DBPEDIA. En effet, DBpedia contient aussi des informations extraites des versions localisées de Wikipédia. Les données venant de ces sources Web sont représentées comme les graphiques RDF différents qui peuvent être facilement choisis via la clause d'une question de SPARQL. La figure \ref{fig12:my_label} décrit les composantes de base du système. 
\begin{figure}[H]
\centering
\includegraphics[width=14cm]{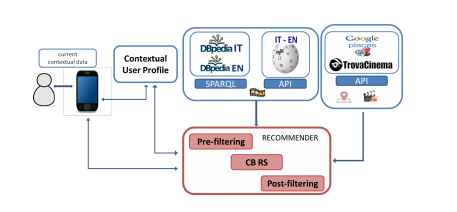}
\caption{Architecture du système \cite{ostuni2012cinemappy}}
\label{fig12:my_label}
\end{figure}
\textbf{Cinemappy} utilise plusieurs approches de recommandation pour suggérer des films et des salles de cinéma à l'utilisateur. Les auteurs tirent parti des informations contextuelles, des techniques de pré-filtrage et de post-filtrage pour les attributs contextuels. En particulier, pour modéliser l'attribut Companion, Ostuni et al utilisent ce que l'on appelle approche de micro-profilage \cite{baltrunas2009towards}, une technique de pré-filtrage particulière. Fondamentalement, avec le micro-profilage, Ostuni et al ont associé un profil différent à l'utilisateur en fonction du compagnon sélectionné.

\item \textbf{Pré-filtrage contextuel} : Avec Cinemappy, il y a une recommandation des films à regarder dans les cinémas. Pour cette raison, les films qui ne seront pas présentés dans le futur ne seront pas suggérés à l'utilisateur. Néanmoins, ces films seront considérés dans le profil de l'utilisateur si ce dernier les a évalués. De plus, pour la position temporelle et spatiale actuelle de l'utilisateur, Ostuni et al ont  contraint l'ensemble des films à recommander des critères géographiques et temporels. Pour chaque utilisateur $u$, l'ensemble des films $M_{u}$ est défini comme le contenant des films programmés dans les prochains jours dans les cinémas dans une plage de $k$ kilomètres autour de la position de l'utilisateur. La liste de recommandations finale sera calculée en considérant uniquement les articles disponibles en $M_{u}$.
Ce genre de restriction sur les éléments, en ce qui concerne le temps, est un pré-filtrage de l'ensemble des articles et non des classements comme cela arrive habituellement dans les approches de pré-filtrage.
\newline 
En ce qui concerne le contexte d'accompagnement, l'approche de micro-profilage est modélisée en considérant un profil spécifique pour utilisateur $u$ et pour chaque  compagnon $cmp$ : 
\newline 
\vspace{-2em}
\begin{center}
\textbf{ \textit{ profile ($u$, $cmp$) = {<$m_j$, $v_j$> | $v_j$ = 1 si $u$ aime $m_j$ avec compagnon  \\ $cmp$ $v_j$ = -1 sinon }}}
\end{center}
\item \textbf{Recommandation basée sur le contenu} : 
l'algorithme de recommandation est basé sur celui proposé dans \cite{di2012linked}, amélioré avec la gestion des micro-profils.
Afin d'évaluer si un film $m_{i}$ $\in$ $M_{u}$ pourrait être intéressant pour $u$ donné $cmp$. Ostuni et al. ont dû de combiner les valeurs de similarité liées à chaque propriété $p$ de $m_{i}$ et de calculer une valeur de similarité globale $\bar{r}_{preF}(u_{cmp}, m_{i})$ :
\vspace{-4em}
\begin{center} 
\begin{equation}
\bar{r}_{preF}(u_{cmp}, m_{i}) = \frac{\sum_{m_{j} \in profile (u, cmp)} v_{j} \times \frac{ \sum_{p}  \alpha_{p} \times sim ^{p}(m_{j} m_{i})}{p} } {|profile(u, cmp)|}
\end{equation}
\end{center}
\vspace{-1em}
\begin{center}
{\setlength {\tabcolsep}{12pt}
\begin{tabular}{|c||p{11cm}|}
\hline 
\textbf{Notation} &  \textbf{Signification} \\
\hline \hline 
\textbf{$P$} & Nombre de propriétés dans DBpedia qui sont considérées comme pertinentes. \\
\hline
\textbf{$profile (u, cmp)$} & La cardinalité du profil d'ensemble ($u$, $cmp$). \\
\hline
\textbf{$sim ^{p}(m_{j} m_{i}) $} & Représente la similarité entre les deux films $m_{i}$ et $m_{j}$ par rapport à une propriété $p$.\\
\hline
\textbf{$\alpha_{p} $} & C'est un poids qui est attribué à chaque propriété représentant sa valeur par rapport au profil de l'utilisateur. \\
\hline 
\end{tabular}
}
\end{center}
\captionof{table}{Notations utilisées dans "la similarité globale"}
\label{tab9}
La valeur est calculée en adaptant l'approche du modèle d'espace vectoriel à un paramètre basé sur RDF.
\item \textbf{Post-filtrage contextuel} : l'application du post-filtrage sur $R_{u, cmp}$ est effectuée pour reclasser ses éléments. En particulier, pour chaque critère, on introduit une variable dont la valeur est définie comme suit:
   \begin{itemize}[label=\textbullet,font=\color{black}]
\item \textbf{\textit{h (hierarchy)}}: Elle est égale à 1 si le cinéma est dans la même ville que la position courante de l'utilisateur, sinon 0.
\item \textbf{\textit{c (cluster)}}: Il est égal à 1 si le cinéma fait partie d'un cinéma multiplex, sinon 0.
\item \textbf{\textit{cl (co-location)}}: Il est égal à 1 si le cinéma est proche d'autres POI, sinon 0.
\item \textbf{\textit{ar (association-rule)}} : Il est égal à 1 si l'utilisateur connaît le prix du ticket, sinon 0. Cette information est implicitement tirée des informations sur le cinéma.
\item  \textbf{\textit{ap (anchor-point proximity)}} : Il est égal à 1 si le cinéma est proche de la maison de l'utilisateur ou du bureau de l'utilisateur, sinon 0.
\end{itemize}
\end{itemize}
\hspace*{3ex} Ces critères géographiques sont combinés avec $\bar{r}_{preF}(u_{cmp}, m_{i})$ pour obtenir un seul score :
\vspace{-3em}
\begin{center} 
\begin{equation}
\bar{r}(u_{cmp}, m_{i}) = \beta_{1} \times \bar{r}_{preF}(u_{cmp}, m_{i}) + \beta_{2} \frac{(a+c+c_{1}+a_{r}+a_{p})}{5}
\end{equation}
\end{center}
\vspace{-1em}
\hspace*{3ex} Où $\beta{1}$+ $\beta{2}$  = 1. Dans la mise en œuvre actuelle de \textbf{Cinemappy}, à la fois $\beta{1}$ et $\beta{2}$ ont été choisis expérimentalement et ont été fixés respectivement à 0,7 et 0,3.
\subsection{Étude expérimentale}
\begin{figure}[H]
\centering 
\includegraphics[width=4cm]{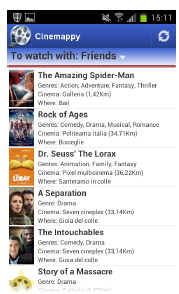}
\caption{Liste des films en fonction du profil utilisateur contextuel actuel \cite{ostuni2012cinemappy}}
\label{fig13:my_label}
\end{figure}

\hspace*{3ex} \textbf{Cinemappy} a été implémenté en tant qu'une application mobile pour les smartphones Android. Lorsque l'utilisateur démarre l'application, \textbf{Cinemappy} affiche une liste de films en fonction du profil utilisateur contextuel actuel (voir la figure\ref{fig13:my_label}).

\begin{figure}[H]
\centering
\includegraphics[width=4cm]{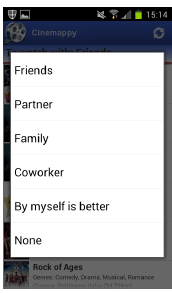}
\caption{Liste des options : "compagnon" \cite{ostuni2012cinemappy}}
\label{fig14:my_label}
\end{figure}

\hspace*{3ex} L'utilisateur peut choisir son compagnon actuel à partir d'une liste de différentes options permettant ainsi son micro-profil (voir la figure\ref{fig14:my_label}).

\begin{figure}[H]
\centering
\includegraphics[width=4cm]{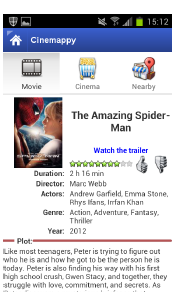}
\caption{Liste des films avec leurs descriptions \cite{ostuni2012cinemappy}}
\label{fig15:my_label}
\end{figure}

\begin{figure}[H]
\centering
\includegraphics[width=4cm]{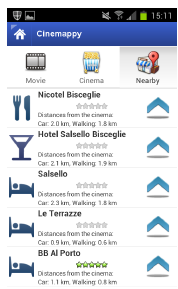}
\caption{Liste des salles de cinéma avec leurs emplacements \cite{ostuni2012cinemappy}}
\label{fig16:my_label}
\end{figure}

\begin{figure}[H]
\centering
\includegraphics[width=8cm]{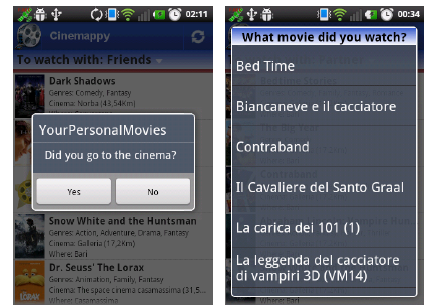}
\caption{La réponse de l'utilisateur et sa évaluation pour l'un des films présentés dans le cinéma \cite{ostuni2012cinemappy}}
\label{fig17:my_label1}
\end{figure}

\hspace*{3ex} \textbf{Cinemappy} est en mesure de proposer des films basés exclusivement sur des informations contextuelles. Pour chaque film de la liste, ses genres et la distance de la salle de cinéma  suggérée par rapport à la position de l'utilisateur sont affichés. Par conséquent, l'utilisateur peut cliquer sur l'un des films suggérés et regarder sa description, regarder sa bande-annonce et exprimer une préférence en termes de "je regarderais" / "je ne regarderais pas" (voir la figure \ref{fig15:my_label}).
\newpage
\hspace*{3ex} En outre, l'utilisateur peut trouver des informations sur la salle de cinéma recommandée ou les autres salles qui diffusent ce film. Selon l'emplacement de la salle de cinéma, l'utilisateur pourrait être intéressé par les endroits où passer du temps avec ses amis, comme les pubs, ou avec partenaire comme les restaurants ou les bars, ou avec sa famille, et dans ce cas l'utilisateur pourrait être intéressé par certains types d'endroits également appropriés pour les enfants. Pour supporter l'utilisateur dans ce choix, l'application suggère des PI en tenant compte des critères contextuels (voir la figure \ref{fig16:my_label}).
\hspace*{3ex} L'application, au moyen d'un service d'arrière-plan, capture la position de l'utilisateur. Si l'utilisateur a été pendant au moins 90 minutes dans une position similaire à proximité d'un cinéma dans une durée correspondante à un ou plusieurs films programmés, Ostuni et al ont supposé que l'utilisateur a regardé un film dans ce cinéma. Alors, dans ce cas, l'application demande à l'utilisateur s'il est allé au cinéma et si une réponse positive s'ensuit, l'utilisateur peut évaluer l'un des films présentés dans ce cinéma (voir la figure \ref{fig17:my_label1}).
\begin{itemize}[font=\color{black} \Large, label=\ding{43}]
\item Ostuni et al ont présenté Cinemappy qui est un système de recommandation contextuel basé sur le contenu pour les suggestions de films et des salles de cinéma. Ce système est alimenté avec des données provenant de graphiques DBpedia localisés et les résultats sont améliorés en exploitant les informations contextuelles sur l'utilisateur. 
\item L'application a été implémentée en tant qu'application Android. Des critères géographiques, qui vont au-delà de la simple distance géographique, ont été mis en œuvre pour exploiter pleinement les informations basées sur la localisation.
\end{itemize}

Dans ce contexte et avec la disponibilité des appareils mobiles et la diversité des fonctionnalités qu'ils offrent, les nouvelles approches en recherche d'information ont eu l'avantage de fournir aux utilisateurs des éléments plus pertinents et plus adaptés à leurs situations courantes \cite{sassi2013recherche}. Citons dans ce cas, quelques travaux qui sont effectués dans le laboratoire \textbf{LIPAH} concernant la prédiction des intérêts des utilisateurs afin d'enrichir leurs requêtes et d'élargir leurs cercles sociaux :  Imen et al. (2012, 2013) \cite{sassi2013recherche} \cite{sassi2012situation}.

\section{Approche de Campos et al}
\hspace*{3ex} Dans cette section, nous allons présenter l'approche de Campos et al. \cite{campos2013context} dans le cadre d'utilisation des systèmes de recommandation contextuelles. Les auteurs ont abordé ce problème en effectuant une comparaison empirique de plusieurs approches de pré-filtrage, de post-filtrage contextuel et de la modélisation contextuelle dans le domaine de recommandation des films.
\subsection{Contexte}
\hspace*{3ex} Les systèmes de recommandation suggèrent des articles aux utilisateurs qui s'appuient sur des préférences généralement exprimées sous la forme d'évaluations numériques des personnes ayant les mêmes idées.

\hspace*{3ex} Les systèmes de recommandation sensibles au contexte prennent également en compte des informations contextuelles (par exemple : le temps, l'emplacement, le compagnon social et l'humeur) associées aux préférences collectées. De cette façon, Ces systèmes peuvent distinguer l'intérêt d'un utilisateur dans différents contextes et situations.
\subsection{Objectif}
\hspace*{3ex} L'objectif de cette recherche est de répondre aux deux questions suivantes : 
\begin{itemize}
\item \textbf{\textit{Question 1}} : Est-ce que les approches des systèmes de recommandation contextuelles (Pré-filtrage contextuel, Post-filtrage contextuel, Modélisation contextuelle) sont capables de mieux prédire la note attribuée à un film dans un contexte particulier ?
\item  \textbf{\textit{Question 2}} : Quelle information contextuelle ou compagnon social (ou une combinaison des deux) fournit des informations plus utiles pour prédire des notes ?
\end{itemize}
\subsection{Les techniques utilisées} 
\hspace{3ex} Plusieurs approches ont été proposées pour traiter correctement les informations contextuelles. Adomavicius et al. \cite{adomavicius2005incorporating,adomavicius2015context} distinguent trois types principaux de CARS: 
 le pré-filtrage contextuel, le post-filtrage contextuel et la modélisation contextuelle. Alors l'évaluation de ces approches se fait comme suit : 
\begin{itemize}
\item \textbf{\textit{Dans le cas du pré-filtrage}} : Campos et al. ont utilisé la stratégie de pré-filtrage exacte proposée par Adomavicius et al\cite{adomavicius2005incorporating}, et la technique de division d'éléments proposée par Baltrunas et Ricci  \cite{baltrunas2014experimental, baltrunas2009context}.
\item \textbf{\textit{Dans le cas du post-filtrage}} : Campos et al. ont utilisé la stratégie de filtrage présentée par Panniello et al. dans \cite{panniello2009experimental}.
\item \textbf{\textit{Dans le cas de modélisation contextuelle}} : Campos et al. ont évalué plusieurs classificateurs développés par la communauté Machine Learning, notamment les algorithmes Naïve Bayes, Random Forest, MultiLayer Perceptron (MLP) et Support Vector Machine (SVM) \cite{nasrabadi2007pattern, breiman2001random}.
Tous les classificateurs ont été construits avec des vecteurs d'attributs basés sur le contenu correspondant aux informations sur l'utilisateur et le genre d'élément, et différents signaux contextuels.
\end{itemize}
\hspace{3ex} Les approches basées sur le pré-filtrage contextuel et celles qui sont basées sur le post-filtrage contextuel utilisent les classificateurs décrits dans le tableau \ref{tab10}.
\begin{center}
 {\setlength {\tabcolsep}{12pt}
    \begin{tabular}{|c||p{11cm}|}
    \hline
    \textbf{Nom du classifier}  & \textbf{Description} \\
    \hline \hline 
    \textbf{knearest
neighbor (KNN)} &  L'algorithme knearest neighbor (kNN) \cite{herlocker1999algorithmic} a été utilisé comme un algorithme de recommandation sous-jacent.\\
\hline
\textbf{Item Splitting (IS)} &  IS est une variante du pré-filtrage contextuel. Cette méthode divise les données de préférence pour les éléments en fonction du contexte  dans lequel ces données ont été générées.\\ 
        \hline 
\textbf{Matrix Factorization (MF)} & Est une factorisation d'une matrice en un produit de matrices.\\
\hline 
\end{tabular} 
}
\captionof{table}{Les classificateurs utilisés dans le pré- et post-filtrage}
\label{tab10}
\end{center}
\hspace*{3ex} Pour la méthode \textbf{Item Splitting (IS)} il y a quelques \textbf{Critères d'impureté} à définir : 
\begin{itemize}
\item \textbf{$ic_{IG}(i, s)$}: \textit{C'est un critère qui mesure le gain d'information donné par $s$ à la connaissance  de l'item $i$}.
\item \textbf{$ic_{M}(i, s)$} : \textit{Estime la signification statistique de la différence dans les moyennes des notations associées à chaque contexte dans s en utilisant le test t}.
\item \textbf{$ic_{p}(i, s)$} : \textit{Estime la signification statistique de la différence entre la proportion d'évaluations élevées et faibles dans chaque contexte de $s$ en utilisant le test z à deux proportions}.
\end{itemize}
\hspace*{3ex} Pour le pré-filtrage contextuel : Campos et al. utilisent les algorithmes de filtrage collaboratif kNN et factorisation matricielle (MF) \cite{koren2010collaborative} séparément comme stratégies de recommandation après IS. Cependant dans le post-filtrage contextuel (PoF), les prédictions d'évaluation sont générées par un algorithme qui ignore le contexte dans une première étape, puis les prédictions sont contextualisées en fonction du contexte cible. Campos et al. ont utilisé le même algorithme de prédiction d'évaluation kNN utilisé avec les approches de pré-filtrage. La contextualisation des prédictions d'évaluation a été réalisée par une stratégie de filtrage présentée dans \cite{panniello2009experimental}, qui pénalise la recommandation des items qui ne sont pas pertinents dans le contexte cible.

\hspace*{3ex} La pertinence d'un item $i$ pour l'utilisateur cible $u$ dans un contexte particulier $c$ est approchée par la probabilité $P_{c}(u, i, c) = \frac{|U_{u,i,c}|}{k}$  

\begin{center}
\begin{tabular}{|c||c|}
\hline 
\textbf{Notation} & \textbf{Description} \\ 
\hline \hline 
\textbf{K} & Nombre de voisins utilisés par kNN \\
\hline 
\textbf{$|U_{u,i,c}|$} & Les voisins de l'utilisateur v dans le voisinage de u, N(u), \\
& qui ont évalué l'item i dans son contexte c.\\
\hline  

\end{tabular}
\end{center}
\captionof{table}{Résultats d'évaluation pour la piste 1}
\label{tab11}
\hspace*{3ex} La pertinence des items est déterminée par une valeur seuil $\tau_{Pc}$ qui est utilisée pour contextualiser les évaluations comme suit : 
\begin{center}
$F(u,i,c)$ = $ \left\{\ \begin{matrix}
 F(u,i) &  si   P_{c}(u,i,c) \geq \tau_{Pc} \\ 
 F(u,i) - 0.5 & si P_{c}(u,i,c) < \tau_{Pc}  )
\end{matrix} \right. $
\end{center}
\hspace*{3ex} Les algorithmes de Machine Learning utilisés pour la modélisation contextuelle fournissent une distribution de score pour une note (étiquette de classe) dans l'espace des valeurs de notation 1, 2, 3, 4 et 5. En effet, les caractéristiques de l'utilisateur et de l'item analysées correspondent aux genres de films. Pour chaque utilisateur $u$, la valeur de l'attribut $a_{m}$ était le nombre d'éléments préférés de $u$ avec le genre $m$. Pour chaque item $i$, la valeur de l'attribut $a_{n}$ était 1 si $n$ est le genre correspondant , et 0 sinon.

\subsection{Étude expérimentale}
\begin{figure}[H]
\centering
\includegraphics[width=11cm]{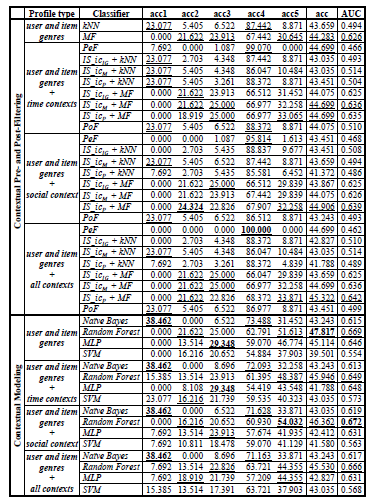}
\caption{Le modèle de PITF et les deux décompositions de Tucker et canonique \cite{campos2013context}}
\label{fig17:my_label}
\end{figure}

\hspace*{3ex} Une validation croisée 10 fois a été effectuée dans toutes les expériences. Dans les cas de pré-filtrage et de post-filtrage, Campos et al ont utilisé les implémentations \textbf{kNN} et \textbf{MF} fournies par le projet Apache Mahout3, avec $K$ = 30 et la corrélation de Pearson pour \textbf{kNN}, et 60 facteurs pour l'algorithme \textbf{MF}. Pour obtenir une couverture complète, dans les cas où un algorithme était incapable de calculer une prédiction, l'évaluation de l'ensemble de données a été fournie comme prévision.
\newline
\hspace*{3ex} Dans les cas de la modélisation contextuelle, Campos et al. ont  utilisé les implémentations des classificateurs fournies dans Weka4.
Ils ont calculé aussi l'exactitude des approches de recommandation évaluées en termes de taux de classification correcte pour chaque valeur de notation (acc1, acc2, acc3, acc4 et acc5), et le taux de classification correct global pondéré (acc) \cite{witten2016data}. Puis , ils ont également calculé la métrique Area under the Curve (AUC) \cite{ling2003auc}. 
\newline 
\hspace*{3ex} Le figure \ref{fig17:my_label} montre les meilleurs résultats obtenus pour chacune des approches testées sur  l'ensemble de données enrichi en contextes. Les résultats sont regroupés selon l'approche de contextualisation (pré-filtrage et post-filtrage ou modélisation contextuelle) et le type de données de profil fourni à chaque algorithme de recommandation. Dans les approches IS, Campos et al. ont testé différentes valeurs de seuil pour les critères d'impureté considérés.
\newline
\hspace*{3ex} Campos et al. ont conclu qu'il n'y a pas de CARS supérieur unique pour améliorer les prédictions d'évaluation dans le domaine des films, et que les améliorations de performance ont une forte dépendance avec l'algorithme de recommandation sous-jacent. De plus, aucune information contextuelle ne semble être plus informative que d'autres pour tous les systèmes de recommandation contextuels évalués. De même que les résultats de recherches antérieures comparant certaines approches CARS sur les applications de commerce électronique \cite{panniello2009experimental}, l'identification de l'approche la plus performante nécessite une évaluation et une comparaison de plusieurs CARS sur les données cibles. Enfin, les auteurs ont conclu que l'utilisation d'un grand nombre des informations  contextuelles ne conduit pas nécessairement à de meilleures performances CARS, et la contribution donnée à une information contextuelle dépend de la combinaison particulière d'approche de contextualisation et d'algorithme de recommandation utilisé.
\section{Approche de Gantner et al}
\hspace*{3ex} Gantner et al. \cite{gantner2010factorization} présentent dans ce qui suit, une méthode appelée "Pairwise Interaction Tensor Factorization (PITF)" qui a été utilisée pour la recommandation d'étiquette personnalisée, pour modéliser le contexte temporel (semaine).
\subsection{Contexte}
\hspace*{3ex} La recommandation contextuelle est considérée comme un cas particulier de recommandation d'item classique. Dans la recommandation des items classiques, le contexte est juste l'utilisateur pour lequel Gantner et al. ont voulu prédire des items (par exemple, des films). Dans une recommandation contextuelle, le contexte contient habituellement plus d'informations que juste l'utilisateur.

\subsection{Objectif}
\hspace*{3ex} L'approche de Gantner et al. suggère d'utiliser \textbf{Pairwise Interaction Tensor Factorization (PITF)} qui est une méthode utilisée pour la recommandation d'étiquette personnalisée et ainsi pour modéliser le contexte temporel (semaine). D'une autre manière, Gantner et al. présentent également une version étendue de \textbf{PITF} qui gère le contexte de la semaine.
\subsection{Les techniques utilisées}
\subsubsection{Encodage du temps en tant que contexte}
\hspace*{3ex} Il existe différentes granularités possibles pour coder les épisodes temporelles en tant que contexte. Dans ce travail, la semaine est l'entité principale. Outre les semaines calendaires normales qui commencent le lundi, il est également possible de laisser démarrer d'autres types de "semaines" les autres jours. En combinant des modèles de différentes "semaines", il est possible d'encoder le contexte de la semaine avec la granularité du jour.
\subsubsection{Pairwise Interaction Tensor Factorization (PITF)} 
\begin{figure}[H]
\centering
\includegraphics[width=11cm]{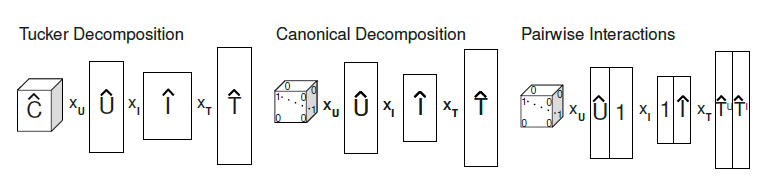}
\caption{Le modèle de PITF et les deux décompositions de Tucker et canonique \cite{gantner2010factorization}}
\label{fig18:my_label}
\end{figure}

\hspace*{3ex} \textbf{PITF} \cite{rendle2010pairwise} est un modèle de factorisation tensorielle initialement développé pour la prédiction d'étiquettes, mais il est également applicable aux d'autres types de tâches de recommandation contextuelle. Le modèle est un cas particulier de la décomposition de Tucker et de la décomposition canonique (voir la figure \ref{fig18:my_label}).\\

\subsubsection{Méthode d'ensemble}
\hspace*{3ex} La combinaison de modèles factoriels \cite {rendle2009factor} avec des régularisations et des dimensions différentes est supposée éliminer la variance des estimations de classement. Il existe essentiellement deux approches simples combinant les prédictions $\widehat{y} ^ {l}_{u,c,i}$  des modèles $l$ : 
\begin{itemize}
\item \textbf{\underline{Ensemble des estimations de valeur }} $\widehat{y} ^ {l}_{u,c,i}$ : 
\vspace{-3em}
\begin{center}
\begin{equation}
\widehat{y} ^ {val}_{u,c,i} := \sum_{l}w_{l} \times  \widehat{y} ^ {l}_{u,c,i} 
\end{equation}
\end{center}
\vspace{-1em}
\begin{center}
\fbox{
 Où $w_{l}$ est le paramètre de pondération pour chaque modèle.
 }
 
\end{center}
 \item \textbf{\underline{Ensemble des estimations des rang}} $\widehat{r} ^ {l}_{u,c,i}$ : 

\vspace{-3em}
\begin{center}
\begin{equation}
\widehat{y} ^ {rank}_{u,c,i} := \sum_{l}w_{l} \times (|I| - \widehat{r} ^ {l}_{u,c,i}) 
\end{equation}
\end{center}
\vspace{-1em}
\end{itemize}
\subsection{Étude expérimentale}
\hspace{3ex}Pour faire une étude expérimentale, Gantner et al. utilisent : 
\begin{itemize}
\item Dataset : original Movie Pilot training
\item Les méthodes comparées : BPR-MF, PITF, item-knn, mp, mp (date/week/event/prior), random.
\end{itemize}

\subsubsection{Christmas}
\hspace*{3ex} Les résultats du problème de prédiction de Noël sont présentés dans le tableau \ref{tab12}. Compte tenu de la précision, BPR-MF est la méthode la plus forte en termes de prédiction, ce qui montre qu'il y a encore des améliorations possibles pour les modèles de factorisation BPR-MF et PITF. 

\hspace*{3ex} PITF-8 est légèrement meilleur que les  modèles PITF-16 et PITF-32, ce qui suggère que les auteurs doivent rechercher de meilleurs paramètres de régularisation pour les modèles plus grands, ainsi la méthode BPR-MF avec 32 facteurs est la plus forte en termes d'AUC.
\newline 
\hspace*{3ex} Les lignes de base contextuelles les plus populaires ne fonctionnent pas mieux que les prévisions les plus populaires, à l'exception des films les plus populaires de la semaine, juste avant la semaine de Noël.
\begin{center}
\begin{tabular}{|c||c|c|c|}
\hline 
  \textbf{Méthode} & \textbf{AUC} & \textbf{Prec@5} & \textbf{Prec@10} \\
  \hline \hline 
  \textbf{Random} & 0.5131 & 0 & 0.00125 \\
  \textbf{mp}  & 0.9568 & 0.1075 & 0.085 \\
  \hline
\textbf{mp (date)}& 0.8634 & 0.09875 & 0.079375 \\
\textbf{mp (week)} & 0.8649 & 0.1000 & 0.081875 \\
\textbf{mp (event)} & 0.8677 & 0.09875 & 0.07875\\
\textbf{mp (prior)} & 0.9533 & 0.11125 & 0.106875\\
\hline 
\textbf{item-knn} & 0.9555 & 0.1325 & 0.12\\
\textbf{BPR-MF-16} & 0.9680 & 0.1418 & 0.1281\\
\textbf{BPR-MF-32} & 0.9711 & 0.1397 & 0.1231\\
\hline
\textbf{PITF-8}  & 0.9511 & 0.13125 & 0.1125\\
\textbf{PITF-16} & 0.9490 & 0.125   & 0.103125\\
\textbf{PITF-32} & 0.9501 & 0.12875 & 0.109375\\
\hline
\end{tabular}
\caption{Résultat de la piste 1 \textbf{" MoviePilot Christmas "}}
\label{tab12}
\end{center}
\subsubsection{Oscar}
\hspace*{3ex} Les résultats pour le problème de prédiction d'Oscar peuvent être vus dans le tableau \ref{tab13}. Ici, PITF se comporte comme prévu, les prédictions s'améliorent avec les modèles qui sont plus grands. Cependant, les auteurs s'attendent à ce que PITF surpasse BPR-MF. Le modèle de prédiction le plus fort en termes de précision est mp (prior).
\begin{center}
 {\setlength {\tabcolsep}{12pt}
\begin{tabular}{|c||c|c|c|}
  \hline 
 \textbf{Méthode} & \textbf{AUC} & \textbf{Prec@5} & \textbf{Prec@10} \\
  \hline \hline
  \textbf{random} & 0.5018 & 0.00136 & 0.00066\\
\textbf{most-popular} & 0.9611 & 0.07895 & 0.08026\\
  \hline
\textbf{mp (date)}& 0.9048 & 0.0684 & 0.0697 \\
\textbf{mp (week)}& 0.8979 & 0.0803 & 0.0631 \\
\textbf{mp (event)} &  0.9001 & 0.075 & 0.0743\\
\textbf{mp (prior)} &  0.9623 & 0.2039 & 0.1822\\
\hline 
\textbf{item-knn} & 0.9597 & 0.1289 & 0.1243\\
\textbf{BPR-MF-16} & 0.9695 & 0.1609& 0.1442\\
\textbf{BPR-MF-32} & 0.9755 & 0.1660 & 0.1498\\
\hline
\textbf{most-popular (recent)} & 0.9656 & 0.0888 & 0.0825\\
\textbf{item-knn} & 0.9644 & 0.135 & 0.1325\\
\textbf{BPR-MF-16 (recent)}& 0.9728 & 0.1634 & 0.1472\\
\textbf{BPR-MF-32 (recent)} & 0.9735 & 0.1574 & 0.1464\\
\hline
\textbf{PITF-8}  & 0.9511 & 0.13125 & 0.1125\\
\textbf{PITF-16} & 0.9490 & 0.125   & 0.103125\\
\textbf{PITF-32} & 0.9501 & 0.12875 & 0.109375\\
\hline
\end{tabular}
}
\caption{Résultat de la piste 1 \textbf{" MoviePilot  Oscar "}}
\label{tab13}
\end{center}
\section{Approche de Biancalana et al}
\hspace*{3ex} Dans cette partie, nous allons aborder les deux approches de Biancalana et al. \cite{biancalana2011context} dans le cadre d'utilisation des systèmes de recommandation sensibles au contexte, qui tentent d'exploiter l'utilisation du contexte afin d'améliorer le processus de génération des recommandations. Ces auteurs ont proposé deux approches contextuelles différentes pour la recommandation des films.
\subsection{Contexte}
\hspace*{3ex} La plupart des moteurs de recommandation existants ne prennent pas en compte les informations contextuelles pour suggérer des items intéressants aux utilisateurs. Des caractéristiques telles que l'heure, l'emplacement ou la météo peuvent affecter les préférences de l'utilisateur pour un élément particulier. 

\hspace*{3ex} Néanmoins, peu de systèmes de recommandation incluent explicitement cette information dans les modèles de préférence. Les systèmes de recommandation contextuels et les systèmes de recommandation classiques sont utilisés pour fournir aux utilisateurs des informations pertinentes: les premiers exploitent les contextes des utilisateurs, les seconds exploitent au moyen les intérêts des utilisateurs.  

\hspace*{3ex} Les systèmes de recommandation contextuels visent à améliorer la satisfaction des utilisateurs en offrant une meilleure suggestion en fonction d'un contexte d'utilisation particulier.

\subsection{Objectif}
\hspace*{3ex} Biancalana et al. ont proposé deux approches contextuelles différentes pour la tâche de recommandation des films : 
\begin{itemize}
\item  Un système de recommandation  hybride qui évalue les facteurs contextuels disponibles liés au temps afin d'accroître le rendement des approches traditionnelles du filtrage collaboratif FC.
\item  La seconde approche vise à identifier les utilisateurs d'un ménage ayant soumis une note donnée. Cette dernière approche est basée sur des techniques d'apprentissage automatique, à savoir des réseaux de neurones et des classificateurs à vote majoritaire.
\end{itemize}
\subsection{Description de l'approche }

\subsubsection{Recommandation de films basée sur le traitement du signal} 
\hspace*{3ex} Une hypothèse liée à la tâche de recommandation actuelle implique que des événements ayant lieu à un moment donné influencent potentiellement les films regardés. Le plus pertinent est le nombre de notes que l'utilisateur soumet habituellement dans une période donnée. La plupart du temps, les utilisateurs rassemblent un certain nombre de préférences et les soumettent au système dans un court laps de temps, généralement autour d'une heure. Un très petit sous-ensemble d'utilisateurs répartit les notations sur plusieurs jours ou semaines.

\hspace*{3ex}\textbf{L'hypothèse de Biancalana et al :} \textbf{\textit{Si un utilisateur est particulièrement intéressé par le visionnage de films sur une période donnée, les films les plus regardés de cette période sont ceux qui devraient gagner en importance au cours de la recommandation collaborative traditionnelle.}}

\hspace*{3ex} Un pré-traitement de l'ensemble de données disponibles a été effectué en regroupant le nombre d'évaluations d'un utilisateur donné selon un intervalle de temps prédéfini, à savoir un jour, une semaine ou un mois. Les auteurs ont  obtenu des échantillons quantifiés composant une représentation numérique d'une quantité variable dans le temps, c'est-à-dire un signal.

\hspace*{3ex} La première mesure du signal correspond au nombre de films visionnés par l'utilisateur donné dans l'intervalle considéré. La deuxième mesure fait référence à l'intervalle suivant, et ainsi de suite. Le même processus a affecté le nombre de vues d'un film "m".
Ensuite, il est possible de corréler les deux signaux dessinant des mesures de similarité. En particulier, Biancalana et al. ont des signaux se rapportant au comportement de l'utilisateur (c'est-à-dire, des films regardés) et un signal associé à chaque film.

\subsubsection{Réseaux de neurones et vote majoritaire pour l'identification de l'utilisateur} 
\begin{enumerate}[label=\alph*)]
\item \textbf{Analyse de la distribution des valeurs données par un utilisateur (classificateur c1)} : 
Biancalana et al. ont regroupé la fréquence des évaluations des utilisateurs en les divisant en cinq classes:
\begin{center}
 {\setlength {\tabcolsep}{12pt}
\begin{tabular}{|c||c|}
\hline
1. & 0-40\\ 
\hline
2. & 41-60\\ 
\hline
3. & 61-80\\
\hline
4. & 81-90\\
\hline 
5. & 91-100\\
\hline 
\end{tabular}
}
\end{center}
Le but de cette analyse est de modéliser chaque utilisateur en fonction des valeurs de ses évaluations. Chaque utilisateur est ainsi représenté par une distribution normalisée par un indice de normalisation Z-score qui est comparé à l'évaluation donnée.

\item \textbf{Analyse de la distribution des temps où la notation a été donnée par un utilisateur (classificateur c2)} : Comme dans le cas précédent, les données d'entrée sont traitées avec les quatre groupes suivants : 
\begin{center}
\begin{tabular}{|c||c|}
\hline
1. & morning (7am to 12pm)\\ 
\hline
2. & afternoon (1pm to 5pm)\\ 
\hline
3. & evening (6pm to 10pm)\\
\hline
4. & night (11pm to 6am)\\
\hline 
\end{tabular}
\end{center}
Ce modèle représente l'habitude de l'utilisateur de donner ses notes à certains moments \cite{lathia2010temporal}.
\item \textbf{Analyse des films pour déterminer si deux utilisateurs ont vu le même film (classificateur c3)}
Biancalana et al. analysent les utilisateurs qui ont vu un film. Ils sont comparés à l'ensemble des utilisateurs candidats via la distance Jaccard suivante:
\begin{center}
\Large {$ \frac{\cap(S_{u}, S_{v})}{\cup (S_{u}, S_{v}) } $ }
\end{center} 
Où $S_{u}$ et $S_{v}$ sont l'ensemble des films évalués par l'utilisateur $u$ et $v$, respectivement.
Le réseau de neurones est entraîné en utilisant les paramètres d'entrée suivants (68 caractéristiques):
\begin{itemize}
\item La distribution du nombre d'utilisateurs qui ont évalué un film par semaine (53 fonctionnalités).
\item La distribution du nombre d'utilisateurs qui ont évalué un film par jour de la semaine (7 fonctions).
\item La distribution des notes attribuées à un film, réparties en cinq groupes (5 fonctions).
\item La date de soumission de la notation, identifiée par semaine de l'année (de 1 à 53) et par jour de la semaine (de 1 à 7) (2 caractéristiques).
\item Le nombre de notes attribuées à un film (1 caractéristique).
\end{itemize}
\begin{center}
\begin{algorithm}
\caption{Algorithme d'empilement pondéré }
\begin{algorithmic} 
\REQUIRE :  Donner un ensemble d'apprentissage ($t_{1}$, $\vec{c_{1}}$), ..., ($t_{n}$, $\vec{c_{n}}$) et un tuple $t$ 
\FOR{i=1 \TO 3 } \STATE {$ TS_{i} \leftarrow $ Sélection de N exemples aléatoires de l'ensemble d'apprentissage} 

$f_{i}  \leftarrow $ Le résultat de l'apprentissage basé sur l'algorithme d'apprentissage sur $TS_{i}$
donne la sortie $y_{c1}$, $y_{c2}$, $y_{c3}$ à partir du réseau de neurones 
\ENDFOR
\STATE Sortie du classificateur combiné
$g(t)$ = majorité ($y_{c1}$ · $f_{1}(t)$, $y_{c2}$ · $f_{2}(t)$, $y_{c3}$ · $f_{3}(t)$)
\end{algorithmic}
\end{algorithm}
\end{center}
\end{enumerate}
\subsection{Les techniques utilisées}
\hspace*{3ex} Les moteurs de recommandation exploitent généralement des rétroactions explicites  ou des données d'utilisation implicites pour déterminer les films que l'utilisateur voudra voir ensuite. Biancalana et al. ont utilisé \textbf{la technique de filtrage collaboratif (FC)} dans la première approche en tenant compte des facteurs contextuels. Par contre, Dans la deuxième approche, Biancalana et al. ont utilisé \textbf{la technique de Machine learning} ou \textbf{Apprentissage automatique} qui permet de prévoir l'utilisateur qui a soumis une évaluation donnée.

\subsection{Étude expérimentale}
\hspace*{3ex} L'évaluation se concentre sur les métriques d'exactitude de classification pour deux pistes différentes comme suit:
\begin{itemize}
\item \textbf{Piste 1 :} il est demandé de générer des recommandations pour chaque ménage. Le nombre de recommandations à suggérer est prédéterminé pour chaque ménage.
\item \textbf{Piste 2 :} le but est d'identifier quel membre d'un ménage a effectué une évaluation donnée. Bien sûr, les membres des ménages sont connus.
\end{itemize}

\subsubsection{Recommandations du ménage}
\hspace*{3ex} Les paramètres utilisés pour l'évaluation qui ont été définis empiriquement sont les suivants:
\begin{center}
\begin{tabular}{|c||c|}
\hline 
\textbf{Notation}& \textbf{Description}\\
\hline \hline
$N_{u}$ = 7 &  Nombre d'utilisateurs dans le neighborhood \\
\hline 
$N_{m}$ = 60 & Les premiers résultats du réducteur utilisé pour le ré-classement \\
\hline 
$b = 1.25$ & Rescorer bosting \\
\hline 
\end{tabular}
\end{center}
\captionof{table}{Les paramètres utilisés pour l'évaluation}
\label{tab14}
\hspace*{3ex} Lorsque le ménage est composé de plus d'une personne, les listes de résultats des algorithmes de chaque utilisateur du ménage sont fusionnées et les résultats les mieux classés sont utilisés pour les tests. Alors, les résultats sont résumés comme suit : 
\begin{center}
\begin{tabular}{|c||c|c|}
\hline 
 & \textbf{CF} & \textbf{CF  w/Rescover}\\
\hline \hline
MAP & 0,002 & 0,298 \\
\hline 
$P@5$ & 0,010 & 0,170\\
\hline 
$P@10$ & 0,006 & 0,124\\
\hline 
\end{tabular}
\end{center}
\captionof{table}{Résultats d'évaluation pour la piste 1}
\label{tab15} 

\subsubsection{Identification des évaluations}
\hspace*{3ex} Afin de déterminer les valeurs optimales pour les poids du réseau, Biancalana et al. ont déjà appliqué un algorithme d'apprentissage supervisé basé sur la descente de gradient et une validation croisée de 10 pour ajuster les poids vers la convergence. Ils ont obtenu une précision de classification élevée de 71,9\%. Les valeurs des autres mesures pertinentes sont résumées dans le tableau \ref{tab16} : 

 \begin{center}
 {\setlength {\tabcolsep}{12pt}
\begin{tabular}{|c||c|}
\hline 
0,03 & \textbf{Mean absolute error (MAE)}\\
\hline \hline
0,26 & \textbf{Root mean squared error (RMSE)} \\
\hline \hline
0,23 & \textbf{Relative absolute error (RAE)}\\
\hline \hline
44 \% & \textbf{Root relative squared error (RRSE)}\\
\hline 
\end{tabular}
}

\captionof{table}{Valeur d'erreur pour " Neural network training "}
\label{tab16}
\end{center}

\hspace*{3ex} Les résultats des trois classificateurs $c1$, $c2$, $c3$ et du classificateur NN, qui adoptent l'approche combinée avec le réseau neuronal, sont résumés dans le tableau \ref{tab17}, en utilisant le taux d'erreur de classification par ménage (Average Error Rate, AER), Taux d'erreur moyen pour le ménage (AERH) pour chaque taille de ménage $\left \{2, 3, 4 \right \}$, zone de ménage sous la courbe ROC (AUC) et  (MAP : Mean Average Precision).
 \begin{center}
 {\setlength {\tabcolsep}{12pt}
\begin{tabular}{|c||c||c||c||c|}
\hline
& c1 & c2 & c3 & NN\\
\hline \hline 
MAP & 0.621 & 0.623 & 0.792 & 0.824\\
\hline
AUC & 0.614 & 0.695 & 0.756 & 0.815\\
\hline 
AER & 0.605 & 0.610 & 0.755 & 0.800\\
\hline 
$AERH_{2}$ & 0.606 & 0.621 & 0.756 & 0.804\\
\hline
$AERH_{3}$ & 0.609 & 0.619 & 0.705 & 0.735\\
\hline 
$AERH_{4}$ & 0.483 & 0.510 & 0.838 & 0.777\\
\hline 
\end{tabular}
}

\end{center}
\captionof{table}{Valeur d'erreur pour " Neural network training "}
\label{tab17}

\hspace*{3ex} Les résultats montrent que les approches $c1$ et $c3$ fournissent des résultats comparables, tandis que l'approche $c2$ présente la pire performance. La combinaison des trois classificateurs à travers un réseau de neurones fournit des valeurs significativement plus élevées de \textbf{MAP} et d'\textbf{AUC} que celles fournies par les classificateurs uniques, représentant ainsi les meilleurs résultats de ces expériences. 

\hspace*{3ex} Dans l'ensemble de tests, il n'y a pas de ménages avec plus de quatre membres effectuant des évaluations. Par conséquent, Biancalana et al. ont décidé de ne pas utiliser les mesures \textbf{P@5} et \textbf{P@10} dans chaque piste, car leurs valeurs ne seraient pas significatives.

\section{Approche de Shi et al}

\hspace*{3ex} Les systèmes recommandation contextuels visent à améliorer les performances de la recommandation en exploitant différentes sources d'informations. Dans cette section, Shi et al. \cite{shi2013mining} ont présenté un nouvel algorithme de recommandation contextuelle de film basé sur \textbf{Joint matrix Factorization (JMF)}.
\subsection{Contexte}
\hspace*{3ex} La recommandation tenant compte du contexte a connu un regain d'intérêt dans la communauté des systèmes de recommandation \cite{said2010putting}. L'intérêt a été stimulé par une prise de conscience croissante du potentiel de l'information contextuelle, si disponible, pour améliorer la qualité des recommandations \cite{adomavicius2005toward},\cite{burke2002hybrid}. Les informations contextuelles peuvent être exploitées pour estimer de manière plus fiable les relations entre les items en compensant les cas dans lesquels les informations de la matrice d'utilisateur-article (en Anglais : user-item) d'origine sont insuffisantes.
\subsection{Objectif}
\hspace*{3ex} Shi et al. \cite{shi2013mining} ont proposé un nouvel algorithme de recommandation contextuelle des films  qui étend le modèle de factorisation matricielle de base (MF) pour prendre en compte les liens induits par le contexte entre les films.
\subsection{Les techniques utilisées}
\begin{figure}[H]
\centering
\includegraphics[width=12cm]{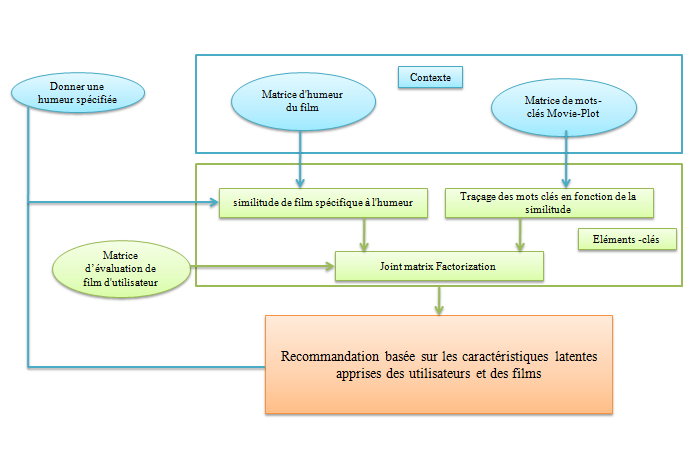}
\caption{Organigramme de l'algorithme proposé \cite{shi2013mining}}
\label{fig19:my_label}
\end{figure}

\hspace*{3ex} Dans cette partie, Shi et al. ont présenté leur algorithme proposé pour la tâche de recommandation contextuelle des films  du défi Moviepilot. 
L'organigramme de l'algorithme proposé est présenté à la figure \ref{fig19:my_label}. 
\subsubsection{Similitude de film spécifique à l'humeur }
\begin{figure}[H]
\centering
\includegraphics[width=8cm]{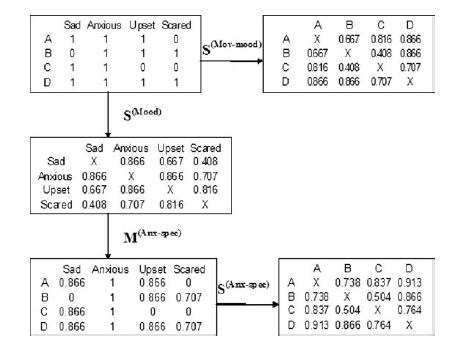}
\caption{Un exemple illustratif du similitude de film spécifique à l'humeur \cite{shi2013mining}}
\label{fig20:my_label}
\end{figure}

\hspace*{3ex} Selon le filtrage collaboratif (FC) basé sur les items, la similarité item à item peut être calculée comme la similarité cosinus entre deux vecteurs d'évaluation d'items \cite{deshpande2004item}.
De même, étant donné la matrice $M$ (composée de $N$ films et de $K_{1}$ marqueurs d'humeur), la similarité d'humeur entre le film $j$ et le film $n$ est calculé comme suit:
\vspace{-2em}
\begin{center}
\begin{equation}
S^{(Mov-mood)}_{jn} = \frac{\sum_{k=1}^{K_{1}} M_{jk} M_{nk}}{\sqrt{\sum_{k=1}^{K_{1}} M_{jk}^{2} }\sqrt{{\sum_{k=1}^{K_{1}} M_{nk}^{2}}}}
\label{eq:equation1}
\end{equation}
\end{center}
\vspace{-1em}
\begin{center}
\fbox{
 Avec $M_{jk}$ = 1 indique que le film $j$ à l'étiquette d'humeur $k$, sinon $M_{jk}$ = 0. }
 \end{center}
\textbf{Exemple 4} Comme le montre la figure \ref{fig20:my_label}, deux films (A et B) partageant différentes propriétés de l'humeur pourraient être également similaires à un autre film (D). Si l'humeur requise d'un film est spécifiée, cette similitude ne permet pas de différencier les films A et B.

\hspace*{3ex} Au lieu d'évaluer la cohérence des marqueurs d'humeur entre deux films, comme dans l'équation \ref{eq:equation1}, le calcul de la cooccurrence normalisée de l'humeur $i$ et de l'humeur $k$ dans la collection de films se fait comme suit:
\vspace{-1em}
\begin{center}
\begin{equation}
S^{(mood)}_{ik} = \frac{\sum_{j=1}^{N} M_{ji} M_{jk}}{\sqrt{\sum_{j=1}^{N} M_{ji}^{2} }\sqrt{{\sum_{j=1}^{N} M_{jk}^{2}}}}
\label{eq:equation2}
\end{equation}
\end{center}
\vspace{-3em}
\hspace*{3ex} Une fois que la matrice de cooccurrence d'humeur \textbf{$S_{Mood}$} est obtenue, La matrice d'humeur cinématographique spécifique à l'humeur est générée, comme l'exprime l'équation \ref{eq:equation3} :
\vspace{-3em}
\begin{center}
\begin{equation}
M_{jk}^{m-spec} = \left\{\begin{matrix}
S_{jk}^{mood} & k \neq m \\ 
M_{jk} & k = m
\end{matrix}\right.
\label{eq:equation3}
\end{equation}
\end{center}
\vspace{-1em}
\hspace*{3ex} Tout en conservant les valeurs originales de matrice de l'humeur du film dans la colonne correspondant à l'humeur $m$, les valeurs de la matrice comme dans l'équation \ref{eq:equation3}, pour toute autre humeur $k$, sont remplacées par les valeurs des similitudes dans l'équation \ref{eq:equation2}, indiquant implicitement dans quelle mesure l'humeur $k$ est informative sur l'humeur $m$. Ensuite, les auteurs ont défini la similarité cinématographique spécifique à l'humeur comme suit : 
\vspace{-3em}
\begin{center}
\begin{equation}
S^{(m-spec)}_{jn} = \frac{\sum_{k=1}^{k_{1}} M_{jk}^{(m-spec)} M_{nk}^{(m-spec)}} {\sqrt{\sum_{k=1}^{k_{1}} M_{jk}^{(m-spec)}}^{2} {\sqrt{\sum_{k=1}^{k_{1}} M_{nk}^{(m-spec)}}^{2}} }
\label{eq:equation4}
\end{equation}
\end{center}
\vspace{-1em}
\subsubsection{Approche de Said et al}
\hspace*{3ex} La similitude entre les films en termes de mots-clés (PK). Puisque les PK représentent le contenu du film, cette similitude peut améliorer les liens basés sur l'humeur entre les films. Tout d'abord, la création d'une matrice $P$ de film-PK binaire composée de $N$ films et $K_{2} PK $.
\begin{center}
\fbox{
Où $P_{jk}$ = 1 si le film $j$ a le $PK k$, et $P_{jk}$ = 0 sinon.
}
\end{center}
\hspace*{3ex} Ensuite, la similarité \textbf{PK-based} entre le film $j$ et le film $n$ peut être définie comme suit :
\vspace{-3em}
\begin{center}
\begin{equation}
S^{(Mov-PK)}_{(jn)} = \frac{\sum_{k=1}^{k_{2}} P_{jk} P_{nk}} {\sqrt{\sum_{k=1}^{k_{2}} P_{jk}^{2} }{\sqrt{\sum_{k=1}^{k_{2}} P_{nk}^{2}}}}
\label{eq:equation5}
\end{equation}
\end{center}
\vspace{-1em}
\subsubsection{Joint Matrix Factorization (JMF)}
\hspace*{3ex} Le MF de base \cite{koren2010collaborative} peut être formulé comme dans l'équation \ref{eq:equation6} : 
\vspace{-3em}
\begin{center}
\begin{equation}
U.V = arg_{U,V} min  \left\{  \frac{1}{2} \sum_{u =1}^{K} \sum_{j =1}^{n} I_{uj}^{R}(R_{uj}-U_{u}^{T} V_{j})^{2} + \frac{\lambda u} {2} \left \| U
 \right \| _{F} ^{2} + \frac{\lambda v} {2} \left \| V \right \| _{F} ^{2} \right\} 
\label{eq:equation6}
\end{equation}
\end{center}
\vspace{-1em}
\begin{center}
{\setlength {\tabcolsep}{12pt}
\begin{tabular}{|c||p{11cm}|}
\hline 
\textbf{Notation}& \textbf{Signification}\\
\hline \hline
\textbf{R} &  La matrice d'évaluation User-item \textbf{R} composée de \textbf{K} utilisateurs et de \textbf{N} éléments.\\
\hline 
\textbf{MF} & La matrice d'évaluation \textbf{R} par les deux matrices de rang inférieur. \textbf{U} et \textbf{V} \\
\hline
\textbf{$U_{u}$} & Une colonne de vecteur de caractéristique d-dimensionnelle de l'utilisateur $u$. \\
\hline 
\textbf{$V_{j}$} & Un vecteur de caractéristique de dimension d de colonne de film $j$. \\
\hline 
\textbf{$R_{uj}$} & L'évaluation de l'utilisateur  $u$ sur le film $j$.\\
\hline
\textbf{$I_{uj}^{R}$} & Désigne une fonction indicatrice égale à 1 lorsque \textbf{$R_{uj}$} > 0, et 0 sinon.\\
\hline
\textbf{$\left \| U \right \| _{F}$}, \textbf{$\left \| V \right \| _{F}$} &  Des normes Frobenius de U et V, qui contribuent à atténuer le sur-apprentissage. \\
\hline 
\textbf{$\lambda v$}, \textbf{$ \lambda u$} & Des paramètres de régularisation. \\
\hline 
\end{tabular}
}
\end{center}
\caption{Les propriétés du MF}
\label{tab18}
\hspace*{3ex} Les auteurs exigeaient que les films soient similaires les uns aux autres, en ce qui concerne le critère de similarité propre à l'humeur dans l'équation \ref{eq:equation4}, partageant des caractéristiques de film similaires. Alors, ces auteurs ont formulé une fonction appelée : \textbf{context-aware loss function $L_{1}(V)$} comme indiqué dans l'équation \ref{eq:equation7} : 
\vspace{-3em}
\begin{center}
\begin{equation}
L_{1}(V) = \frac{1}{2} \sum_{j=1}^{N} \sum_{n=1}^{N} I_{jn}^{MS}(S_{jn}^{(m-spec)} - V_{j}^{T} V_{n} )^{2}
\label{eq:equation7}
\end{equation}
\end{center}
\vspace{-1em}
\begin{center}
\fbox{
 Où $I_{jn}^{MS}$ désigne une fonction d'indicateur qui est égale à 1 lorsque $S_{jn}^{(m-spec)}$ > et 0 sinon. 
 }
\end{center}
\hspace*{3ex} les auteurs ont supposé que les films similaires les uns aux autres (voir l'équation \ref{eq:equation5}), devraient également partager des caractéristiques de films similaires, ce qui implique que la similitude des graphiques est informative pour un film spécifique à l'humeur. Par conséquent, les auteurs ont formulé une autre fonction de perte contextuelle $L_{2}(V)$ comme indiqué dans l'équation suivante \ref{eq:equation8} : 
\vspace{-3em}
\begin{center}
\begin{equation}
L_{2}(V) = \frac{1}{2} \sum_{j=1}^{N} \sum_{n=1}^{N} I_{jn}^{PK}(S_{jn}^{(Mov-PK)} - V_{j}^{T} V_{n} )^{2}
\label{eq:equation8}
\end{equation}
\end{center}
\vspace{-1em}
\begin{center}
\fbox{
 Où $I_{jn}^{PK}$ désigne une fonction d'indicateur qui est égale à 1 lorsque $S_{jn}^{(Mov-PK)}$ > et 0 sinon.}
 \end{center} 
\hspace*{3ex} En prenant en compte les fonctions de perte contextuelle en tant que termes de régularisation dans le modèle de base \textbf{MF}, un modèle de \textit{(JMF : joint matrix factorization)} peut être formulé comme suit:
\begin{center}
\begin{multline*}
 L(U, V) = \frac{1}{2} \sum_{u=1}^{K} \sum_{j=1}^{N} I_{uj}^{R} (R_{uj} - V_{u}^{T} V_{j} )^{2}\\
+\frac{\alpha}{2} \sum_{j=1}^{N} \sum_{n=1}^{N} I_{jn}^{MS}(S_{jn}^{(m-spec)} - V_{j}^{T} V_{n} )^{2}\\
+ \frac{\beta }{2} \sum_{j=1}^{N} \sum_{n=1}^{N} I_{jn}^{PK}(S_{jn}^{(Mov-PK)} - V_{j}^{T} V_{n} )^{2} + \frac{\lambda }{2}(\left \| U \right \| _{F}^{2} + \left \| V \right \| _{F}^{2} )
\end{multline*}
\label{eq:equation9}
\end{center}
\hspace*{3ex} Dans ce modèle, $\alpha$ et $\beta $ sont des paramètres utilisés pour pondérer les contributions de la régularisation par la similitude de film spécifique à l'humeur et par la similarité de film \textbf{PK}, respectivement.
\begin{center}
\fbox{
Lorsque $\alpha$ = 0 et $\beta $ = 0, le modèle \textbf{JMF} converge vers le modèle de base \textbf{MF}. }
\end{center}
\begin{algorithm}[H]
\caption{JMF-MS-PK}
\textbf{Entrée :} Matrice d'évaluation de film d'utilisateur \textbf{$R$}, similarité de film à film spécifique à l'humeur \textbf{$S^{(m-spec)}$}, similarité de film à film à base de \textbf{PK} \textbf{$S^{(Mov-PK)}$}, paramètres de tradeoff $\alpha$ et $\beta$, paramètre de régularisation $\lambda$, condition d'arrêt $\epsilon$.
\newline 
\textbf{Sortie :} Compléter la matrice de pertinence du film-utilisateur \textbf{$\hat{R} $}.
\newline
Initialiser $U^{(0)}$, $V^{(0)}$ avec des valeurs aléatoires;
\newline 
$t$ = 0;
\newline 
$f$ = 0;
\newline 
Calculer $L^{(t)}$ comme dans l'équation \ref{eq:equation9}.
\newline 
\textbf{Répéter} 
\newline 
$\eta$ = 1;
\newline 
Calculer $\frac{\partial L}{\partial U^{(t)}}$, $\frac{\partial L}{\partial V^{(t)}}$ comme dans l'équation \ref{eq:equation10} et \ref{eq:equation11};
\newline 
\hspace*{3ex} \textbf{Répéter} 
\newline 
$\eta = \frac{\eta}{2}$; // Maximiser la taille de l'apprentissage.
\newline 
\hspace*{3ex} \textbf{jusqu'à }
\newline 
$L(U^{(t)}-\eta \frac{\partial L}{\partial U^{(t)}} , V^{(t)} - \frac{\partial L}{\partial V^{(t)}}) < L^{(t)}$ ;
\newline 
$U^{(t+1)} = U^{(t)} - \frac{\partial L}{\partial U^{(t)}} , V^{(t+1)} = V^{(t)} - \frac{\partial L}{\partial V^{(t)}}$
Calculer $L^{(t + 1)}$ comme dans l'équation \ref{eq:equation9};
\newline 
\textbf{Si} 1-$L^{(t+1)}$ / $L^{(t)}$ $\leq$ $\epsilon$ Alors  
\newline  
$f = 1$ ; // indicateur de convergence
\newline
\textbf{Fin Si}
\newline 
$t = t+1 $ ; 
\newline
\textbf{jusqu'à} $f = 1$
\newline 
$\hat{R} = U^{(t)T} V^{(t)}$;
\end{algorithm} 

\hspace*{3ex} Les gradients de \textbf{$L(U, V)$} par rapport à \textbf{U} et \textbf{V} peuvent être calculés comme suit : 
\vspace{-3em}
\begin{center}
\begin{equation}
\frac{\partial L}{\partial U_{u}} = \sum_{j=1}^{N}I_{uj}^{R}(U_{u}^{T} V_{j} - R_{uj}) V_{j} + \lambda U_{u}
\label{eq:equation10}
\end{equation}
\end{center}
\vspace{-1em}
\hspace*{3ex} Selon l'équation \ref{eq:equation11}, les auteurs exploitent la symétrie de $S^{(m-spec)}$ et $S^{(Mov-PK)}$. L'algorithme \textbf{JMF-MS-PK} est décrit en détail dans l'algorithme 2.
\vspace{-3em}
\begin{center}
\begin{multline}
\frac{\partial L}{\partial V_{j}} = \sum_{u=1}^{K}I_{uj}^{R}(U_{u}^{T} V_{j} - R_{uj}) U_{u}  + 2 \alpha \sum_{n=1}^{N} I_{jn}^{MS}(V_{j}^{T} V_{n} - S_{jn}^{(m-spec)})V_{n} \\
+ 2 \beta \sum_{n=1}^{N} I_{jn}^{PK}(V_{j}^{T} - S_{jn}^{(Mov-PK)}) V_{n}+\lambda V_{j} 
\label{eq:equation11}
\end{multline}
\end{center}
\vspace{-1em}
\subsection{Étude expérimentale}
\hspace*{3ex} Selon Shi et al., les expériences montrent que l'algorithme surpasse plusieurs autres approches de recommandation.  Une amélioration peut être obtenue en exploitant les similarités de séquences contextuelles, parmi lesquelles la similarité de film spécifique à l'humeur est montrée pour apporter la contribution majeure à la performance de recommandation et la similitude de film basée \textbf{PK} pourrait augmenter la contribution. De plus, ces auteurs ont validé spécifiquement l'utilité de la similarité de film spécifique à l'humeur par rapport à la similarité de film basée sur l'humeur générale, ce qui conduit en effet à une amélioration des performances. Ils ont montré également le \textbf{JMF} avec similarité de film spécifique de l'humeur et la similarité de film en termes de mots-clés de la parcelle pourrait être l'option la plus bénéfique pour les utilisateurs de profils contenant différents nombres de films classés, par rapport aux autres variantes.
\begin{center}
\begin{tabular}{|c||c|c|c|c|}
\hline 
& \textbf{P@1} & \textbf{P@5} & \textbf{P@10} & \textbf{MAP} \\
\hline \hline 
\textbf{PopRec} & 0.213 & 0.248 & 0.251 & 0.264\\
\hline
\textbf{RWR} & 0.238 & 0.253 & 0.274 & 0.281\\
\hline 
\textbf{MF} & 0.325 & 0.305 & 0.241 & 0.252\\
\hline 
\textbf{JMF-MB} & 0.338 & 0.328 & 0.286 & 0.273\\
\hline
\textbf{JMF-MS} & 0.350 & 0.335 & 0.295 & 0.289 \\
\hline 
\textbf{JMF-MS-PK} & 0.363 & 0.335 & 0.306 & 0.290\\
\hline  
\end{tabular}
\caption{Comparaison des performances de la recommandation entre l'algorithme proposé et d'autres approches de base}
\label{tab21}
\end{center}
\begin{center}
\begin{tabular}{|c||c|c|c|c|}
\hline 
\textbf{Num. films classés (Num. Utilisateurs)}& \textbf{MF} & \textbf{JMF-MB} & \textbf{JMF-MS} & \textbf{JMF-MS-PK} \\
\hline \hline 
\textbf{1 $\sim$ 50 (19)} & 0.305 & 0.326 & 0.374 & 0.379\\
\hline
\textbf{51 $\sim$ 100 (16)} & 0.250 & 0.269 & 0.313 & 0.313\\
\hline 
\textbf{101 $\sim$ 150 (13)} & 0.208 & 0.223 & 0.238 & 0.238\\
\hline 
\textbf{151 $\sim$ 200 (12)} &  0.242 & 0.317 & 0.250 & 0.267\\
\hline
\textbf{>200 (20)} & 0.195 & 0.285 & 0.270 & 0.300 \\
\hline  
\end{tabular}

\end{center}
\caption{Comparaison des performances de P@10 entre l'algorithme proposé et d'autres approches de référence par rapport aux utilisateurs ayant différents nombres de films classés}
\label{tab22}
\section{Approche de Said et al}
\subsection{Contexte}
\hspace*{3ex} L'importance du contexte et des données contextuelles d'un utilisateur pour des recommandations précises a été largement reconnue \cite{adomavicius2005incorporating, berkovsky2006predicting}. Cependant, la grande majorité des techniques de recommandation existantes se concentrent sur la recommandation des items les plus pertinents aux utilisateurs et ne tiennent pas compte du contexte. De plus, il en résulte des recommandations qui ont été statiques. Alors, les systèmes de recommandation sensibles au contexte sont arrivés pour résoudre ce problème puisque ces systèmes utilisent la notion du contexte pour fournir des informations et/ou des services pertinents à l'utilisateur \cite{dey2001understanding}.
\subsection{Objectif}
\hspace*{3ex} Le défi sur la recommandation contextuelle des films(CAMRa20101) visant à stimuler la recherche de la conscience du contexte dans le système de recommandation. Deux Datasets, rassemblés par les communautés de recommandation des films en ligne \textbf{Moviepilot} et \textbf{Filmtipset}, ont été lancés exclusivement pour le défi. 
\subsection{Les techniques utilisées}
\hspace*{3ex} Le défi a servi à stimuler la recherche sur la connaissance du contexte dans les systèmes de recommandation. Ceci a été réalisé en rassemblant des chercheurs travaillant sur des systèmes de recommandation et en les laissant étudier les mêmes défis, tout en évaluant les solutions proposées en utilisant les mêmes Datasets et les mêmes métriques d'évaluation.

\hspace{3ex} Cet environnement semi-contrôlé permettrait une comparaison juste des solutions et des algorithmes développés par les participants. De plus, les participants devaient utiliser uniquement les datasets fournis.

\hspace*{3ex} Les datasets utilisés dans le défi ont été publiés exclusivement par les sites de recommandation de films \textbf{Moviepilot2} et \textbf{Filmtipset3}. Pour des raisons de confidentialité, les datasets ont été anonymisés avant la publication. Quatre versions des datasets ont été créées: deux ensembles \textbf{Moviepilot} et deux ensembles \textbf{Filmtipset}, c'est-à-dire une version pour chaque piste et sous-piste. Les datasets ont été générés à l'aide des algorithmes aléatoires.

\begin{figure}[H]
\centering
\includegraphics[width=16cm]{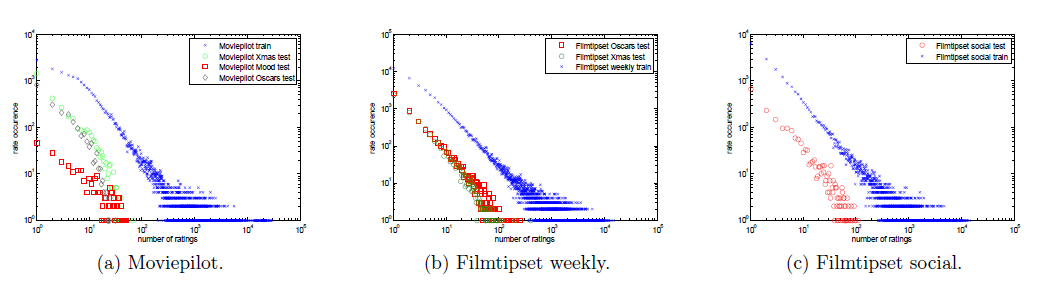}
\caption{Nombre d'évaluations par rapport à l'occurrence de taux des datasets d'apprentissage et de test \cite{said2010putting}}
\label{fig21:my_label1}
\end{figure}

\hspace*{3ex} La figure \ref{fig21:my_label1} montre le nombre d'évaluations par rapport à l'occurrence de taux dans les datasets. La distribution des évaluations dans tous les datasets suit la distribution de la loi de puissance, démontrant qu'il n'y avait pas d'anomalie liée aux évaluations dans les datasets.

\subsubsection{Moviepilot}
\hspace*{3ex} \textbf{Moviepilot} est la première communauté de recommandation de films et de télévision en ligne en Allemagne. Il compte plus de 100 000 utilisateurs enregistrés et une base de données de plus de 40 000 films avec environ 7,5 millions d'évaluations. Les détails du dataset sont présentés dans le tableau \ref{tab23:mytab}. \textbf{Moviepilot} étaient basés sur les évaluations qui ont été divisées en sept parties: 
\begin{itemize}
\item Ensemble d'apprentissage (Train-set).
\item L'ensemble de test pour la semaine de Noël.
\item L'ensemble d'évaluation pour la semaine de Noël.
\item L'ensemble de test pour la semaine des Oscars.
\item L'ensemble d'évaluation pour la semaine des Oscars.
\item L'ensemble de test pour la piste d'humeur.
\item L'ensemble d'évaluation pour la piste d'humeur.
\end{itemize}
\begin{center}
    \begin{tabular}{|c||c|c|c|}
    \hline
  &  \textbf{Utilisateurs}  & \textbf{Films} & \textbf{Évaluations} \\
    \hline \hline 
    Train-set & 105,137 & 25,058 & 4,544,409 \\
    \hline 
    Ensemble de test pour la semaine de Noël &  160 & 3,377 & 16,174\\
    \hline 
    Ensemble d'évaluation pour la semaine de Noël & 80 & 2,153 & 6,701\\
    \hline 
    Ensemble de test pour la semaine des Oscars  & 160 & 2,144 & 8,277\\
    \hline 
    Ensemble d'évaluation pour la semaine des Oscars & 80 &  1,520 & 4,169\\
    \hline 
    Ensemble de test pour la piste d'humeur & 160 & 251 & 2,656\\
    \hline 
    Ensemble d'évaluation pour la piste d'humeur & 80 & 220 & 1,421 \\
   
\hline 
\end{tabular} 
\caption{Le nombre de films, d'utilisateurs et d'évaluations dans \textbf{Moviepilot}}
\label{tab23:mytab}
\end{center}

\hspace*{3ex} En plus des évaluations, les datasets comprenaient plusieurs autres caractéristiques, telles que les étiquettes d'humeur cinématographique, les étiquettes d'audience prévues, les films préférés/détestés et le lieu et l'heure du film. Un diagramme entité-relation abstraite des ensembles de données et de leurs caractéristiques est présenté dans la figure \ref{fig21:my_label}:
\begin{figure}[H]
\centering
\includegraphics[width=10cm]{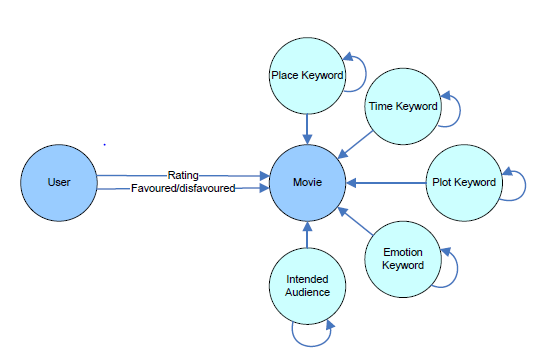}
\caption{Un diagramme entité-relation abstraite de dataset de \textbf{Moviepilot} \cite{said2010putting}}
\label{fig21:my_label}
\end{figure}

\hspace*{3ex} Toutes les données se rapportant aux paires utilisateur-film trouvées dans les ensembles de test et d'évaluation et dans les semaines correspondantes (Noël 2009 et Oscars 2010 \cite{said2013introduction}) ont été supprimées. Afin de ne pas révéler des informations non disponibles au moment des recommandations. Les statistiques d'attribution d'étiquettes dans l'ensemble d'apprentissage sont présentées dans le tableau \ref{tab24}.
\begin{center}
    \begin{tabular}{|c||c|c|c|}
    \hline
    & \textbf{Affectations} \\
    \hline \hline 
\textbf{Humeur} &  6,712 \\
\hline 
\textbf{Plot} & 92,124\\
\hline
\textbf{Time} & 3,687\\
\hline 
\textbf{Place} & 8,586\\
\hline 
\textbf{Audience} & 2,436\\
\hline 
\end{tabular} 
\caption{Affectations des tags dans \textbf{Moviepilot}}
\label{tab24}
\end{center}

\subsubsection{Filmtipset}
\hspace*{3ex} \textbf{Filmtipset} est la plus grande communauté de recommandation de films en Suède. Il compte plus de 90 000 utilisateurs enregistrés et une base de données de plus de 20 millions d'évaluations. Ces datasets ont été divisés en plusieurs fichiers, chaque piste ayant un ensemble d'apprentissage, un ensemble de test et un ensemble d'évaluation.

\begin{figure}[H]
\centering
\includegraphics[width=10cm]{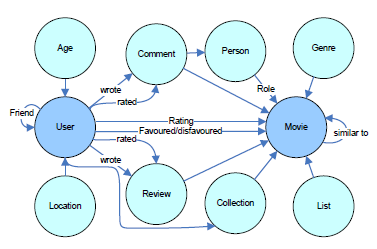}
\caption{Un diagramme entité-relation abstraite de dataset de \textbf{Filmtipset} \cite{said2010putting}}
\label{fig22:my_label}
\end{figure}

\hspace*{3ex} Ces données d'évaluation ont été regroupées avec d'autres fonctionnalités, telles que les commentaires, les relations avec les amis, les informations sur les acteurs/auteurs/réalisateurs et les détails. Un diagramme d'entité-relation abstraite du dataset et de leurs caractéristiques est présenté à la figure \ref{fig22:my_label}. 
\hspace*{3ex} Le tableau \ref{tab25} montre certaines collections non liées à l'évaluation dans Filmtipset.
\begin{center}
    \begin{tabular}{|c||c|c|}
    \hline
    & \textbf{hebdomadaire} & \textbf{Social} \\
    \hline \hline 
\textbf{Collection} &  307,131 & 102\\
\hline 
\textbf{Favorites} &  44,765 & 15,283 \\
\hline
\textbf{Friends} &  83,966 & 12,171\\
\hline 
\textbf{Genres} &  143,316 & 67,997\\
\hline 
\textbf{Lists} &  519,515 & 438,643\\
\hline 
\textbf{Movie comments} &  289,586 & 146,510\\
\hline 
\textbf{People in movies} & 452,074 & 224,410\\
\hline
\textbf{Person comments} & 322,555 & 2,822\\
\hline 
\textbf{Review ratings} &  37,491 & 2,423\\
\hline 
\textbf{Reviews} &  1,341 & 1,044\\
\hline 
\textbf{Movie similarities} &  35,925 & 28,372\\
\hline
\end{tabular} 
\caption{Le nombre de collections, de favoris, etc. dans \textbf{Filmtipset}}
\label{tab25}
\end{center}

\hspace*{3ex} L'ensemble de test pour la piste des recommandations hebdomadaires était basé sur toutes les évaluations fournies entre le premier février 2008 et le 25 février 2010. Les ensembles de test et d'évaluation pour la sous-piste de la semaine de Noël étaient basés sur les évaluations fournies entre le 21 décembre 2009 et le 27 décembre 2009.  Les détails de ces datasets sont présentés dans le tableau \ref{tab26}.
\begin{center}
    \begin{tabular}{|c||c|c|c|}
    \hline
    & \textbf{Utilisateurs} & \textbf{Films} & \textbf{Évaluations} \\
    \hline \hline 
\textbf{Ensemble d'apprentissage} &   34,857 & 53,600 & 5,862,464\\
\hline 
\textbf{Ensemble de test pour la semaine de Noël} & 2,500 & 5,110 & 23,393   \\
\hline
\textbf{Ensemble d'évaluation pour la semaine de Noël} &  1,000 & 3,450 & 9,250\\
\hline 
\textbf{Ensemble de test pour la semaine des Oscars} & 2,500 & 5,670 & 33,548 \\
\hline 
\textbf{Ensemble d'évaluation pour la semaine des Oscars} &  848 & 3,235 & 11,486 \\
\hline
\end{tabular} 
\caption{Nombre d'utilisateurs, de films et d'évaluations dans le dataset hebdomadaires \textbf{Filmtipset}}
\label{tab26}
\end{center}
\hspace*{3ex} Les données hebdomadaires ont été supprimées du dataset de recommandations sociales, créant pratiquement deux ensembles de données disjoints par un utilisateur. Les détails de ces datasets sont présentés dans le tableau \ref{tab27}.
\begin{center}
    \begin{tabular}{|c||c|c|c|}
    \hline
    & \textbf{Utilisateurs} & \textbf{Films} & \textbf{Évaluations} \\
    \hline \hline 
\textbf{Train-set} & 16,473 & 24,222 & 3,075,346  \\
\hline
\textbf{Test} & 439 & 1,915 & 15,729  \\
\hline
\textbf{Évaluation} &  153 & 1,449 & 6,224\\
\hline 
\end{tabular} 
\caption{Nombre d'utilisateurs, de films et d'évaluations dans le dataset sociales \textbf{Filmtipset}}
\label{tab27}
\end{center}

\subsection{Étude expérimentale}
\hspace*{3ex} Les pistes hebdomadaires et les datasets ont été couverts par \cite{liu2010adapting} où les auteurs ont mis en œuvre un modèle de filtrage collaboratif prenant en compte le temps en utilisant la factorisation matricielle. La piste hebdomadaire utilisant les données de \textbf{Moviepilot} a été couverte par \cite{gantner2010factorization} où les auteurs ont utilisé une approche de la recommandation de tag, PITF (Pairwise Interaction Tensor Factorization) où des semaines ont été utilisées pour former des tenseurs de film pour des utilisateurs. Deux essais sur la piste hebdomadaire utilisant les données de \textbf{Filmtipset} ont été présentés par les auteurs de \cite{brenner2010predicting, campos2010simple}, les documents ont présenté une recommandation utilisant kNN basé sur le temps \cite{campos2010simple} et une approche basée sur des modèles de régression \cite{brenner2010predicting}.
\newline
\hspace*{3ex} La piste d'humeur a été couverte par \cite{shi2010mining, wu2010novel} où les approches utilisées étaient une moyenne pondérée kNN basée sur l'humeur et l'utilisateur \cite{wang2010new}, un modèle de factorisation matricielle étendue incluant des informations sur l'humeur \cite{shi2010mining} et un algorithme de filtrage collaboratif utilisant des utilisateurs experts \cite{wu2010novel}.
Enfin, la piste sociale a été couverte par les auteurs de \cite{diez2010movie, liu2010incorporating,rahmani2010three}. L'approche sociale de \cite{liu2010adapting} était, de même que l'approche hebdomadaire couverte par cet article, basée sur la factorisation matricielle. L'approche dans \cite{diez2010movie} était un modèle de random-walk utilisant l'information implicite dans les amitiés \cite{liu2010incorporating}, présentée et étendue du filtrage collaboratif traditionnel où les données sociales étaient prises en compte, et \cite{rahmani2010three} présentait deux approches: une approche kNN basée sur des combinaisons linéaires de mesures de similarité entre des utilisateurs, et une approche basée sur la programmation logique inductive. Les résultats de chaque approche sont présentés dans le tableau \ref{tab28}.
  \begin{center}
    \begin{tabular}{|c||c|c|c|c|c|}
    \hline
    & \textbf{paper} & \textbf{P@5} & \textbf{P@10} &\textbf{MAP} & \textbf{AUC} \\
    \hline \hline 
\textbf{$MP_{Xmas}$ } & $\cite{liu2010adapting}_{TWMF}$  &   0.3637 &  0.3168 & 0.1654 & 0.9212 \\

& \cite{gantner2010factorization}  & 0.1418 & 0.1281 & & 0.9680 \\
\hline 
\textbf{$MP_{Oscar}$ } & $\cite{liu2010adapting}_{TWN}$ & 
0.2775 & 0.2237 & 0.1362 & 0.9556 \\
& \cite{gantner2010factorization} & 0.2039 & 0.1822& & 0.9623 \\
\hline 
\textbf{$FT_{Xmas}$ } & $\cite{liu2010adapting}_{SMF}$ & 
0.0817 & 0.0596 & 0.0902 & 0.9283  \\
&  \cite{campos2010simple} & 0.0070 & 0.0044 & 0.0405 & 0.4552\\
& \cite{brenner2010predicting} & 0.0795 & 0.0821 & 0.0973 & 0.9231\\
\hline 
\textbf{$FT_{Oscar}$ } & $\cite{liu2010adapting}_{SMF}$ & 
0.1087 & 0.0708 & 0.0911 & 0.9467  \\
&  \cite{campos2010simple} & 0.0034 & 0.0028 & 0.0359 & 0.4161\\
& \cite{brenner2010predicting} &  0.0942 & 0.0655 & 0.0849 & 0.9295\\
\hline 
\end{tabular} 
\end{center}
\begin{center}
    \begin{tabular}{|c||c|c|c|c|c|}
    \hline
\textbf{Mood} & \cite{shi2010mining} & 0.3380 & 0.2970 & 0.2940 & 0.8690\\
& \cite{wu2010novel} & 0 & 0 & 0.0037 & 0.6548\\
\hline 
\textbf{Social} & $\cite{liu2010adapting}_{NRMF}$ & 
0.5144 & 0.4185 & 0.3103 & 0.9782 \\
& \cite{diez2010movie} & 0.0802 & 0.0704 & 0.0596 & 0.4276 \\
& \cite{liu2010incorporating} & & & 0.4167 & \\
& \cite{rahmani2010three} & 0.1480 & 0.1230 & 0.0970 & 0.9880 \\
\hline 
\end{tabular} 
\caption{Les résultats obtenus dans chaque piste par les participants (les valeurs manquantes n'ont pas été fournies)}
\label{tab28}
\end{center}

\section{Approche de Wang et al}
\subsection{Contexte}
\hspace*{3ex} Les technologies de personnalisation et les systèmes de recommandation ont été largement utilisés dans plusieurs domaines (tels que la recherche d'informations) pour atténuer le problème de "surcharge d'informations". Le but de tels systèmes est d'aider les utilisateurs à trouver des articles (tels que des films, des pages Web et des services).
\newline
\hspace*{3ex} La grande majorité des systèmes de recommandation basiques se concentrent sur la recommandation des éléments les plus pertinents aux utilisateurs et ne tiennent pas compte du contexte. Alors, les systèmes de recommandation sensibles au contexte arrivent pour résoudre ce problème. Dans ce cadre,  l'humeur s'est avérée être une caractéristique contextuelle importante dans les systèmes de recommandation contextuels par certaines études.

\subsection{Objectif}
\hspace*{3ex} Wang et al. \cite{wang2010new} ont proposé deux nouvelles approches du filtrage collaboratif hybride (FC) basé sur l'humeur afin d'améliorer encore la précision des performances et la satisfaction des utilisateurs en utilisant le contexte émotionnel dans les systèmes de recommandation contextuels.
\subsection{Les techniques utilisées}
\subsubsection{Le filtrage collaboratif}
\hspace*{3ex} Le filtrage collaboratif (FC) (\cite{adomavicius2005toward}, \cite{candillier2009state}) est l'approche la plus populaire dans le domaine des systèmes de recommandation. Alors, Wang et al. ont utilisé le filtrage collaboratif traditionnel basé sur l'utilisateur comme étant l'approche de base qui peut être divisée en deux étapes: 
\begin{itemize}
\item L'évaluation de la similarité de l'utilisateur.
\item La prédiction d'évaluation.
\end{itemize}
\hspace*{3ex} L'approche du filtrage collaboratif traditionnel basée sur l'utilisateur repose sur des utilisateurs similaires ayant des profils d'évaluation similaires, c'est-à-dire que la prédiction d'une évaluation $r_{c, m}$ pour l'utilisateur $c$ et l'élément $m$ est calculée comme un agrégat des utilisateurs pour le même élément $m$ \cite{adomavicius2005toward}. 
\newline
\hspace*{3ex} Les $k$ utilisateurs les plus similaires sont généralement appelés $k$ plus proches voisins (KNN). Ainsi, une mesure de similarité $sim (x, y)$ entre l'utilisateur $x$ et $y$ doit être définie et calculée avec le coefficient de corrélation de Pearson, le cosinus ou d'autres méthodes. Dans ce cadre, les auteurs ont utilisé 
le coefficient de corrélation de Pearson pour mesurer la similarité de $x$ et $y$ comme suit \cite{adomavicius2005toward}:
\vspace{-3em}
\begin{center}
\begin{equation}
Sim(x, y) = \frac{\sum_{s \in S_{xy}}(r_{x, s} - \bar{r}_{x}) (r_{y, s} - \bar{r}_{y})} {\sqrt{\sum_{s \in S_{xy}}(r_{x, s} - \bar{r}_{x})^{2} \sum_{s \in S_{xy}} (r_{y, s} - \bar{r}_{y})^{2}}}
\label{eq:equation12}
\end{equation}
\end{center}
\vspace{-1em}
 \begin{center}
 {\setlength {\tabcolsep}{12pt}
    \begin{tabular}{|c||p{9cm}|}
    \hline
    \textbf{Notation} & \textbf{Propriétés} \\
    \hline \hline 
$ S_{xy} = \begin{Bmatrix}
s \in S|r_{x, s} \neq nul, r_{y, s} \neq nul
\end{Bmatrix} $ & Ensemble de tous les éléments co-évalués par les deux utilisateurs $x$ et $y$ \\
\hline 
$|S_{xy}|$ & Nombre d'éléments co-évalués \\
\hline
\end{tabular} 
}
\caption{Les propriétés de la corrélation de Pearson}
\end{center}
\label{tab29}
\hspace*{3ex} L'approche du filtrage collaboratif traditionnelle introduit généralement l'évaluation moyenne de l'utilisateur. Ensuite, l'estimation prédite $\bar{r_{x}}$, $m$ peut être calculée comme suit :
 \vspace{-3em}
\begin{center}
\begin{equation}
r_{c, m} = \frac{\bar{r_{c}} + \sum_{c' \in \hat{C}} Sim(c, c')_{tran-pearson} \times (r_{c', m} - \bar{r_{c'}})} {\sum_{c' \in \hat{C}}|sim(c, c')|_{tran-pearson}}
\label{eq:equation13}
\end{equation}
\end{center}
\vspace{-1em}
\subsubsection{Recommandation tenant compte de l'humeur}
\hspace*{3ex} La plupart des travaux sur la recommandation d'humeur ont abordé les recommandations d'humeur et de musique, mais il y avait peu d'études liées au cinéma. Winoto et al. \cite{winoto2010role} ont étudié le rôle de l'humeur de l'utilisateur dans les recommandations de films en proposant une approche du \textbf{FC} sensible à l'humeur et ont comparé aussi les performances par rapport à l'approche classique.
\subsection{Description de l'approche}
Dans cette partie, nous allons aborder de l'approche de Wang et al. Dans le contexte d'utilisation de des systèmes de recommandation contextuels, l\\
es auteurs ont présenté une nouvelle approche FC basée sur l'humeur, puis, ils ont  proposé deux approches hybrides basées sur l'approche décrite ci-dessus afin d'améliorer la précision des performances dans les systèmes de recommandation.
\subsubsection{Approche du filtrage collaboratif basée sur l'humeur} 
\hspace*{3ex} La logique de la piste \textbf{Moviepilot} implique que l'humeur d'un film pourrait implicitement refléter l'humeur d'un utilisateur au moment de regarder le film \cite{said2010putting}. L'approche proposée du filtrage collaboratif basée sur l'humeur repose sur des utilisateurs similaires qui ont des profils de préférence d'humeur similaires, c'est-à-dire ceux qui ont un intérêt similaire pour le contexte émotionnel. De plus, l'approche proposée prend en considération l'effet de la volatilité de l'évaluation d'humeur sur les préférences de l'utilisateur.
 
\hspace*{3ex} L'humeur choisie pour cette piste était "Eigenwillig", c'est-à-dire "bizarre". Cependant, les participants n'ont pas reçu la description textuelle de l'humeur, mais plutôt une représentation numérique anonymisée. Par conséquent, Wang et al. ont utilisé le modèle d'espace vectoriel (VSM) pour décrire le contexte émotionnel :
\vspace{-3em}
\begin{center}
\begin{equation}
r_{c, m} = \frac{\bar{r_{c}} + \sum_{c' \in \hat{C}} Sim(c, c')_{tran-pearson} \times (r_{c', m} - \bar{r_{c'}})} {\sum_{c' \in \hat{C}}|sim(c, c')|_{tran-pearson}}
\label{eq:equation13}
\end{equation}
\end{center}
\vspace{0.5em}
\begin{itemize}
\item \textbf{\underline{Cette approche est divisée en trois étapes :}}
\end{itemize}
 
\begin{itemize}[font=\color{black} \Large, label=\ding{43}]
\item Construire une matrice des préférences de l'humeur basée sur la matrice d'évaluation \textbf{user-movie}, les relations entre les films et les émotions et le facteur de volatilité d'évaluation émotionnelle. Chaque rangée de la matrice construite est représentée comme le vecteur émotionnel d'un utilisateur :
\vspace{-3em}
\begin{center}
\begin{equation}
P_{c, E} = \begin{Bmatrix}
p_{c, e} | e\in E, p_{c, e} \in [0, 100]
\end{Bmatrix}
\end{equation}
\end{center}
Où $p_{c,e}$ est le poids qui mesure la préférence de l'utilisateur $c$ à une émotion spécifique $e \in  E$.
\newline 
Un ensemble des calculs est effectué comme suit : 
\begin{itemize}
\item $r_{c}^{\bar{e}} = \frac{1}{N_{e}} \sum _{e \in E} r_{c, e}$ : mesurer la préférence moyenne de l'utilisateur $c$ pour toutes les émotions, $C_{1}$ et $C_{2}$ sont des seuils constants, et $\Theta $ est un coefficient exponentiel.
\item Si $vol_{c, e}$ est inférieur à un certain seuil ($C_{1}$), l'émotion $e$ a un effet significatif
sur les préférences de l'utilisateur et $p_{c, e}$ peut être représenté comme $r_{c, e}$.
\item Si $vol_{c, e}$ est au-delà d'un certain seuil ($C_{1}$), l'émotion $e$ a peu ou pas d'effet sur les préférences de l'utilisateur et $p_{c, e}$ peut être représenté comme la valeur moyenne $r_{c, e}$. 
\end{itemize}
\item Calcul de la similarité entre l'utilisateur $x$ et $y$ :
\vspace{-3em}
\begin{center}
\begin{equation}
sim(x, y)_{mood-pearson} = \frac{\sum_{e \in E} (p_{x, e} - \bar{p}_{x})(p_{y, e} - \bar{p}_{y})}{\sqrt{\sum_{e \in E}(p_{x, e} - \bar{p}_{x})^{2} \sum_{e \in E} (p_{y, e} - \bar{p}_{y})^{2}}}
\end{equation}
\end{center}
\vspace{-1em}
\begin{center}
\fbox{
Où $p_{x}$ signifie les préférences émotionnelles moyennes de l'utilisateur $x$. }
\end{center}
\item La valeur prédite $r_{c, m}$ peut être calculée comme dans l'équation \ref{eq:equation14} et des recommandations basées sur l'humeur peuvent alors être générées.
\vspace{-3em}
\begin{center}
\begin{equation}
r_{c, m} =\bar{r}_{c} + \frac{\sum_{c' \in \hat{C}} sim(c, c')_{mood-pearson} \times (r_{c', m}-\bar{r}_{c'})}{\sum_{c' \in \hat{C}}|sim(c, c')_{mood-pearson}|} 
\label{eq:equation14}
\end{equation}
\end{center}
\vspace{-1em}
\end{itemize}
\subsubsection{Approche du FC hybride basée sur l'humeur}
\begin{itemize}
\item \textbf{Approche de fusion de similarité} : Pour rechercher des voisins plus proches, Wang et al. ont proposé une méthode de recherche KNN en plusieurs étapes. Ensuite, les auteurs ont prédit les films non évalués pour les utilisateurs. Le processus d'algorithme est décrit comme suit :
\begin{enumerate}
\item Mesure des similarité de chaque utilisateur $x$ et des autres utilisateurs en fonction de l'approche du \textbf{FC} traditionnelle, et construire un vecteur $sim_ tran_{x}$ pour l'utilisateur $x$.
\vspace{-3em}
\begin{center}
\begin{equation}
sim-tran_{x} = \begin{Bmatrix}
sim (x, y)_tran-pearson \mid \\
\mid S_{xy}\mid \geq min-corated-num
\end{Bmatrix}
\end{equation}
\label{eq:equation15}
\end{center}
\vspace{-0.05em}
\begin{center}
\fbox{
Où min-corated-num signifie le nombre minimum de séquences vidéo.
}
\end{center}
\item Mesure des similarité de chaque utilisateur $x$ et d'autres utilisateurs en fonction du FC basé sur l'humeur, et construire une ambiance de simulation vectorielle $sim_mood_{x}$ pour l'utilisateur $x$.
\vspace{-3em}
\begin{center}
\begin{equation}
sim-tran_{x} = \begin{Bmatrix}
sim (x, y)_mood-pearson \mid \\
\mid S_{xy}\mid \geq min-corated-num
\end{Bmatrix}
\end{equation}
\label{eq:equation15}
\end{center}
\vspace{-1em}
\item Trouver les plus proches voisins pour l'utilisateur $x$ basé sur $sim_tran_{x}$ et $sim_mood_{x}$. 
\newline
Trouver d'abord la k-ème valeur maximale $sim_tran_{xk}$ et les valeurs qui ne sont pas inférieures à D\% $0 \leq D \leq 100$ à partir du vecteur $sim_tran_{x}$, en les considérant comme les plus proches voisins candidats. Ensuite, sélectionner le même nombre de valeurs maximales à partir du vecteur $sim_mood_{x}$ avec la  construction de deux groupes de voisins les plus proches pour l'utilisateur $x$.
\newline 
$G1 = \begin{Bmatrix}
y \mid sim{x, y}_{tran-pearson} \in [sim-tran_{xk} \times (2- D \%), sim-tran_{x1}] 
\end{Bmatrix}$
\vspace{1em}
$ \\ G2 = \begin{Bmatrix}
y \mid sim{x, y}_{tran-pearson} \in [sim-tran_{xk} \times D \%, \\ 
 sim-tran_{xk} \times (2- D \%)], sim(x, y)_{mood-pearson} \geq sim-mood_{xk} \times D\%
\end{Bmatrix} $ 
\newpage 
\vspace{2em}
Les utilisateurs de ces deux groupes sélectionnés sont  considérés comme les plus proches voisins de l'utilisateur $x$ : $ KNN_{x} = \begin{Bmatrix}
y \mid y \in G1 \cup G2
\end{Bmatrix} $ 

\item Les évaluations prédites $r_{c, m}$ peuvent être calculées pour l'utilisateur $c$ comme dans l'équation \ref{eq:equation13}
\end{enumerate}
\end{itemize}
\subsection{Étude expérimentale}
\begin{figure}[H]
\centering
\includegraphics[width=9cm]{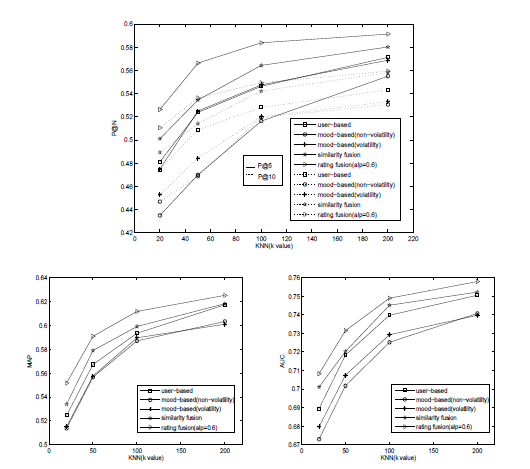}
\caption{Comparaison des performances de différentes approches FC pour le dataset Moviepilot sélectionné en ce qui concerne $k$ le plus proche voisin \cite{wang2010new}}
\label{fig23:my_label}
\end{figure}

\hspace*{3ex} Dans cette sous-section, les auteurs ont présenté les résultats expérimentaux moyens pour 160 utilisateurs de test de la figure \ref{fig23:my_label}, respectivement pour $MAP$, $P@N$ et $AUC$. Alors, ils définissent $C_{1}$ = 0.1, $C_{2}$ = 0.5, $\Theta $ = 2 et $\alpha$ = 0.6 en raison de meilleurs résultats avec ces valeurs.
\newline 
\hspace*{3ex} Les résultats de la figure  \ref{fig23:my_label} montrent que les deux approches (c'est-à-dire la fusion prédictive, la fusion par similarité) surpassent les autres approches du FC en termes des trois métriques d'évaluation. Ainsi, en incorporant l'humeur dans FC en utilisant la fusion de similitude et fusion prédictive.
\newline 
\hspace*{3ex} Les stratégies de fusion des évaluations peuvent améliorer la précision des recommandations sur la piste \textbf{Moviepilot}. Cependant, les deux autres solutions proposées de la FC axée sur l'humeur qui ne prennent pas en compte la volatilité des évaluations d'humeur, ne sont pas meilleures que les approches du FC traditionnelles. 
Ce résultat indique que seuls les états d'humeur ou les préférences d'humeur de l'utilisateur peuvent ne pas suffire à générer des recommandations précises.
\newline 
\hspace*{3ex} De plus, la figure \ref{fig23:my_label} montrvoisin le plus proche [e seulement les résultats expérimentaux moyens, tandis que pour quelques certains utilisateurs, l'approche à base d'humeur ne fonctionne pas plus mauvais que FC traditionnel. À partir des résultats de l'expérience, que la volatilité (basée sur l'humeur) permet d'obtenir une meilleure précision des recommandations que celle basée sur l'humeur (non-volatilité) pour les trois mesures d'évaluation. Il est donc significatif de présenter la volatilité dans l'approche de filtrage collaboratif basé sur l'humeur.
Par ailleurs, $P@5$ est meilleur que $P@10$ pour toutes les approches, alors que ce n'est pas très intéressant puisque la performance semble augmenter de façon monotone avec la valeur de $k$.

\section{Étude comparative}
\hspace*{3ex} Dans cette section, nous allons mener une comparaison théorique entre les approches que nous avons détaillées auparavant. Le tableau \ref{tab:19} présente une comparaison théorique entre les approches des systèmes de recommandation sensibles au contexte par rapport aux films. La comparaison est faite suivant les axes suivants : 
\begin{itemize}
\item \textbf{Type de contexte} : cet axe décrit les types de contexte utilisés dans les différentes approches des systèmes de recommandation sensibles au contexte.
\item \textbf{Source de données} : cet axe précise les sources de données utilisées dans les systèmes de recommandation contextuels.
\item \textbf{Techniques utilisées} : cet axe indique les différents techniques utilisées dans les systèmes de recommandation contextuels par rapport aux films. 
\end{itemize}
\begin{center}
{\setlength {\tabcolsep}{12pt}
\begin{tabular}{|p{3cm}||p{3cm}|p{3.9cm}|p{4.5cm}|}
  \hline 
  \textbf{Approches} & \textbf{Type de contexte} & \textbf{Source de données} & \textbf{Techniques utilisées} \\ 
  \hline \hline
Ostuni et al. \cite{ostuni2012cinemappy} & Emplacement  & Système DBPEDIA & Pré-filtrage contextuel   \\
& &  & Post-filtrage contextuel \\
& &  & Recommandation basée \\
& &  &  sur le contenu \\
\hline
Campos et al. \cite{campos2013context} & Social et temporel & Les sites web & Pré et post-filtrage contextuel \\
& &  & La modélisation contextuelle \\
\hline 
Gantner et al. \cite{gantner2010factorization} & Temporel & MoviePilot & Pairwise Interaction Tensor Factorization (PITF) \\
  \hline 
Biancalana et al. \cite{biancalana2011context} & Social & Réseau de neurones & Filtrage collaboratif  \\
& &  & Machine learning \\
&  &  & Réseau de neurones \\
\hline
Shi et al. \cite{shi2013mining} & humeur & Les mots clés & Filtrage collaboratif \\
&  &  & Joint Matrix Factorization (JMF) \\
\hline 
Said et al. \cite{said2010putting} & Social et humeur  & Les réseaux sociaux & Pairwise Interaction Tensor Factorization (PITF) \\
&   &   & KNN \\
&  &  &  Factorisation matricielle \\
&  &  & Filtrage collaboratif \\
\hline 
Wang et al. \cite{wang2010new} & humeur & Les sites web & 
Filtrage collaboratif \\
\hline 
\end{tabular}
\caption{Comparaison théorique des approches dans la littérature}
\label{tab:19}
}
\end{center}
\section{Limites et discussion}
\hspace*{3ex} les systèmes de recommandation ont été développés en vue de faciliter l'accès aux items pertinents. Leur objectif est d'anticiper les besoins de l'utilisateur en lui fournissant des recommandations d'items jugés pertinents par rapport à ses goûts. À la lumière de ce qui a été dans ce chapitre et à partir du tableau \ref{tab:19}, nous pouvons conclure que toutes les approches citées dans la littérature sont différentes aux niveaux de type de contexte et les techniques utilisées pour effectuer une recommandation. Parmi les limites des systèmes de recommandation, nous avons identifié la non-prise en compte du contexte dans lequel l'utilisateur décide de faire une recommandation.
\section{Conclusion}
\hspace*{3ex} Dans ce chapitre, nous avons présenté quelques approches des systèmes de recommandation sensibles au contexte dans le domaine des films. Par ailleurs tout le long de ce chapitre, nous avons constaté que les approches ont presque toutes l'objective de construire un système de recommandation contextuelle. Pour notre travail de mémoire, nous nous intéressons à la proposition d'une approche de recommandation contextuelle en se basant sur la méthode d'Analyse Hiérarchique des Procédés (AHP).
\chapter{Nouvelle approche de recommandation contextuelle en se basant sur la méthode d'Analyse Hiérarchique des Procédés (AHP)}

\section{Introduction}
\hspace*{3ex} Dans un système de recommandation, les éléments d'intérêt et les préférences de l'utilisateur sont représentés sous diverses formes, e.g utiliser un ou plusieurs attributs pour décrire un article. Particulièrement dans les systèmes où les recommandations sont basées sur l'opinion d'autrui, il est crucial de prendre en considération les multiples critères qui affectent les opinions des utilisateurs afin de faire des recommandations plus efficaces. Bien que les systèmes de recommandation utilisent déjà plusieurs attributs pour la production de recommandations, la recherche sur la façon dont les méthodes de prise des décisions multi-critères \textbf{(MCDM)} peuvent faciliter le processus de création d'une recommandation, peut être encore considérée comme sporadique. 
\newline
\hspace*{3ex} Après avoir étudié l'état de l'art des systèmes de recommandation sensibles au contexte dans le domaine des films. Dans ce chapitre, nous commençons par la description des méthodes 
\newpage
\vspace*{0.6em}
de prise de décision multi-critères \textbf{(MCDM)}. Par la suite, nous allons  présenter la méthode d'Analyse Hiérarchique des Procédés et l'intégration de cette méthode dans le processus de recommandation. Enfin, nous allons finir par une conclusion.
\section{Description des méthodes de prise de décision multicritères (MCDM) }
\hspace*{3ex} Dans cette section, nous intéressons aux  méthodes de prise de décision multi-critères en donnant leurs définitions, les étapes de prise de décision et nous finissons par le processus de fonctionnement des \textbf{MCDM}.

\subsection{Définition des méthodes de prise de décision à critères multiples (MCDM)}
\textbf{Définition 37}
\textbf{MCDM} est l'une des méthodologies de décision les plus utilisées dans divers domaines tels que : énergie et environnement, affaires, économie, production, etc. Les techniques et approches des \textbf{MCDM} améliorent la qualité des décisions en rendant le développement plus efficace, rationnel et explicite \cite{mardani2015multiple}.
\vspace*{3mm}
\newline
\textbf{Définition 38} 
la prise de décision multi-critères \textbf{(MCDM)} est une technique de prise de décision basée sur plusieurs alternatives existantes ou une théorie qui explique le processus de prise de décision en considérant plusieurs critères. Afin de modéliser les problèmes de recommandation en tant que MCDM, il faut suivre quatre étapes générales de la méthodologie de modélisation pour prendre une décision sur le problème \cite{manouselis2007analysis} : 
\begin{itemize}
\item \textbf{Objectif de décision} : il définit l'objectif sur lequel les décisions doivent être prises et les raisons de la recommandation de décision. 
\item \textbf{Famille de critères} : il définit l'identification et la modélisation d'un ensemble de critères influençant la décision, ainsi qu'une recommandation complète et non redondante.
\item \textbf{Modèle de préférence globale} : il définit la fonction d'agrégation pour la préférence marginale sur chaque critère à la préférence globale du décideur pour chaque élément.
\item \textbf{Processus d'aide à la décision} : il définit l'étude des différentes catégories et types de systèmes de recommandation qui peuvent être utilisés pour appuyer les recommandations des décideurs, conformément aux résultats des étapes précédentes.
\end{itemize}
\hspace*{3ex} La mise en œuvre de la méthode de prise de décision multi-critères \textbf{(MCDM)} dans un système de recommandation n'a pas encore été explorée systématiquement. Un système de recommandation est capable d'expliquer certaines contributions du système qui impliquent plusieurs méthodes MCDM. 
\newline 
\hspace*{3ex} \textbf{MCDM} est une théorie de prise de décision qui considère un ensemble limité d'options alternatives par rapport à de nombreux critères. Le problème dans \textbf{MCDM} peut être formulé comme suit:
\begin{itemize}
\item Supposons qu'il existe $M$ critères et $N$ alternatives. Nous devons choisir une partie ou une série d'alternatives répondant à des critères aussi élevés que possible \cite{fulop2005introduction}.
\end{itemize}
\hspace*{3ex} Le problème \textbf{MCDM} peut être modélisé dans la matrice de décision ci-dessous : 
\begin{center}
{\setlength {\tabcolsep}{12pt}
\begin{tabular}{|c||c|c|c|c|c|}
  \hline 
   
 \textbf{Critères avec leurs poids}  &  $ C_{1}$ & $ C_{2}$ & $ C_{3}$ & ... & $ C_{N}$    
  \\
  \hline \hline 
\textbf{Alternatives} &  $ W_{1}$ & $ W_{2}$ & $ W_{3}$ & ... & $ W_{N}$ \\
\hline 
$A_{1}$ & $a_{11}$ & $a_{12}$ & $a_{13}$  & ... & $a_{1N}$   
\\
$A_{2}$ & $a_{21}$ & $a_{22}$ & $a_{23}$  & ... & $a_{2N}$   
\\
$A_{3}$ & $a_{31}$ & $a_{32}$ & $a_{33}$  & ... & $a_{3N}$   
\\
. &  .  &  .  &  . & ... & . \\
. &  .  &  .  &  . & ... & . \\
. &  .  &  .  &  . & ... & . \\
$A_{M}$ & $a_{M1}$ & $a_{M2}$ & $a_{M3}$  & ... & $a_{MN}$   
\\
\hline 
\end{tabular}
\caption{Matrice de décision \cite{fulop2005introduction}}
\label{tab:20}
}
\end{center}
\hspace*{3ex} La matrice de décision est une matrice de taille $M \times  N$ où l'élément $a_{ij}$ indique la performance de l'alternative $A_{i}$ lorsqu'elle est évaluée par rapport au critère $C_{j}$ (pour $i$ = 1,2,3, ..., $M$ et $j$ = 1,2,3, ..., $N$).

\subsection{Les étapes de prise de décision}
\hspace*{3ex} Un processus de prise de décision implique les étapes suivantes à suivre:
\begin{center}
{\setlength {\tabcolsep}{12pt}
\begin{tabular}{|c||c|}
  \hline 
 \textbf{Étapes} & \textbf{Description} \\
  \hline \hline
\textbf{1} &   Identifier l'objectif / le but du processus de prise de décision \\
\hline 
\textbf{2} & Sélection des critères / paramètres / facteurs / décision \\
\hline 
\textbf{3} & Sélection des alternatives \\
\hline 
\textbf{4} & Sélection des méthodes de pondération pour représenter l'importance des critères \\
\hline 
\textbf{5} & Méthode d'agrégation \\
\hline 
\textbf{6} & Prise de décision basée sur les résultats d'agrégation \\
\hline 
\end{tabular}
\caption{Les étapes de la prise de décision}
\label{tab:21}
}
\end{center}
\subsection{Principe de fonctionnement}
\hspace*{3ex} Dans cette partie, nous nous intéressons au principe de fonctionnement des méthodes de prise de décision à critères multiples \textbf{(MCDM)}. Le processus du \textbf{MCDM} suit un principe de fonctionnement commun comme décrit dans le tableau \ref{tab:222}

\begin{center}
\begin{tabular}{|c||p{6cm}|p{8cm}|}
  \hline 
  & \textbf{Étapes} &  \textbf{ \centering Propriétés}\\
\hline  \hline
\textbf{1} & Sélection des critères & 
Les critères sélectionnés doivent être : cohérent avec la décision, indépendant les uns des autres, représenté dans la même échelle et mesurable. \\
\hline 
\textbf{2} &  Sélection des alternatives & 
Les alternatives sélectionnées doivent être :
Disponible, comparable, vrai pas idéal et pratique / faisable \\
\hline
\end{tabular}
\end{center}
\begin{center}
\begin{tabular}{|c||p{6cm}|p{8cm}|}
\hline 
\hline 
\textbf{3} & Sélection des méthodes de pondération pour représenter l'importance des critères &   
Exemple des méthodes compensatoires : 
\begin{itemize}
\item Analyse Hiérarchique des Procédés
\item Processus  décisionnel multi-critères flous \textbf{(FDM)}, etc
\end{itemize} \\
\hline 
& & Exemple des méthodes de classement :
\begin{itemize}
\item Élimination et Choix Exprimant la Réalité \textbf{(ELECTRE)} 
\item Classement des Préférences
\item Méthode d'organisation pour l'enrichissement des évaluations  \textbf{(PROMETHUS)}
\end{itemize} \\
\hline 
\textbf{4} & Méthode d'agrégation &  Peut être un produit, une moyenne ou une fonction \\
\hline
\end{tabular}
\end{center}
\caption{Principe de fonctionnement de \textbf{MCDM}}
\label{tab:222}
\begin{itemize}[font=\color{black} \Large, label=\ding{43}]
\item Pour les méthodes d'agrégation : le résultat de cette agrégation séparera la meilleure alternative des options disponibles.
\end{itemize}
 
\hspace*{3ex} \textbf{AHP} et \textbf{FLDM} sont deux exemples les plus populaires de \textbf{MCDM} qui sont largement utilisés pour résoudre des problèmes de prise de décision et dans divers systèmes d'aide à la décision dans le monde. Dans la section suivante, nous allons intéresser à \textbf{la méthode d'Analyse Hiérarchique des Procédés (AHP)}.
\section{Méthode d'Analyse Hiérarchique des Procédés (AHP)}
\hspace*{3ex} Dans cette section, nous définissons la méthode d'Analyse Hiérarchique des Procédés. Nous commençons, tout d'abord, par l'explication du principe de la méthode AHP ainsi que sa mise en œuvre. Ensuite, nous préciserons les caractéristiques de cette méthode de prise de décision et nous finirons par une discussion et quelques critiques.
\subsection{Définition de la méthode d'analyse hiérarchique des procédés (AHP)}
\textbf{Définition 39} \textbf{AHP} introduit par Thomas Saaty (1980), est un outil efficace pour traiter les décisions complexes et pouvoir aider le décideur à établir des priorités et à prendre la meilleure décision. En réduisant les décisions complexes à une série de comparaisons par paires, puis en synthétisant les résultats, AHP aide à saisir les aspects subjectifs et objectifs d'une décision. De plus, AHP intègre une technique utile pour vérifier la cohérence des évaluations du décideur, réduisant ainsi les biais dans le processus décisionnel.
\subsection{Fonctionnement de la méthode AHP}
\hspace*{3ex} AHP considère un ensemble des critères d'évaluation et un ensemble d'options alternatives parmi lesquelles la meilleure décision doit être prise. Il est important de noter que certains des critères peuvent être contrastés et il n'est pas vrai que la meilleure alternative est celle qui optimise chaque critère unique, plutôt celle qui réalise le compromis le plus approprié parmi les différents critères. 
\newline
\hspace*{3ex} AHP génère un poids pour chaque critère d'évaluation selon les comparaisons par paires des critères du décideur. Plus le poids est élevé, plus le critère correspondant est important. Ensuite, pour un critère fixe, la méthode AHP attribue un score à chaque alternative en fonction des comparaisons par paires des alternatives basées sur ce critère. Plus le score est élevé, la performance de l'option est meilleure par rapport au critère considéré. Enfin, cette méthode combine les pondérations de critères et les scores des évaluations, déterminant ainsi un score global pour chaque option et un classement qui résulte. Le score global pour une option donnée est une somme pondérée des scores obtenus par rapport à tous les critères.

\subsection{Caractéristiques de l'AHP}
\hspace*{3ex } AHP est un outil très flexible et puissant car les scores, et donc le classement final, sont obtenus sur la base des évaluations relatives du comparaison par paires des critères et des alternatives fournies par l'utilisateur. Les calculs effectués par AHP sont toujours guidés par l'expérience du décideur et cette méthode peut donc être considérée comme un outil capable de traduire les évaluations (qualitatives et quantitatives) faites par le décideur dans un classement multi-critère. De plus, AHP est simple car il n'est pas nécessaire de construire un système expert complexe avec les connaissances du décideur intégré.
\newline
\hspace*{3ex} D'autre part, AHP nécessite un grand nombre d'évaluations par l'utilisateur, en particulier pour les problèmes avec de nombreux critères et alternatives. Bien que chaque évaluation unique est très simple, puisqu'elle n'exige que le décideur d'exprimer comment deux alternatives ou critères se comparent, la charge de la tâche d'évaluation peut devenir déraisonnable. En fait, le nombre de comparaisons par paires croît quadratiquement avec le nombre de critères et des alternatives.
\newline 
\textbf{Exemple 4} 
En comparant 10 alternatives sur 4 critères, 
$ 4 \times \frac{3}{2} = 6$ comparaisons sont demandées pour construire le vecteur de poids, et 
$4 \times (10 \times \frac{9}{2}) = 180$ comparaisons par paires sont nécessaires pour construire la matrice de score.
\newline 
\hspace*{3ex} Toutefois, afin de réduire la charge de travail du décideur, AHP peut être complètement ou partiellement automatisée en spécifiant des seuils appropriés pour décider automatiquement certaines comparaisons par paires.
\subsection{Mise en œuvre de la méthode AHP}
\hspace*{3ex} La méthode AHP consiste à représenter un problème de décision par une structure hiérarchique reflétant les interactions entre les divers éléments du problème, à procéder ensuite à des comparaisons par paires des éléments de la hiérarchie, et enfin à déterminer les priorités des actions. La méthodologie de l'AHP peut être expliqué dans les étapes suivantes : 
\begin{itemize}
\item \textbf{Étape 1} : Décomposer le problème en une hiérarchie d'éléments inter-reliés. Au sommet de la hiérarchie, on trouve l'objectif et dans les niveaux inférieurs on trouve les éléments contribuant à atteindre cet objectif. Ceci est la partie la plus importante et créative de la prise de décision. 
\begin{figure}[H]
\centering
\includegraphics[width=16cm]{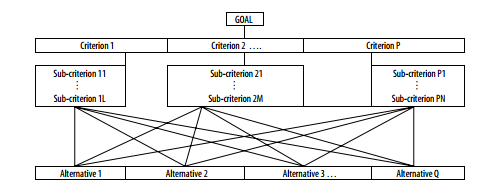}
\caption{Structure hiérarchique générique \cite{saaty1988analytic}}
\label{figure24:my_label}
\end{figure}
\begin{figure}[H]
\centering
\includegraphics[width=16cm]{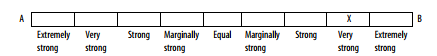}
\caption{Format pour les comparaisons par paires \cite{saaty1988analytic}}
\label{figure241:my_label}
\end{figure}
\hspace*{3ex} La classification des hiérarchies se distingue en deux classes : 
\begin{itemize}
\item \textbf{\underline{Hiérarchie structurelle}} : 
consiste à décomposer les objets que nous percevons que nos sens, en groupes, en sous-groupes et en ensembles encore plus petits.
\item \textbf{\underline{Hiérarchie fonctionnelle}} : 
consiste à décomposer un système complexe en éléments en fonction de leur relation essentielle.
\end{itemize}
\hspace*{3ex} Chaque élément dans une hiérarchie occupe un niveau. Le sommet de la hiérarchie est constitué d'un élément unique : \textbf{l'objectif à atteindre}. Les autres niveaux contiennent plusieurs éléments. Saaty et al \cite{saaty2004decision} propose entre 5 et 9 éléments sous le sommet pour effectuer une bonne analyse. Les éléments d'un même niveau doivent être comparés entre eux en fonction d'un élément du niveau supérieur et doivent être du même ordre de grandeur. 

\item \textbf{Étape 2} : Procéder à des comparaisons par paires des éléments de chaque niveau hiérarchique par rapport à un élément du niveau hiérarchique supérieur. Cette étape permet de construire des matrices de comparaisons. Les valeurs de ces matrices sont obtenues par la transformation des jugements en valeurs numériques selon l'échelle de Saaty (Échelle de comparaisons binaires), tout en respectant le principe de réciprocité : $P_{c}(E_{A}, E_{B}) = \frac{1}{P_{c}(E_{B}, E_{A})}$
\begin{center}
\begin{tabular}{|c||p{5cm}|p{5cm}|}
  \hline 
 \textbf{Degré d'importance} &  \textbf{Définition} & \textbf{Explication}\\
    
\hline  \hline
\textbf{1} & Importance égale des deux éléments. & Deux éléments contribuent autant à la propriété.\\
\hline 
\textbf{3} &  Faible importance d'un élément par rapport à un autre.  & L'expérience et l'appréciation   personnelles favorisent légèrement un élément à un autre. \\
\hline
\textbf{5} &  Importance forte ou déterminante d'un élément par rapport à un autre. & 
L'expérience et l'appréciation personnelles favorisent fortement un élément à un autre. \\
\hline
\textbf{7} &  Importance attestée d'un élément par rapport à un autre. &  Un élément est fortement  favorisé et sa dominance est attestée dans la pratique. \\ 
\hline  
\textbf{9} &  Importance absolue d'un élément par rapport à un autre. &  Les preuves favorisant un élément par rapport à un autre sont aussi convaincantes que possible. \\
 \hline 
\textbf{2, 4, 6, 8} &  Valeurs intermédiaires entre deux appréciations voisines. & Un compromis est nécessaire entre deux appréciations. \\
\hline
\textbf{Réciprocité} & \multicolumn{2}{c|}{Si l'élément $i$ se voit attribuer l'un des chiffres précédents } \\
& \multicolumn{2}{c|}  { lorsqu'elle est comparée à l'élément $j$, $j$ aura donc la valeur} \\
&  \multicolumn{2}{c|}   { inverse lorsqu'on la compare à $i$.} \\
\hline
\end{tabular}
\caption{Échelle de Saaty \cite{saaty1977scaling}}
\label{tab:23}
\end{center}
\newpage
\item \textbf{Étape 3} : Déterminer l'importance relative des éléments en calculant les vecteurs propres correspondants aux valeurs propres maximales des matrices de comparaisons. 
\item \textbf{Étape 4}: Vérifier la cohérence des jugements. 
\begin{itemize}[label=\textbullet,font=\color{black}]
\item On calcule d'abord, l'indice de cohérence $IC$ : 

\vspace{-3.5em}
\begin{center} 
\begin{equation}
IC = \frac{\lambda _{max} - n}{n - 1}  
\end{equation}
\end{center}
\vspace{-1em}

Où : $\lambda_{max}$ est la valeur propre maximale correspondant à la matrice des comparaisons par paires et $n$ est le nombre d'éléments comparés.
\item On calcule le ratio de cohérence $(RC)$ définit par : 
\vspace{-3.5em}
\begin{center} 
\begin{equation}
RC = 100 . \frac{IC}{ACI}
\end{equation}
\end{center}
\vspace{-1em}
Où $ACI$ est l'indice de cohérence moyen obtenu en générant aléatoirement des matrices de jugement de même taille.
\end{itemize}
\begin{center}
\begin{tabular}{|p{3cm}||c|c|c|c|c|c|c|c|c|c|}
  \hline 
\textbf{Dimension de la matrice} & 1 & 2 & 3 & 4 & 5 & 6 & 7 & 8 & 9 & 10 \\
\hline 
\textbf{Cohérence aléatoire $(ACI)$ } & 0.00 & 0.00 & 0.58 & 0.90 & 1.12  & 1.24 & 1.32 & 1.41 & 1.45 & 1.49
\\
\hline
\end{tabular}
\caption{Indice de cohérence moyen}
\label{tab:24}
\end{center}
\item \textbf{Étape 5} : Établir la performance relative de chacune des actions. 
\vspace{-3em}
\begin{center} 
\begin{equation}
P_{k}(e^{k}_{i}) = \sum_{j=1}^{n_{k-1}} P_{k-1}(e^{k-1}_{i}) . P_{k}(e^{k}_{i} / e^{k-1}_{i}) \hspace{3ex} avec \hspace{1ex}  \sum_{j=1}^{n_{k}}P_{k}(e^{k}_{i}) = 1
\end{equation}
\end{center}
\vspace{-1em}
Où $n_{k-1}$ est le nombre d'éléments du niveau hiérarchique $k-1$, et $P_{k}(e^{k}_{i} )$ est la priorité accordée à l'élément $e_{i}$ au niveau hiérarchique $k$. 
\end{itemize}
\textbf{\underline {Exemple}} : Une organisation cherche un nouveau directeur pour l'un de ses départements. 
\hspace*{3ex} Une entreprise doit sélectionner un partenaire pour externaliser un processus de fabrication. Le comité de direction a décidé d'exploiter la méthode AHP pour faire son choix.
\begin{itemize}
\item \textbf{Étape 1} : Décomposition du problème en une hiérarchie d'éléments inter-reliés.
Le groupe de décision a retenu la hiérarchie suivante : L'objectif est de sélectionner un partenaire pour sous-traiter le produit en question.
Les critères de sélection retenus sont la qualité du produit fourni, la fiabilité du partenaire et l'économie engendrée par cette relation de partenariat : 4 alternatives sont possibles (4 partenaires ont déposé des offres).  
\begin{figure}[H]
\centering
\includegraphics[width=13.5cm]{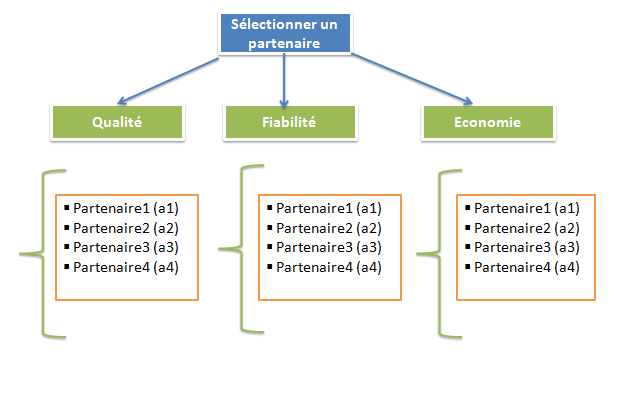}
\caption{Décomposition hiérarchique}
\label{figure25:my_label}
\end{figure}
\item \textbf{Étapes 2 et 3} : Comparaisons par paires des éléments de chaque niveau hiérarchique et détermination de l'importance relative des éléments. Après discussion, le groupe de décision s'est mis d'accord que : 
\begin{itemize}[label=\textbullet,font=\color{black}]
\item La fiabilité est légèrement plus importante que la qualité.
\item La qualité a une importance faible par rapport à l'économie.
\item La fiabilité a une importance modérée par rapport à l'économie.
\end{itemize}
Ces jugements sont traduits dans la matrice suivante : 
\begin{center}
\begin{tabular}{|c||c|c|c|}
  \hline 
  & \textbf{Qualité} &  \textbf{Fiabilité} & \textbf{Économie}\\
\hline  \hline
\textbf{Qualité} & 1 & 1/2 & 3 \\
\hline 
\textbf{Fiabilité} & 2 & 1 & 4 \\
\hline 
\textbf{Économie} & 1/3 & 1/4 & 1  \\
\hline
\end{tabular}
\caption{Matrice des critères de décision}
\label{tab:25}
\end{center}
\newpage
Après la création de la matrice des critères de décision, Le traitement de cette matrice donne : 
\newline 
$\lambda _{max} = 3.0183 $ ; $ w_{1} = \begin{pmatrix}
0.4481 \\
0.8527 \\
0.1862 \\  
\end{pmatrix} $ ; après normalisation, on obtient : 
$\bar{w}_{1} = \begin{pmatrix}
0.3196 \\ 
0.5584 \\
0.1220\\ 
\end{pmatrix}$

\begin{center}
\begin{tabular}{|c||c|c|}
  \hline 
  \textbf{Critères} &  \textbf{Résultats} & \textbf{Description}\\
\hline  \hline
\textbf{Qualité} & 0.3196 & Le second critère important\\
\hline 
\textbf{Fiabilité} & 0.5584 & Le critère le plus important \\
\hline 
\textbf{Économie} & 0.1220  &  Le critère le moins important\\
\hline
\end{tabular}
\caption{Description des critères}
\label{tab:26}
\end{center}
En termes de qualité et de fiabilité, les comparaisons par paires des actions sont récapitulées dans les matrices suivantes : 

\begin{figure}[H]
\centering
\includegraphics[width=13cm]{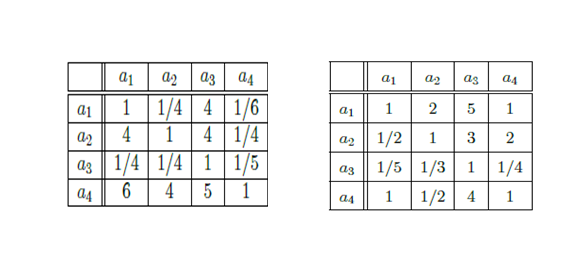}
\caption{Matrices des comparaisons par paires}
\label{figure26:my_label}
\end{figure}

\begin{center}
\begin{tabular}{|c||c|}
  \hline 
  \textbf{En termes de qualité} &  \textbf{En termes de fiabilité} \\
\hline  \hline
$\lambda _{max} = 4.4347 $ & $\lambda _{max} = 4.1913 $ \\ 
\hline 
 $ \bar{w_{2}} = \begin{pmatrix}
0.1160 \\
0.2470 \\
0.0600 \\
0.5770 \\  
\end{pmatrix} $  & 

$ \bar{w_{3}} = \begin{pmatrix}
0.3790 \\
0.2900 \\
0.0740 \\
0.2570 \\  
\end{pmatrix} $  \\
\hline
\end{tabular}
\caption{Normalisation des matrices pour les deux critères : qualité et fiabilité}
\label{tab:27}
\end{center}
Donc, si on ne tient compte que du critère qualité, c'est le partenaire 4 qui sera sélectionné. Toutefois, si on ne prend compte que le critère fiabilité, c'est le partenaire 1 qui sera retenu. 
L'information concernant l'économie engendrée par l'association avec un partenaire est récapitulée dans le tableau suivant : 
\begin{center}
\begin{tabular}{|c||c|c|}
  \hline 
  & \textbf{Économie} &  \textbf{Normalisation} \\
\hline  \hline
\textbf{$a_{1}$} & 34 & 34/113 = 0.3010 \\
\hline 
\textbf{$a_{2}$} & 27  & 27/113 = 0.2390  \\
\hline 
\textbf{$a_{3}$} & 24  & 24/113 =  0.2120  \\
\hline
\textbf{$a_{4}$} & 28  & 28/113 =   0.2480   \\
\hline 
\textbf{Total} & 113 & 1\\
\hline 
\end{tabular}
\caption{Normalisation de matrice pour le critère : Économie}
\label{tab:28}
\end{center}

La normalisation des données relatives à l'économie nous a permis de déterminer le vecteur de priorité $\bar{w_{4}}$ :
\begin{center}
 $ \bar{w_{4}} = \begin{pmatrix}
0.3010 \\
0.2390 \\
0.2120\\
0.2480 \\  
\end{pmatrix} $ 
\end{center} 
\item \textbf{Étape 4 } : Évaluation des cohérences des jugements : 
\begin{center}
\begin{tabular}{|c||c|c|c|c|c|}
  \hline 
  \textbf{Matrice} &  \textbf{M1} & \textbf{M2} & \textbf{M3} & \textbf{M4} \\
\hline  \hline
\textbf{$\lambda _{max}$} & 3.0183 & 4.4347 & 4.1913 & ***  \\
\hline 
\textbf{IC } & 0.0092 & 0.1449 & 0.0638 & ***   \\
\hline 
\textbf{ACI } & 0.58 & 0.90 & 0.90 & 0.90  \\
\hline
\textbf{RC} & 1.6 $\%$ & 16.1 $\%$ &  7.1 $\%$ & ***\\
\hline 
\end{tabular}
\caption{Évaluation des cohérences des jugements}
\label{tab:29}
\end{center}
\newpage
\item \textbf{Étape 5} : Détermination de la performance relative de chacune des actions. 
\begin{figure}[H]
\centering
\includegraphics[width=15cm]{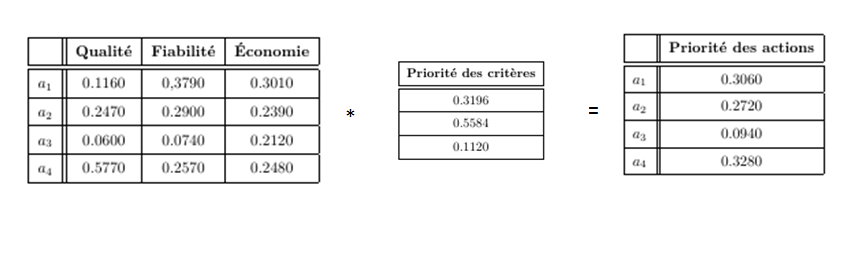}
\caption{Détermination de la performance relative des actions}
\label{figure27:my_label}
\end{figure}
D'après le vecteur des priorités des actions, on conclut que le partenaire 4 ($a_{4}$) a
proposé la meilleure offre, suivi du partenaire 1 ($a_{1}$), puis des partenaires 2 ($a_{2}$) et 3 ($a_{3}$). 
\end{itemize}
\subsection{Critiques}
\hspace*{3ex} Les points forts de la méthode AHP sont la modélisation du problème de décision par une
structure hiérarchique et l'utilisation d'une échelle sémantique pour exprimer les préférences du décideur. 
Bien qu'elle soit très populaire, la méthode AHP a fait l'objet de plusieurs critiques :
\begin{itemize}
\item Un grand nombre d'éléments dans le problème de décision fait exploser le nombre de comparaisons par paires.
\item Le problème de renversement de rang (deux actions peuvent voir leur ordre de priorité s'inverser suite à une modification (ajout ou suppression d'une ou de plusieurs actions) de l'ensemble des actions.
\item L'association d'une échelle numérique à l'échelle sémantique est restrictive et introduit des biais.  
\end{itemize}
\section{Approche proposée de recommandation contextuelle se basant sur la méthode AHP }
\subsection{Description de l'approche}
\hspace*{3ex} La plupart des systèmes de recommandation sensibles au contexte se concentrent d'avantage sur la recommandation de documents pertinents à un utilisateur tout en intégrant des données contextuelles  pour autant considérer le problème de l'évolution des contenus \cite{bouneffouf2014recommandation}. En se basant sur cette caractéristique, nous allons proposer une nouvelle approche de recommandation contextuelle en utilisant la méthode d'Analyse Hiérarchique des Procédés (AHP). Cette méthode est considérée comme un outil efficace pour traiter les décisions complexes et peut aider le décideur à établir des priorités et à prendre la meilleure décision \cite{saaty1988analytic}. Dans ce qui suit, nous allons présenter l'architecture de filtrage et l'intégration de la méthode d'Analyse Hiérarchique des Procédés (AHP) dans le processus de recommandation.
\subsection{Architecture de filtrage}
\hspace*{3ex} L'architecture de filtrage est composée de pré-filtrage contextuel qui permet d'éliminer les données d'évaluation (rating data) qui ne sont pas pertinentes pour la situation ciblée. Les évaluations restantes sont utilisées pour apprendre un modèle local de prédiction d'évaluation et de recommandation. Puis, la méthode des systèmes de recommandation, le filtrage collaboratif (FC), est lancée sur la base de
donnée réduite afin d'obtenir les recommandations liées au contexte C. 
\begin{figure}[H]
\centering
\includegraphics[width=10cm]{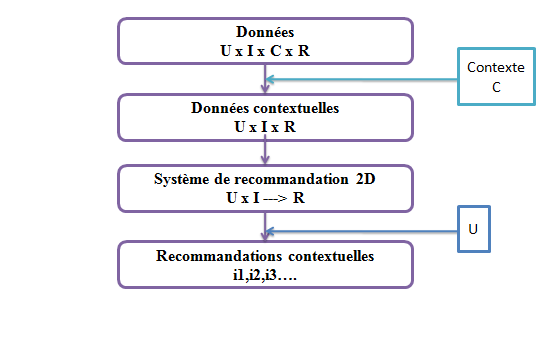}
\caption{Pré-filtrage contextuel \cite{adomavicius2015context}}
\label{figure8:my_label}
\end{figure}
\subsection{Méthodologie de recommandation}
\hspace*{3ex} La méthodologie de recommandation proposée utilise principalement la méthode d'Analyse Hiérarchique des Procédés (AHP). Cette méthodologie se fait en quatre étapes : 
\begin{itemize}
\item Préparation et pré-traitement des données.
\item Extraction de données contextuelles.
\item Intégration de la méthode AHP.
\item Recommandation contextuelle des films.
\end{itemize}
\begin{figure}[H]
\centering
\includegraphics[width=14cm]{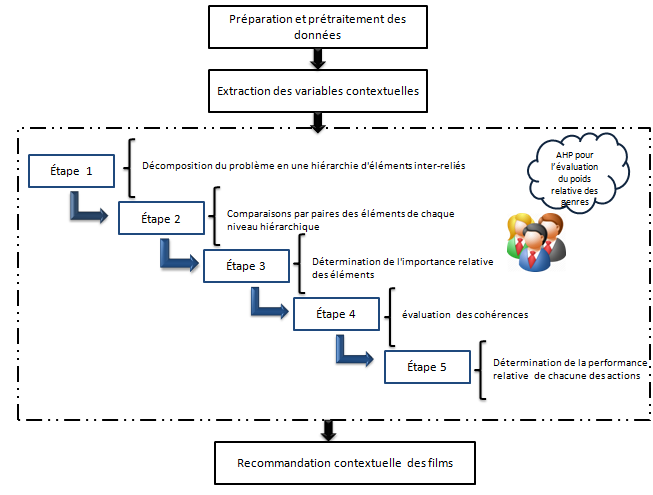}
\caption{Méthodologie de recommandation}
\label{figure9:my_label}
\end{figure}
\subsubsection{Préparation et pré-traitement des données}
\hspace*{3ex} La préparation et pré-traitement des données se fait à partir du dataset LDOS-CoMoDa qui est un dataset de recommandation de films riche en contexte et composée de douze variables contextuelles décrivant la situation dans laquelle les films ont été consommés.
\subsubsection{Extraction de données à partir du dataset \textbf{LDOS-CoMoDa}}
\hspace*{3ex} Dans notre nouvelle approche, nous nécessitons des variables contextuelles pour assurer une recommandation des films sensibles au contexte. Les paramètres de contexte qui ont été utilisés dans le dataset LDOS-CoMoDa sont : le temps, type de jour, saison, emplacement, météo, social, endEmotions, émotions dominantes, ambiance, physique, décision, interaction.
\begin{itemize}
\item \textbf{Temps (time)} : comprend quatre valeurs (Matin, après midi, soir, la nuit).
\item \textbf{Type de jour (datatype)} : comprend trois valeurs (jour ouvrable, week-end, vacances).
\item \textbf{Saison (season)} : comprend quatre valeurs (printemps, été, automne, hiver).
\item \textbf{Emplacement (location)} : comprend trois valeurs (maison, lieu public, maison d'ami).
\item \textbf{Météo (weather)} : comprend cinq valeurs (ensoleillé/clair, pluvieux, orageux, neigeux, nuageux).
\item \textbf{Social} : comprend sept valeurs (seul, partenaire, amis, collègues, parents, public, famille).
\item \textbf{endEmotions (endEmo)} : comprend sept valeurs (triste, heureux, effrayé, surpris, fâché, dégoûté, neutre).
\item \textbf{Émotions dominantes (dominantEmo)} : comprend sept valeurs (triste, heureux, effrayé, surpris, fâché, dégoûté, neutre).
\item \textbf{Ambiance (mood)} : comprend trois valeurs (positif, neutre, négatif).
\item \textbf{Physique (physical)} : comprend seulement deux valeurs (En bonne santé, malade (healthy, ill)).
\item \textbf{Décision (decision)} : comprend deux valeurs (l'utilisateur a choisi l'article, l'article suggéré par d'autres).
\item \textbf{Interaction} : comprend aussi deux valeurs 
(premier, nième).
\end{itemize}
\hspace*{3ex} La figure \ref{figure10:my_label} décrit les douze variables contextuelles utilisées dans LDOS-CoMoDa, le nombre de classes catégoriels ou ordinales pour la variable (taille de son rang) $|Rg|$ et le ratio de valeur manquante MVR : 
\begin{figure}[H]
\centering
\includegraphics[width=7cm]{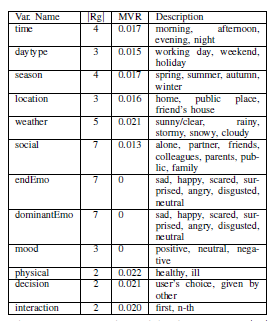}
\caption{Variables contextuelles et leurs propriétés de base \cite{kovsir2011database}}
\label{figure10:my_label}
\end{figure}
\subsubsection{Intégration de la méthode AHP}
\hspace*{3ex} AHP a été utilisé pour déterminer l'importance relative (poids) de genre des films. Les trois étapes principales de l'AHP sont les suivantes : 
\begin{itemize}
\item \textbf{Étape 1 : Décomposition du problème en une hiérarchie d'éléments inter-reliés}
\newline 
Le problème est décomposé en une hiérarchie de buts, critères, sous-critères et alternatives. Ceci est la partie la plus importante et créative de prise de décision. Le but alors est de structurer le problème de décision comme une hiérarchie qui est fondamentale pour le processus de l'AHP. La hiérarchie indique une relation entre les éléments d'un niveau avec ceux du niveau immédiatement inférieur. Cette relation percole vers le bas pour les niveaux les plus bas de la hiérarchie et de cette manière chaque élément est reliée à tous les autres, du moins d'une manière indirecte.

\begin{figure}[H]
\centering
\includegraphics[width=12cm]{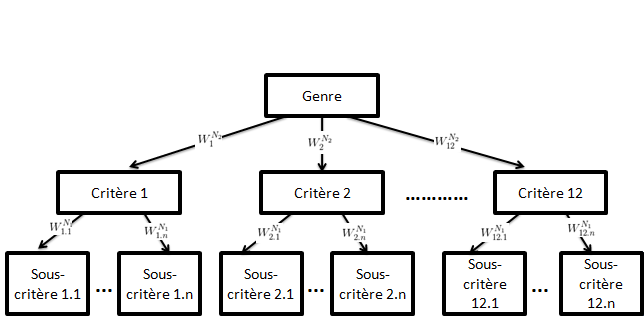}
\caption{Décomposition du problème en arbre hiérarchique}
\label{figure11:my_label}
\end{figure}
\item \textbf{Étape 2 : Comparaisons par paires des éléments de chaque niveau hiérarchique}
\newline 
Les données sont recueillies correspondant à la structure hiérarchique, dans la comparaison par paires des alternatives sur une échelle qualitative comme décrit ci-dessous (voir le tableau \ref{tab:50}). Nous avons évalué la comparaison comme égale, modérée, forte, très forte, et extrêmement forte.
\begin{center}
\begin{tabular}{|c||c|}
  \hline 
  \textbf{Degré d'importance} & \textbf{Description} \\
  \hline  \hline
  \textbf{1} & Importance égale \\
\hline
\textbf{2} & Importance modérée \\
\hline 
\textbf{3} & Importance forte \\
\hline 
\textbf{4} & Importance très forte \\
\hline
\textbf{5} & Importance extrême\\
\hline 
\multicolumn{2}{|c|}{Ratio = importance de critère sur la ligne / importance du critère dans la colonne} \\
\hline 
\end{tabular}
\caption{Importance relative des éléments  }
\label{tab:50}
\end{center}
\item \textbf{Étape 3 : Détermination de l'importance relative des éléments}
\newline
Les comparaisons par paires de différents critères générés à l'étape 2 sont organisées en une matrice carrée. Les éléments de la diagonale de la matrice sont $1$. Le critère dans la rangée $i$ est mieux que critère dans la colonne $j$ si la valeur de l'élément ($i$, $j$) est supérieure à 1 sinon le critère de la colonne $j$ est meilleur que celle dans la ligne $i$. L'élément ($j$, $i$) de la matrice est l'inverse de l'élément ($i$, $j$).
\item \textbf{Étape 4 : Évaluation des cohérences}
\newline 
La principale valeur propre et le vecteur propre droit normalisé correspondant de la matrice de comparaison donnent l'importance relative des différents critères étant comparée. Les éléments du vecteur propre normalisé sont appelés poids par rapport aux critères ou sous-critères,
\item \textbf{Étape 5 : Détermination de la performance relative de chacune des actions}  
\newline 
La consistance de la matrice d'ordre $n$ est évaluée. Les comparaisons faites par cette méthode sont subjectives et l'AHP tolère incohérence par la quantité de redondance dans l'approche. Si cet indice de consistance ne parvient pas à atteindre un niveau requis, les réponses aux comparaisons peuvent être réexaminées. Le classement de chaque alternative est multiplié par les poids des sous-critères et agrégées pour obtenir des classements locaux par rapport à chaque critère. Les évaluations locales sont ensuite multipliées par les poids des critères et agrégées pour obtenir les classements globaux.
\end{itemize}
\subsubsection{Recommandation contextuelle des films}
\hspace*{3ex} L'intégration du contexte dans les systèmes de recommandation se fait selon les trois types suivants : le pré, post-filtrage contextuel et la modélisation contextuelle. Dans notre approche, nous nous  intéressons au pré-filtrage contextuel. De plus, en se basant sur les douze critères contextuels fournis par LDOS-CoMoDa, la recommandation des films se fait après l'intégration de la méthode AHP pour arriver à faciliter le choix pour l'utilisateur et lui recommander les films les plus pertinents.

\section{Conclusion}
\hspace*{3ex} Dans ce chapitre, nous avons présenté les méthodes de prise de décision multicritères (MCDM). En effet, notre intérêt est porté à la méthode d'Analyse Hiérarchique des Procédés (AHP) qui a été intégrée dans notre approche. Enfin, nous avons présenté notre nouvelle approche de recommandation contextuelle en détaillant l'architecture utilisée et la méthodologie de recommandation. Dans le chapitre suivant, nous exposerons l'étude expérimentale de notre approche ainsi que sa évaluation.

\chapter{Étude expérimentale }
\section{Introduction}
\hspace*{3ex} Pendant le chapitre précédant, nous avons présenté les méthodes de prise de décision multi-critères (MCDM) et nous nous sommes intéressés à la méthode d'Analyse Hiérarchique des Procédés (AHP) que nous l'avons déjà utilisée dans notre approche. 
\newline 
\hspace*{3ex} Dans ce chapitre, Nous allons présenter une étude expérimentale de notre approche sur les base de tests \textbf{Movielens} et \textbf{LDOS-CoMoDa} pour évaluer ses performances. Dans un premier lieu, nous allons présenter l'environnement d'expérimentation pour évaluer notre approche. Dans un deuxième lieu, nous allons présenter les résultats de l'application de l'approche de recommandation contextuelle utilisant la méthode d'Analyse Hiérarchique des Procédés (AHP). 
\section{Environnement d'expérimentation}
\hspace*{3ex} Dans cette section, nous allons commencer tout d'abord par la présentation de l'environnement expérimental sur lequel nous avons travaillé pour évaluer et tester notre approche de recommandation contextuelle basant sur la méthode d'Analyse Hiérarchique des Procédés (AHP).
\subsection{Environnement matériel et logiciel}
\hspace*{3ex} Toutes les expérimentations ont été réalisées sur PC muni d'un processeur Intel Core i3 ayant une fréquence 2.53 GHz et 4 Go de mémoire tournant sous la plate-forme Windows 7. Alors, nous avons implémenté notre approche de recommandation contextuelle en utilisant la méthode d'Analyse Hiérarchique des Procédés (AHP) en JAVA. 
\subsection{Base de test}
\subsubsection{LDOS-CoMoDa}
\hspace*{3ex} Même si le problème de la recommandation contextuelle est devenu un problème de recherche dans le système de recommandation, l'ensemble de données publiques le plus approprié pour l'étude de la prédiction d'évaluation et la sélection des caractéristiques de contexte est l'ensemble de données LDOS-CoMoDa \cite{kovsir2011database}. Les statistiques de cet ensemble de données sont montrées dans le tableau \ref{tab:59}, il contient 1665 évaluations de 961 articles par 95 utilisateurs sous 12 variables de contexte. 

\begin{center}
\begin{tabular}{|c||c|}
  \hline 
  \textbf{Dataset} &  \textbf{LDOS-CoMoDa} \\
\hline  \hline
\textbf{\#Utilisateurs} & 95 \\
\hline 
\textbf{\#Items} & 961 \\
\hline 
\textbf{\#Variables contextuelles} & 12 \\
\hline
\textbf{\#Évaluations} & 1665 \\
\hline 
\end{tabular}
\caption{Statistiques de dataset LDOS-CoMoDa}
\label{tab:59}
\end{center}

\hspace*{3ex} Le dataset \textbf{LDOS-CoMoDa} est un dataset de recommandation de film riche en contexte. Certaines des propriétés importantes de l'ensemble de données sont les suivantes:
\begin{itemize}
\item Les évaluations et les informations contextuelles sont acquises explicitement auprès des utilisateurs immédiatement après que l'utilisateur a consommé l'élément.
\item Les informations contextuelles décrivent la situation dans laquelle l'utilisateur ait consommé l'article.
\item Les évaluations et les informations contextuelles proviennent d'une véritable interaction utilisateur-article et non d'une situation hypothétique ou de la mémoire d'interactions passées de l'utilisateur.
\item Les utilisateurs peuvent évaluer le même article plus d'une fois s'ils consomment l'article plusieurs fois.
\end{itemize} 
\begin{figure}[H]
\centering
\includegraphics[width=16cm]{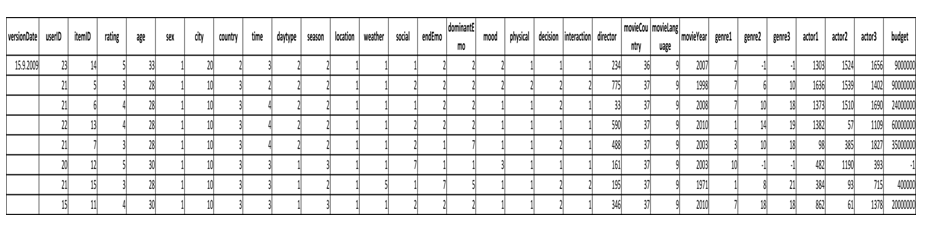}
\caption{Dataset \textbf{LDOS-CoMoDa}}
\label{figure11:my_label}
\end{figure}

\subsubsection{MovieLens}
\hspace*{3ex} Nous avons utilisé dans nos expérimentations la base de données MovieLens afin d'évaluer notre approche. L'ensemble de données MovieLens est composé de 943 utilisateurs et 1682 items avec une échelle d'évaluation comprise entre [1-5], où chaque utilisateur dispose de plus de 20 votes. Les statistiques de cet ensemble de données sont montrées dans le tableau \ref{tab:60}.
\begin{center}
\begin{tabular}{|c||c|}
  \hline 
  \textbf{Dataset} &  \textbf{MovieLens} \\
\hline  \hline
\textbf{\#Utilisateurs} & 943 \\
\hline 
\textbf{\#Items} & 1682 \\
\hline 
\textbf{\#Votes} & 20 \\
\hline 
\end{tabular}
\caption{Statistiques de dataset MovieLens}
\label{tab:60}
\end{center}
\subsection{Expérimentations}
\subsubsection{Expérimentation 1}
\hspace*{3ex} Dans cette section, nous allons présenter les résultats de l'application de notre approche (c'est à dire l'intégration de la méthode AHP dans le processus de recommandation pour calculer les poids des genres pour chaque  film). Le dataset LDOS-CoMoDa contient la liste des films avec leur genres. Il existe 22 genres différents pour la liste des films proposés dans LDOS-CoMoDa : (Romance, Adventure, Comedy, Biography, Drama,
Horror, Documentary, Mystery, Sci-Fi, Action, War, Sport, Musical, Film-Noir, Animation, History, Thriller,
Music, Family, Fantasy, Crime, Western). Un film peut appartenir à l'un des types de genre ou les trois types de genre à la fois :
\newline
\textbf{\underline{Exemple : }}
\begin{center}
{\setlength {\tabcolsep}{12pt}
\begin{tabular}{|c||c|c|c|c|}
  \hline 
\textbf{Item-ID} & \textbf{Titre de film }& \textbf{Genre 1}  & \textbf{Genre 2} & \textbf{Genre 3}  \\
  \hline \hline 
\textbf{15} & DIRTY HARRY & Action  & Crime &  Thriller \\ 
\hline
\textbf{23} & MACGRUBER & Action & Comedy & -1\\
\hline
\textbf{14} & DEATH AT A FUNERAL (UK) & comedy & -1 & -1\\
\hline 
\end{tabular}
\caption{Les différents types de genre pour quelques films} 
\label{tab:61}
}
\end{center}
\hspace*{3ex} Le but de l'intégration de la méthode AHP est de générer les genres avec leurs poids relatifs pour arriver à aider l'utilisateur à faire le bon choix ou bien la bonne décision. Le tableau \ref{tab:21} décrit les différents types des genres des films avec leurs poids relatifs dans le dataset \textbf{LDOS-CoMoDa} :
\begin{center}
{\setlength {\tabcolsep}{12pt}
\begin{tabular}{|c||c|c|c|}
  \hline 
 \textbf{Classement}  & \textbf{Type de genre} & \textbf{Poids relatifs} & 
   \textbf{Poids relatifs en \%} \\
   \hline 
 \textbf{Genre 1 (G1)} & Romance & 0,5666 & 56,66  
\\
\hline
\textbf{Genre 2 (G2)} & Comedy & 0,3100 & 31  \\
\hline
\textbf{Genre 3 (G3)} & Adventure & 0,2991 & 29,91  \\
\hline 
\textbf{Genre 4 (G4)} & Biography & 0,0764 &  7,64  \\
\hline
\textbf{Genre 5 (G5)} & Drama & 0,0740 &  7,40  \\
\hline 
\textbf{Genre 6 (G6)} & Action & 0,0729 &  7,29  \\
\hline 
\textbf{Genre 7 (G7)} & Horror & 0,0637 &  6,37  \\
\hline
\textbf{Genre 8 (G8)} & Documentary & 0,0621 &  6,21  \\
\hline
\textbf{Genre 9 (G9)} & Mystrey & 0,0505  &  5,05 \\
\hline
\textbf{Genre 10 (G10)} & Western & 0,0490 & 4,90  \\
\hline 
\textbf{Genre 11 (G11)} & Sport & 0,0487 & 4,87  \\
\hline 
\textbf{Genre 12 (G12)} & War & 0,0487 & 4,87  \\
\hline 
\textbf{Genre 13 (G13)} & Musical & 0,0484 & 4,84  \\
\hline 
\textbf{Genre 14 (G14)} & Film-Noir & 0,0483 & 4,83  \\
\hline 
\textbf{Genre 15 (G15)} & Sci-Fi & 0,0483 & 4,87 \\
\hline
\textbf{Genre 16 (G16)} & Animation & 0,0480 & 4,80  \\
\hline 
\textbf{Genre 17 (G17)} & History & 0,0475 & 4,75  \\
\hline 
\textbf{Genre 18 (G18)} & Thriller & 0,0473 & 4,73  \\
\hline

\end{tabular}
}
\end{center}
\newpage
\begin{center}
{\setlength {\tabcolsep}{12pt}
\begin{tabular}{|c||c|c|c|} 
\hline 
\textbf{Classement}  & \textbf{Type de genre} & \textbf{Poids relatifs} & 
   \textbf{Poids relatifs en \%} \\
   \hline 
\textbf{Genre 19 (G19)} & Music & 0,0473 & 4,73  \\
\hline 
\textbf{Genre 20 (G20)} & Family & 0,0470 & 4,70 \\
\hline 
\textbf{Genre 21 (G21)} & Fantasy & 0,0465 & 4,65  \\
\hline 
\textbf{Genre 22 (G22)} & Crime & 0,0461 & 4,61  \\
\hline
\end{tabular}
}
\caption{Description des genres avec leurs poids relatifs dans \textbf{LDOS-CoMoDa}}
\label{tab:21}
\end{center}
\begin{figure}[H]
\centering 
\includegraphics[width=12cm]{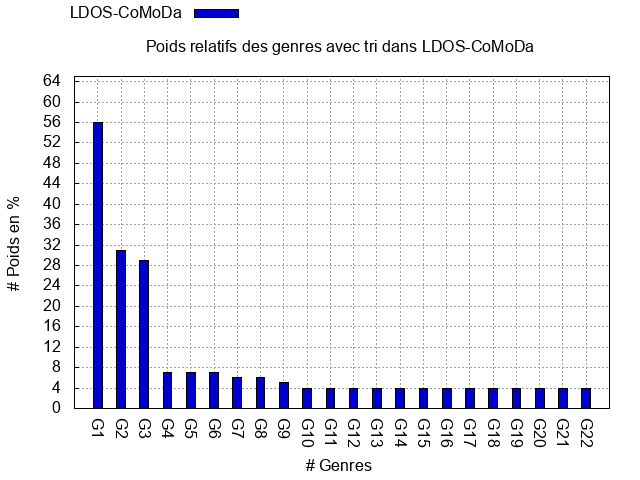}
\caption{Histogramme des poids relatifs des genres avec tri décroissant}
\label{figure8:my_label}
\end{figure}

\subsubsection{Expérimentation 2}
\begin{figure}[H]
\centering
\includegraphics[width=15cm]{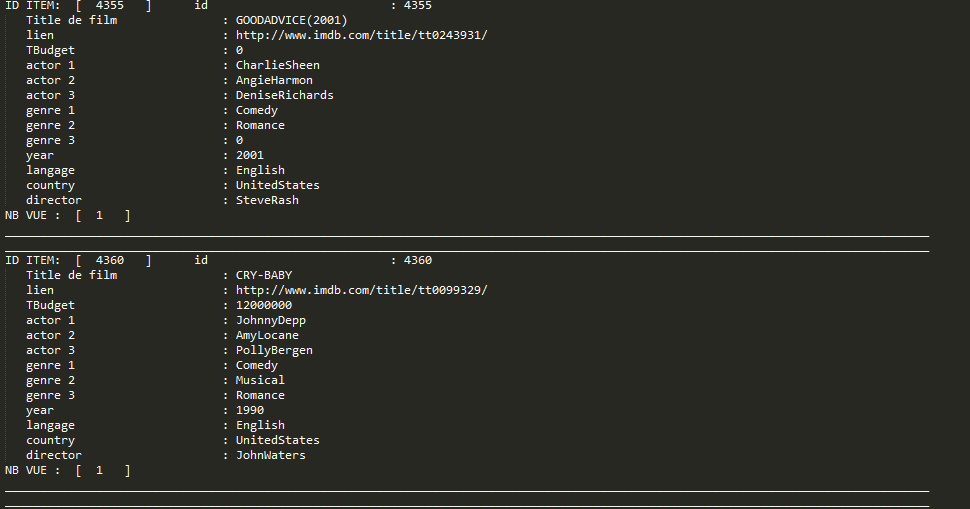}
\caption{Recommandation de N-Top films}
\label{figure4:my_label}
\end{figure}

\begin{figure}[H]
\centering
\includegraphics[width=15cm]{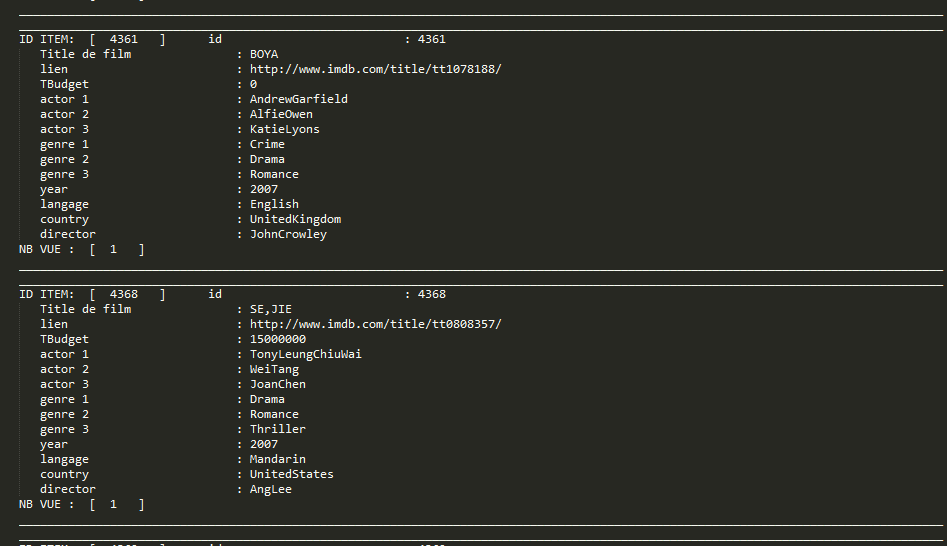}
\caption{Recommandation de N-Top films}
\label{figure5:my_label}
\end{figure}

\begin{figure}[H]
\centering
\includegraphics[width=15cm]{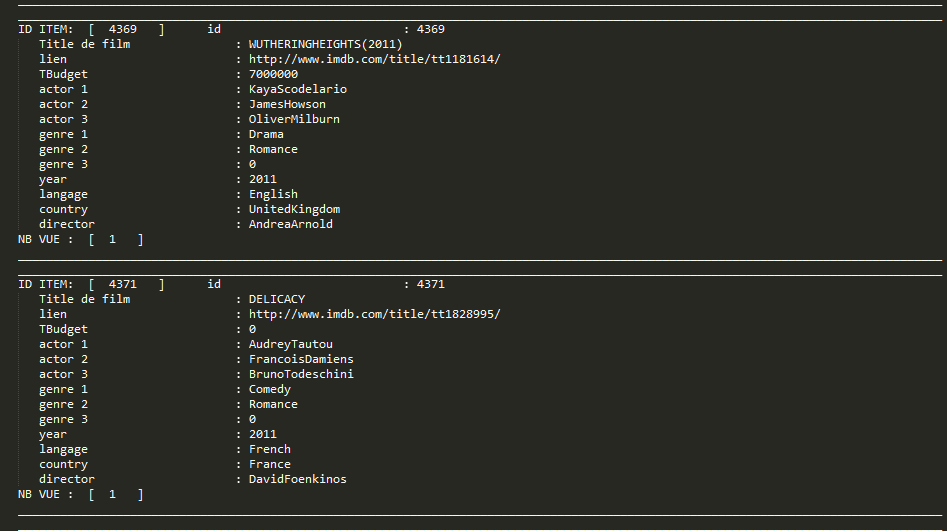}
\caption{Recommandation de N-Top films}
\label{figure6:my_label}
\end{figure}

\begin{figure}[H]
\centering
\includegraphics[width=15cm]{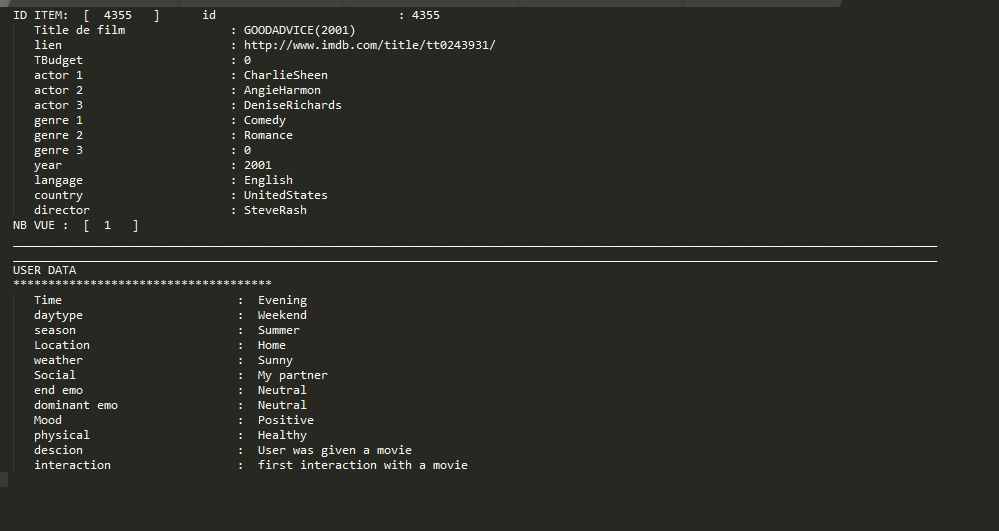}
\caption{Film recommandé avec les informations contextuelles}
\label{figure7:my_label}
\end{figure}

\hspace*{3ex} L'étape qui suit permet de générer les films recommandés en se basant sur le calcul des poids effectué dans l'étape précédente. Le genre "Romance" est classé numéro 1 selon le calcul de poids effectué alors la recommandation des films met en considération le genre "Romance" et la situation de l'utilisateur au cours du processus de recommandation. Les figures \ref{figure4:my_label}, \ref{figure5:my_label}, \ref{figure6:my_label} montre les N-top films recommandés dans lesquels le genre romance est apparu.
\hspace*{3ex} L'intégration de la notion du contexte dans le processus de recommandation a donné plus de confiance dans les recommandations. La figure \ref{figure7:my_label} décrit le film recommandé avec la situation de l'utilisateur (les 12 critères contextuels).
\subsubsection{Expérimentation 3}
\hspace*{3ex} Dans cette section, nous allons présenter les résultats obtenus à partir du calcul des poids relatifs des genres. Le dataset MovieLens contient la liste des films avec leurs genres. Il existe 18 genres différents pour la liste des films proposés dans MovieLens : (Unknown, Action, Adventure, Animation, Children's, Comedy, Crime, Documentary, Drama, Fantasy,
Film-Noir, Horror,  Musical, Romance, Sci-Fi, Thriller, War, Western). Un film peut appartenir à l'un des types de genre ou à tous les types à la fois.
L'intégration de la méthode AHP permet de  générer les genres avec leurs poids relatifs pour arriver à aider l'utilisateur à faire le bon choix. Le tableau \ref{tab:221} nous montre le résultat suite au calcul de poids relatifs pour chaque genre ainsi que le tri par ordre décroissant dans le dataset \textbf{MovieLens} :

\begin{center}
{\setlength {\tabcolsep}{12pt}
\begin{tabular}{|c||c|c|c|}
  \hline  
\textbf{Classement}   & \textbf{Type de genre} & \textbf{Poids relatifs} & 
   \textbf{Poids relatifs en \%} \\
   \hline 
 \textbf{Genre 1 (G1)} & Thriller & 0,6766 &  67,66 \% 
\\
\hline
\textbf{Genre 2 (G2)} & Comedy & 0,5666 & 56,66 \% \\
\hline
\textbf{Genre 3 (G3)} & Fantasy & 0,3910 & 39,10 \% \\
\hline 
\textbf{Genre 4 (G4)} & War & 0,3250 & 32,50 \% \\
\hline
\textbf{Genre 5 (G5)} & Crime & 0,3120 &  31,20 \% \\
\hline 
\textbf{Genre 6 (G6)} & Documentary & 0,3000 &  30 \% \\
\hline 
\textbf{Genre 7 (G7)} & Film-Noir & 0,2980&  29,80 \% \\
\hline
\textbf{Genre 8 (G8)} & Drama & 0,2876 &  28,76 \% \\
\hline
\textbf{Genre 9 (G9)} & Western & 0,2789 &  27,89 \% \\
\hline
\textbf{Genre 10 (G10)} & Adventure & 0,2545 & 25,45 \% \\
\hline 
\textbf{Genre 11 (G11)} & Sci-Fi & 0,2501 & 25,01 \% \\
\hline 
\textbf{Genre 12 (G12)} & Animation & 0,2345 & 23,45 \% \\
\hline 
\textbf{Genre 13 (G13)} & Horror & 0,2213 &  22,13 \% \\
\hline 
\textbf{Genre 14 (G14)} & Children & 0,2000 & 20 \% \\
\hline 
\end{tabular}
}
\end{center}
\begin{center}
{\setlength {\tabcolsep}{12pt}
\begin{tabular}{|c||c|c|c|} 
  \hline 
  \textbf{Classement}  & \textbf{Type de genre} & \textbf{Poids relatifs} & 
   \textbf{Poids relatifs en \%} \\
   \hline 
\textbf{Genre 15 (G15)} & Romance & 0,1987 & 19,87 \% \\
\hline
\textbf{Genre 16 (G16)} & Musical & 0,1798 & 17,98 \% \\
\hline
\textbf{Genre 17 (G17)} & Action & 0,1760 & 17,60 \% \\
\hline 
\textbf{Genre 18 (G18)} & Mystrey & 0,1679 & 16,79 \% \\
\hline 
\textbf{Genre 19 (G19)} & Unknown & 0,0769 &   7,69\% \\
\hline
\end{tabular}
\caption{Description des genres avec leurs poids relatifs dans \textbf{MovieLens}}
\label{tab:221}
}
\end{center}
\begin{figure}[H]
\centering
\includegraphics[width=12cm]{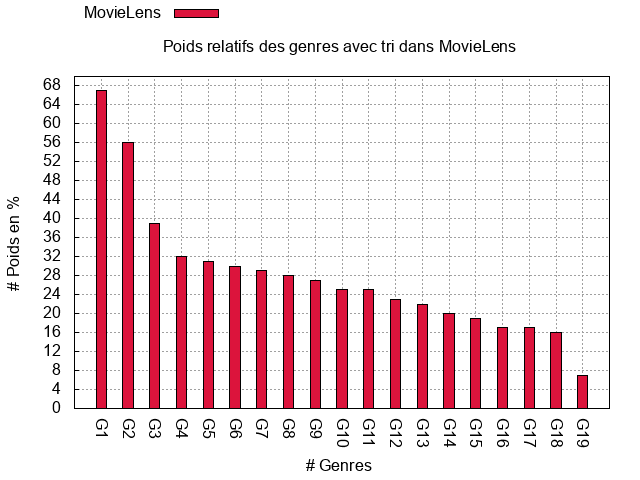}
\caption{Poids relatifs des genres avec tri dans \textbf{MovieLens}}
\label{figure10:my_label1}
\end{figure}
\begin{figure}[H]
\centering
\includegraphics[width=14cm]{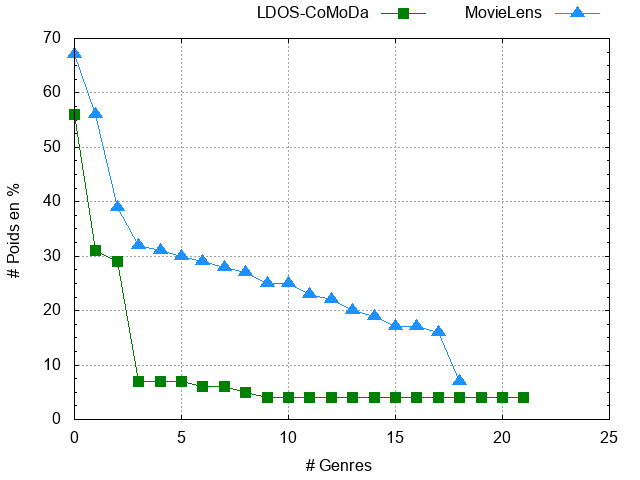}
\caption{Les poids relatifs des genres communs dans \textbf{LDOS-CoMoDa} et \textbf{MovieLens}}
\label{figure11:my_label1}
\end{figure}
\hspace*{3ex} La figure \ref{figure12:my_label1} décrit le film recommandé avec les informations personnels de l'utilisateur. Le film recommandé est de genre "Thriller" et "Sci-fi".
\begin{figure}[H]
\centering
\includegraphics[width=18cm]{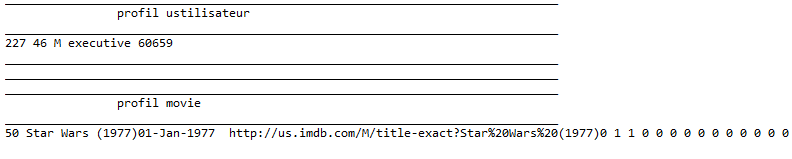}
\caption{Film recommandé avec \textbf{MovieLens}}
\label{figure12:my_label1}
\end{figure}
\begin{center}
{\setlength {\tabcolsep}{12pt}
\begin{tabular}{|c||c|c|c|c|}
  \hline
\textbf{ID utilisateur} & \textbf{Âge} & \textbf{genre}  & \textbf{profession} & \textbf{Code postal} \\
   \hline 
 \textbf{227} & 46 & M &  executive & 60659 \\
\hline
\end{tabular}
\caption{Description des informations personnelles de l'utilisateur}
\label{tab:24}
}
\end{center}
\subsection{Évaluation expérimentale}
\hspace*{3ex} Nous allons présenté dans cette partie les différentes mesures d'évaluation utilisées dans nos expérimentations, pour tester la performance de notre approche. La performance des systèmes de recommandation est testée en termes de la précision des listes de recommandation générées. Les mesures d'évaluation de la performance des recommandations sont : \textbf{la précision, le rappel et F-mesure}.
\textbf{\textsc{Les mesures de la précision}}
\hspace*{3ex} Afin d'évaluer la performance de notre approche de recommandation, nous avons utilisé les métriques de rappel, précision et F-mesure. Où, F-mesure est calculée comme suit :
\vspace{-3em}
\begin{center}
\begin{equation}
F-mesure = \frac{2PR}{P+R}
\label{eq1:equation}
\end{equation}
\end{center}
\vspace{-1em}
\hspace*{3ex} Les deux paramètres, précision et rappel, sont couramment utilisés pour mesurer la qualité d'une recommandation. Ce sont également des mesures utilisées dans la recherche d'information .
\hspace*{3ex} Où P et R sont la précision et le rappel respectivement, et ils sont calculés comme suit :
\vspace{-3em}
\begin{center}
\begin{equation}
P = \frac{N_{t}}{N}
\label{eq2:equation}
\end{equation}
\end{center}
\begin{center}
\vspace{-3em}
\begin{equation}
R = \frac{N_{t}}{N_{p}}
\label{eq3:equation}
\end{equation}
\end{center}
\vspace{-1em}

\begin{center}
{\setlength {\tabcolsep}{12pt}
\begin{tabular}{|c||c|}
  \hline 
 \textbf{Notation} &  \textbf{Signification} \\
\hline  \hline
\textbf{$N_{p}$} & Nombre total des items pertinents \\
\hline 
\textbf{$N_{t}$} &  Nombre des items pertinents trouvés \\
\hline
\textbf{$N$} &  Nombre total des items   \\
\hline 
\end{tabular}
\caption{Notations utilisées dans les mesures de précision}
\label{table:1}
}
\end{center}
\begin{itemize}
\item \textbf{Précision} : une mesure de l'exactitude, détermine la fraction des éléments pertinents récupérés sur tous les éléments récupérés. Par exemple, la proportion de films recommandés qui sont réellement pertinents. Dans notre cas, la précision est égale au quotient de la liste des films pertinents recommandés par rapport à la liste totale des films.
\item \textbf{Rappel} : une mesure d'exhaustivité, détermine la fraction des éléments pertinents extraits de tous les éléments pertinents. Par exemple, la proportion de tous les films recommandés qui sont pertinents. Dans notre cas, le rappel est égale au quotient de la liste des films pertinents recommandés par rapport au total des films pertinents.
\item \textbf{F-mesure} : 
F-mesure tente de combiner Précision et Rappel en une seule valeur à des fins de comparaison. Elle peut être utilisé pour obtenir une vision plus équilibrée des performances. Après le calcul de la précision et le rappel, nous allons passer au calcul de F-mesure qui nous donne la valeur : 0.394.
\end{itemize} 
 \begin{center}
{\setlength {\tabcolsep}{12pt}
\begin{tabular}{|c||c|c|}
  \hline 
 \textbf{ Précision} &  \textbf{Rappel}  & \textbf{F-mesure}\\
\hline  \hline
\textbf{$0.729966 \simeq 0.73$} & \textbf{$0.270034 \simeq 0.27$} & \textbf{$0.394$} \\
\hline 
\end{tabular}
\caption{Valeurs de métriques pour LDOS-CoMoDa}
\label{table:2}
}
\end{center}
\begin{figure}[H]
\centering
\includegraphics[width=11cm]{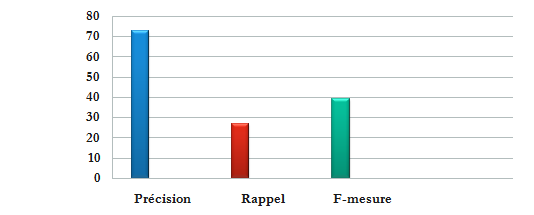}
\caption{Les valeurs de précision, rappel et F-mesure pour \textbf{LDOS-CoMoDa}}
\label{figure1000:my_label1}
\end{figure}
\hspace*{3ex} Dans l'étape suivante, nous allons présenter une comparaison entre les résultas de notre nouvelle approche et l'approche de Bader et al. \cite{bader2011context}. L'approche de Bader et al, utilise plusieurs méthodes multi-critères telles-que (AHP, WSM, WPM et TOPSIS) afin de calculer la précision. 
nous nous intéressons seulement au niveau statique dans la structure hiérarchique lors de l'utilisation de méthode AHP dans l'approche de Bader et al. afin de faire une comparaison avec notre approche. Nous nous intéressons seulement à la première ligne du tableau \ref{tab:tab12} : 
\begin{table}[h!]
\begin{center}
\begin{tabular}{|p{2cm}|p{2cm}|p{2cm}|p{2cm}|p{1cm}|p{1cm}|p{1cm}|p{1cm}|p{1.5cm}|}
\hline
\multicolumn{2}{|c|}{Level 1} & \multicolumn{2}{|c|}{Level 2} & 
\multicolumn{5}{|c|}{POI set}\\
\hline
MCDM & Norm  & MCDM & Norm & 1 & 2 & 3 & 4 & AVG \\
\hline
\cellcolor[HTML]{AA0044} AHP & \cellcolor[HTML]{AA0044} 1 &\cellcolor[HTML]{AA0044}  AHP & \cellcolor[HTML]{AA0044} 1 & \cellcolor[HTML]{AA0044} 0.62 & \cellcolor[HTML]{AA0044} 0.69 & \cellcolor[HTML]{AA0044} 0.74 & \cellcolor[HTML]{AA0044} 0.58 & \cellcolor[HTML]{AA0044} 0.66 \\
\hline 
TOPSIS & 0 & TOPSIS & 0 & 0.82 &  0.55 & 0.63 & 0.58 & 0.65 \\
\hline 
TOPSIS & 1 & TOPSIS & 0 & 0.82 &  0.55 & 0.63 & 0.58 & 0.65 \\
\hline 
WSM & 0 & TOPSIS & 0 & 0.77 & 0.63 & 0.60 & 0.54 & 0.64 \\
\hline 
AHP & 1 & TOPSIS & 1 & 0.59 & 0.61 & 0.73 & 0.55 & 0.62 \\
\hline
TOPSIS & 1 & AHP & 1 & 0.59 & 0.61 & 0.73 & 0.55 & 0.62 \\ 
\hline 
WSM & 0 & AHP & 1 & 0.60 & 0.59 & 0.72 & 0.50 & 0.60 \\
\hline
AHP & 1 & WPM & 1  & 0.62  & 0.55  & 0.68 & 0.48 & 0.58 \\
\hline 
AHP  &  1 & WSM & 1 & 0.57 & 0.59  & 0.74 & 0.38 & 0.57 \\
\hline 
WSM & 0 & WSM & 0 & 0.64 & 0.60 & 0.57 & 0.44 & 0.56 \\
\hline 
\end{tabular}
\caption{Précision des scores globaux \cite{bader2011context}}
\label{tab:tab12}
\end{center}
\end{table}
\newpage
\hspace*{3ex} Le tableau \ref{table:3} donne les valeurs de précision ainsi le rappel pour les deux approches : 

 \begin{center}
{\setlength {\tabcolsep}{12pt}
\begin{tabular}{|c||c|c|}
  \hline 
 \textbf{} &  \textbf{Précision}  & \textbf{Rappel}\\
\hline  \hline
\textbf{Notre Approche} & \textbf{0.73} & \textbf{0.27}\\
\hline 
\textbf{Approche de Bader et al. \cite{bader2011context} } & \textbf{0.66} & --- \\
\hline 
\end{tabular}
\caption{Les valeurs de précision et de rappel pour notre approche et l'approche de Bader el al.}
\label{table:3}
}
\end{center}
\begin{figure}[H]
\centering
\includegraphics[width=11cm]{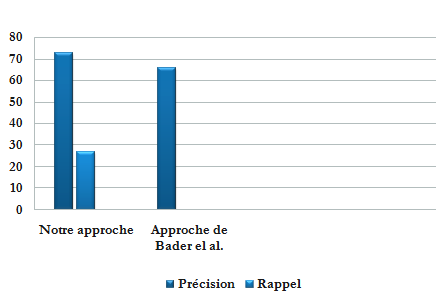}
\caption{Les valeurs de précision et rappel pour les deux datasets \textbf{LDOS-CoMoDa} et \textbf{MovieLens}}
\label{figure1003:my_label1}
\end{figure}

\hspace*{3ex} Nous calculons la précision de notre approche, cette dernière nous donne une valeur est égal à 0.73, alors nous pouvons constater que si la précision est importante (est égale à 0.73) alors notre système de recommandation est classé comme un système précis.  

\section{Conclusion}
\hspace*{3ex} Au cours de ce chapitre, nous avons évoqué une évaluation de notre nouvelle approche de recommandation contextuelle en se basant sur la méthode d'Analyse Hiérarchique des Procédés (AHP).
En effet, les résultats des expérimentations ont montré que notre approche a donné de bon résultats en termes de précision et rappel vs l'approche de Bader et al.

\chapter*{Conclusion et Perspectives}
\addcontentsline{toc}{chapter}{Conclusion}

\markboth{CONCLUSION ET PERSPECTIVES}{}
\hspace*{3ex} 
L'expansion de l'Internet et du nombre d'applications basées sur le Web, tels que les portails d'entreprise, est associée à une prolifération d'informations ou d'items dont le volume ne cesse de croître. Devant cette profusion et cette surcharge d'items, l'utilisateur peine à repérer l'information pertinente qui correspond le plus à ses besoins. Dans ce contexte, les systèmes de recommandation ont été développés en vue de faciliter l'accès à ces items pertinents. Leur objectif est d'anticiper les besoins de l'utilisateur en lui fournissant des recommandations d'items jugés pertinents par rapport à ses goûts. Parmi les limites des systèmes de recommandation classiques, nous avons identifié la non-prise en compte du contexte dans lequel  l'utilisateur décide de faire une recommandation. 
\newline 
\hspace*{3ex} Dans ce mémoire, nous avons proposé une nouvelle approche de recommandation contextuelle en utilisant la méthode d'Analyse Hiérarchique des Procédés (AHP). En effet, cette approche repose essentiellement sur la prédiction des films en tenant compte de contexte.
\newline 
\hspace*{3ex} Dans un premier lieu, nous avons entamé ce mémoire par la présentation des notions de base utiles, à savoir les systèmes de recommandation classiques et sensibles au contexte, leurs définitions et leurs techniques, et leurs architectures. Dans un second lieu, nous avons effectué une étude synthétique sur les travaux proposés les plus récents dans la littérature pour la recommandation sensible au contexte. La principale limite de ces approches, réside dans le fait que certaines approches ne sont pas prises en considération la décision de l'utilisateur.
 \newline 
\hspace*{3ex} Nous avons permis à cette approche  la possibilité d'appliquer la méthode AHP pour aider l'utilisateur à effectuer le bon choix. Ainsi, cette étape simplifie le processus de recommandation.
\newline 
\hspace*{3ex} Nous avons mené une étude expérimentale des performances de notre approche. Par la suite nous avons élaboré une étude comparative entre notre approche et les approches de la littérature. Par conséquent, notre approche a permit de donner de bons résultats en termes de précision.

\hspace*{3ex} Les systèmes de recommandation sensibles au contexte présentent un domaine plus riche dans plusieurs  pistes non encore explorées. Nous pouvons envisager comme perspectives de recherche :  
\begin{itemize}
\item Il serait intéressant d'exploiter les scores multi-critères contenant des informations contextuelles, pour améliorer les recommandations.
\item L'intégration de la méthode TOPSIS ( Technique for Order of Preference by Similarity to Ideal Solution), qui permet de classer par ordre de choix un certain nombre d'alternatives sur la base d'un ensemble de critères favorables ou défavorables, dans le processus de génération des recommandations, peut améliorer les recommandations contextuelles.

\end{itemize}

\appendix
\appendix
\chapter*{Annexes}
\addstarredchapter{Annexes} 
\chapter{LDOS-CoMoDa DATASET}	
\section{Introduction}
\hspace*{3ex} Dans cette annexe, nous allons présenter brièvement certains concepts utilisés dans ce mémoire pour l'évaluation de notre nouvelle approche de recommandation sensible au contexte en utilisant la méthode d'analyse hiérarchique des procédés (AHP). Nous présentons en premier lieu le dataset \textbf{LDOS-CoMoDa}. La deuxième section sera consacrée à présenter le dataset  \textbf{Movielens}. 
\subsection{Présentation de dataset LDOS-CoMoDa } 
\hspace*{3ex} \textbf{LDOS-CoMoDa} est un dataset de recommandation de film riche en contexte. Il contient des notes pour les films et les douze informations contextuelles décrivant la situation dans laquelle les films ont été consommés. 
\subsection{Description détaillée des fichiers de données}
\begin{enumerate}[label=\fbox{\arabic*}] 
\item \textbf{Champs de données} : 

versionDate : date of the dataset version
userID (15 - 200, some missing)
itemID (1 -4138, some missing)
rating (1-5)
user's age
user's sex (1= male, 2 = female)
user's city
user's country
time (1-4)
daytype (1-3)
season	(1-4)
location (1-3)	
weather	(1-5)
social (1-7)	
endEmo(1-7)
dominantEmo (1-7)	
mood (1-3)	
physical (1-2)	
decision (1-2)	
interaction (1-2)
movie director
movie's country
movie's language
movie's year
genre1
genre2
genre3
actor1
actor2
actor3
movie's budget
\item \textbf{Variables contextuelles} : 

\begin{center}
{\setlength {\tabcolsep}{12pt}
\begin{tabular}{|c||p{7cm}|}
  \hline 
 \textbf{Variables contextuelles} &  \textbf{Description}\\
\hline  \hline
\textbf{time} & Morning, Afternoon, Evening, Night \\
\hline 
\textbf{daytype} & Working day, Weekend, Holiday \\
\hline 
\textbf{season } & Spring, Summer, Autumn, Winter \\
\hline 
\textbf{location} & Home, Public place, Friend's house \\
\hline 
\textbf{weather} & Sunny / clear, Rainy, Stormy, Snowy, Cloudy \\
\hline 
\textbf{social} & Alone, My partner, Friends, Colleagues, Parents, Public, My family \\
\hline 
\textbf{endEmo} & Sad, Happy, Scared, Surprised, Angry, Disgusted, Neutral \\
\hline 
\textbf{dominantEmo} & Sad, Happy, Scared, Surprised, Angry, Disgusted, Neutral \\
\hline 
\textbf{mood} & Positive, Neutral, Negative \\
\hline 
\textbf{physical} & Healthy, Ill \\
\hline 
\textbf{decision} & User decided which movie to watch, User was given a movie \\
\hline 
\textbf{interaction} & first interaction with a movie, n-th interaction with a movie \\
\hline 
\end{tabular}
\caption{Description des variables contextuelles}
\label{tab:36}
}
\end{center}
\end{enumerate}

\hspace*{3ex} Les valeurs de contexte dans la base de données correspondent à cet ordre.
\vspace{-1em}
\begin{center}
\fbox{
Par exemple : daytype-> 1 = Working day, 2 = Weekend, 3 = Holiday 
Valeur manquante = -1
}
\end{center}
\vspace{-1em}
\section{Conclusion}
\hspace*{3ex} Dans cette annexe, nous avons présenté le dataset \textbf{LDOS-CoMoDa}. Nous avons décrit en détails les fichiers de données.

\chapter{Movielens DATASET}

\section{Présentation de dataset MovieLens}
\hspace*{3ex} Les ensembles de données MovieLens ont été recueillies par le Projet de recherche GroupLens6 à l'Université du Minnesota. Cet ensemble de données comprend 100.000 notes dans une échelle de (1-5) à partir de 943 utilisateurs sur 1682 films. Où, chaque utilisateur a évalué au moins 20 films. Les données ont été collectées à travers le site MovieLens (movielens.umn.edu) au cours de la période de sept mois à partir du 19 Septembre 1997 au 22 Avril, 1998. Ces données ont été nettoyées, où les utilisateurs qui avaient moins de 20 évaluations ou qui n'ont pas eu des informations démographiques complètes ont été retirés de cet ensemble de données. Cet ensemble de données a été utilisé pour la première fois par Herlocker et al., \cite{herlocker1999algorithmic}.
\subsection{Informations sur le projet de recherche  GroupLens}
\hspace*{3ex} Le projet de recherche GroupLens est un groupe de recherche au Département d'informatique et de génie de l'Université de Minnesota. Les Membres du projet de recherche GroupLens sont impliqués dans de nombreux projets de recherche liés aux domaines du filtrage d'information, filtrage collaboratif, et les systèmes de recommandation. Le projet est dirigé par les professeurs John Riedl et Joseph Konstan. Le projet a commencé à explorer le filtrage collaboratif automatisé en 1992, mais est surtout connu pour son vaste essai mondial d'un système de filtrage collaboratif automatisé pour les nouvelles Usenet en 1996. La technologie développée dans le procès Usenet a formé la base pour la formation des perceptions nettes, Inc., qui a été fondée par des membres de recherche GroupLens. 
\hspace*{3ex} Depuis, le projet a élargi son champ d'application à la recherche de solutions globales de filtrage de l'information,l'intégration dans les méthodes basées sur le contenu ainsi que l'amélioration de la technologie actuelle de filtrage collaboratif. De plus amples informations sur le projet GroupLens recherche, y compris les publications de recherche, peut être trouvé sur le site Web suivant: \textit{http://www.grouplens.org/}
GroupLens recherche exploite actuellement un recommandeur de film basé sur le filtrage collaboratif: \textit{http://www.movielens.org/}

\subsection{Description détaillée des fichiers de données}
\hspace*{3ex} La base de données MoviLens contient plusieurs fichiers pour représenter les différents ensembles de données et pour décrire les films, les utilisateurs, la relation entre utilisateurs et items, ainsi que les scripts de génération des différents ensembles de données. Nous présentons dans les sections suivantes une description détaillée des différents fichiers, ceux disponibles avec la base et ceux que nous avons créés et ajouté afin d'accomplir les expérimentations. 
\subsubsection{Les fichiers disponibles avec la base}
\begin{enumerate}[label=\fbox{\arabic*}] 
\item \textbf{ml-data.tar.gz}: Fichier compressé de type .tar qui contient toute la base de données téléchargée.
\item \textbf{u.data} : L'ensemble de données u.data est l'ensemble plein, c.à.d. l'ensemble qui contient les 100 000 évaluations des 943 utilisateurs sur les 1682 films, où, chaque utilisateur a évalué au moins 20 films. Dans cet ensemble, les utilisateurs et les articles sont numérotés consécutivement à partir de 1, mais les relations entre les utilisateurs et les items sont arrangés dans l'ordre de leur collecte. La relation entre un utilisateur et un item dans l'ensemble de données u.data est représentée avec une tab séparée comme suit :
\begin{center} 
 \fbox {ID utilisateur | Id item | Note | Horodatage (timestamp)} \end{center}
Un exemple de cet ensemble de données est le suivant : \textbf{1 1 5 874965758} : où, cette ligne indique que l'utilisateur 1 a évalué l'item 1 avec une note (vote) égale à 5 au moment 874965758s, qui est le temps de l'évaluation en secondes depuis le 1/1/1970.
\item \textbf{u.info} : Ce fichier contient le nombre d'utilisateurs, d'items, et de notes dans l'ensemble de données \textit{u.data}, et il est constitué de 3 lignes:
\begin{itemize}
\item 943 users
\item 1682 items
\item 100000 ratings
\end{itemize}
\item \textbf{u.item} : Ce fichier contient les informations sur les items (films), représentés sous forme de tab séparée comme suit :
\newline 
\vspace{-1em}
 \fbox {\begin{minipage}{16cm}\setlength{\parskip}{5mm}
Film id | Titre du film | date de diffusion | date de sortie de la vidéo | IMDb URL | inconnu | Action | Aventure | Animation | Enfant | Comédie | Crime | Documentaire | Drama | Fantaisie | Film-Noir | Horreur | Musique | Mystère | Romance | science Fiction | Thriller | guerre | Western | 
\end{minipage}}
\vspace{1em}

\hspace*{3ex} Où, les identifiants des films sont ceux utilisés dans la base \textit{u.data}. Comme exemple nous citons la description du premier film $i1$:

\vspace{1em}
 \fbox {\begin{minipage}{16cm}\setlength{\parskip}{5mm}
1|Toy Story (1995)|01-Jan-1995||http://us.imdb.com/M/title-exact?Toy$\%$20Story$\%$20 (1995)|0|0|0|1|1|1|0|0|0|0|0|0|0|0|0|0|0|0|0
Où, les 19 derniers champs présentent les genres des films, avec $i1$ indique que le film est de ce genre, un 0 indique qu'il n'est pas; et les films peuvent être en plusieurs genres à la fois, ce qui est évidemment le cas avec le film $i1$ qui appartient à 3 genres.
\end{minipage}}
\vspace{-1em}
\item \textbf{u.genre} : 
\begin{itemize}
\item unknown|0
\item Action|1
\item Adventure|2
\item Animation|3
\item Children's|4
\item Comedy|5
\item Crime|6
\item Documentary|7
\item Drama|8
\item Fantasy|9
\item Film-Noir|10
\item Horror|11
\item Musical|12
\item Mystery|13
\item Romance|14
\item Sci-Fi|15
\item Thriller|16
\item War|17
\item Western|18
\end{itemize}
\item \textbf{u.user} : Ce fichier contient les données démographiques sur les utilisateurs, sous forme d'une tab séparée de données démographiques simples pour les utilisateurs comme suit :
\begin{center}
\fbox {ID utilisateur | âge | genre | profession | code postal}
\end{center}
Où, les identifiants des utilisateurs (ID utilisateur) sont ceux utilisés dans l'ensemble de données u.data. À titre d'exemple, l'utilisateur ayant l'identifiant 7, qui est un homme âgé de 57 ans et qui travaille comme administrateur avec un code postal 91344, est représenté dans la base comme suit :
\begin{center}
\fbox {7|57|M|administrator|91344}
\end{center}
\item \textbf{u.occupation} : Ce fichier contient la liste des professions des différents utilisateurs : administrator, artist, doctor, educator, engineer, entertainment, executive, healthcare, homemaker, lawyer, librarian, marketing, none, other, programmer, retired, salesman, scientist, student, technician, writer.
\item \textbf{u1.base, u1.test, u2.base, u2.test, u3.base, u3.test, u4.base, u4.test, u5.base, u5.test} : Les ensembles de données u1.base et u1.test jusqu'à u5.base et u5.test sont des sous ensembles de données de u.data avec 80 $\%$ de données dans .base et 20 $\%$ dans .test. Les ensembles de u1, ..., u5 sont des ensembles de test disjoints et ils sont construits pour être utilisés dans la validation croisée. Ces ensembles de données peuvent être générés à partir u.data par le fichier mku.sh.
\item \textbf{ua.base, ua.test, ub.base, ub.test} : Les ensembles de données ua.base, ua.test, ub.base et ub.test divisent l'ensemble de données u.data en un ensemble d'entrainement et un ensemble de test, avec 10 notes par utilisateur dans les ensembles de test qui sont disjoints. Les quatre ensembles de données peuvent être générés à partir u.data par le fichier mku.sh.
\item \textbf{mku.sh} : Ce fichier contient un script pour générer tous les ensembles de données de u.data.
\item \textbf{allbut.pl} : Ce fichier contient le script qui génère les ensembles d'apprentissage et de test, avec contrainte de sélection de seulement n évaluations des utilisateurs dans les données d'apprentissage.
\begin{figure}[H]
\centering
\includegraphics[width=10cm]{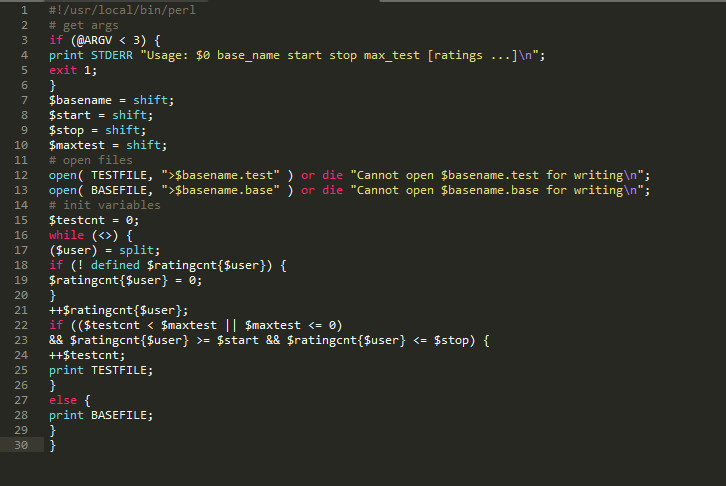}
\caption{Script avec les ensembles d'apprentissage et de test}
\label{figure20:my_label}
\end{figure}
\end{enumerate}
          
\section{Conclusion}
\hspace*{3ex} Dans cette annexe, nous avons présenté le dataset \textbf{MovieLens}. Nous avons présenté également le projet de recherche \textbf{"GroupLens"} ainsi qu'une description détaillée des fichiers de données.

\include{allbiblio}
\bibliography{allbiblio}

\begin{thebibliography}{10}

\bibitem{berrut2003filtrage}
Catherine Berrut and Nathalie Denos.
\newblock Filtrage collaboratif, 2003.

\bibitem{adomavicius2015context}
Gediminas Adomavicius and Alexander Tuzhilin.
\newblock Context-aware recommender systems.
\newblock In {\em Recommender systems handbook}, pages 191--226. Springer,
  2015.

\bibitem{ostuni2012cinemappy}
Vito~Claudio Ostuni, Tommaso Di~Noia, Roberto Mirizzi, Davide Romito, and
  Eugenio Di~Sciascio.
\newblock Cinemappy: a context-aware mobile app for movie recommendations
  boosted by dbpedia.
\newblock In {\em Proceedings of the 2012 International Conference on Semantic
  Technologies Meet Recommender Systems \& Big Data-Volume 919}, pages 37--48.
  CEUR-WS. org, 2012.

\bibitem{campos2013context}
Pedro~G Campos, Ignacio Fern{\'a}ndez-Tob{\'\i}as, Iv{\'a}n Cantador, and
  Fernando D{\'\i}ez.
\newblock Context-aware movie recommendations: an empirical comparison of
  pre-filtering, post-filtering and contextual modeling approaches.
\newblock In {\em International Conference on Electronic Commerce and Web
  Technologies}, pages 137--149. Springer, 2013.

\bibitem{gantner2010factorization}
Zeno Gantner, Steffen Rendle, and Lars Schmidt-Thieme.
\newblock Factorization models for context-/time-aware movie recommendations.
\newblock In {\em Proceedings of the Workshop on Context-Aware Movie
  Recommendation}, pages 14--19. ACM, 2010.

\bibitem{shi2013mining}
Yue Shi, Martha Larson, and Alan Hanjalic.
\newblock Mining contextual movie similarity with matrix factorization for
  context-aware recommendation.
\newblock {\em ACM Transactions on Intelligent Systems and Technology (TIST)},
  4(1):16, 2013.

\bibitem{said2010putting}
Alan Said, Shlomo Berkovsky, and Ernesto~W De~Luca.
\newblock Putting things in context: Challenge on context-aware movie
  recommendation.
\newblock In {\em Proceedings of the Workshop on Context-Aware Movie
  Recommendation}, pages 2--6. ACM, 2010.

\bibitem{wang2010new}
Licai Wang, Xiangwu Meng, Yujie Zhang, and Yancui Shi.
\newblock New approaches to mood-based hybrid collaborative filtering.
\newblock In {\em Proceedings of the Workshop on Context-Aware Movie
  Recommendation}, pages 28--33. ACM, 2010.

\bibitem{saaty1988analytic}
Thomas~L Saaty.
\newblock What is the analytic hierarchy process?
\newblock In {\em Mathematical models for decision support}, pages 109--121.
  Springer, 1988.

\bibitem{kovsir2011database}
Andrej Ko{\v{s}}ir, Ante Odic, Matevz Kunaver, Marko Tkalcic, and Jurij~F
  Tasic.
\newblock Database for contextual personalization.
\newblock {\em Elektrotehni{\v{s}}ki vestnik}, 78(5):270--274, 2011.

\bibitem{fulop2005introduction}
J{\'a}nos F{\"u}l{\"o}p.
\newblock Introduction to decision making methods.
\newblock In {\em BDEI-3 workshop, Washington}, 2005.

\bibitem{saaty1977scaling}
Thomas~L Saaty.
\newblock A scaling method for priorities in hierarchical structures.
\newblock {\em Journal of mathematical psychology}, 15(3):234--281, 1977.

\bibitem{bader2011context}
Roland Bader, Eugen Neufeld, Wolfgang Woerndl, and Vivian Prinz.
\newblock Context-aware poi recommendations in an automotive scenario using
  multi-criteria decision making methods.
\newblock In {\em Proceedings of the 2011 Workshop on Context-awareness in
  Retrieval and Recommendation}, pages 23--30. ACM, 2011.

\bibitem{adomavicius2005incorporating}
Gediminas Adomavicius, Ramesh Sankaranarayanan, Shahana Sen, and Alexander
  Tuzhilin.
\newblock Incorporating contextual information in recommender systems using a
  multidimensional approach.
\newblock {\em ACM Transactions on Information Systems (TOIS)}, 23(1):103--145,
  2005.

\bibitem{sassi2017context}
Imen~Ben Sassi, Sehl Mellouli, and Sadok~Ben Yahia.
\newblock Context-aware recommender systems in mobile environment: On the road
  of future research.
\newblock {\em Information Systems}, 72:27--61, 2017.

\bibitem{bobadilla2013recommender}
Jes{\'u}s Bobadilla, Fernando Ortega, Antonio Hernando, and Abraham
  Guti{\'e}rrez.
\newblock Recommender systems survey.
\newblock {\em Knowledge-based systems}, 46:109--132, 2013.

\bibitem{burke2002hybrid}
Robin Burke.
\newblock Hybrid recommender systems: Survey and experiments.
\newblock {\em User modeling and user-adapted interaction}, 12(4):331--370,
  2002.

\bibitem{jelassi2013personalized}
Mohamed~Nader Jelassi, Sadok Ben~Yahia, and Engelbert Mephu~Nguifo.
\newblock A personalized recommender system based on users' information in
  folksonomies.
\newblock In {\em Proceedings of the 22nd International Conference on World
  Wide Web}, pages 1215--1224. ACM, 2013.

\bibitem{jelassi2014vers}
Mohamed~Nader Jelassi, Sadok~Ben Yahia, and Engelbert~Mephu Nguifo.
\newblock Vers des recommandations plus personnalis{\'e}es dans les
  folksonomies.
\newblock In {\em IC-25{\`e}mes Journ{\'e}es francophones d'Ing{\'e}nierie des
  Connaissances}, pages 187--198, 2014.

\bibitem{jelassi2015towards}
Mohamed~Nader Jelassi, Sadok~Ben Yahia, and Engelbert~Mephu Nguifo.
\newblock Towards more targeted recommendations in folksonomies.
\newblock {\em Social Network Analysis and Mining}, 5(1):68, 2015.

\bibitem{jelassi2016etude}
Mohamed~Nader Jelassi, Sadok Benyahia, and Mephu~Nguifo Engelbert.
\newblock {\'E}tude du profil utilisateur pour la recommandation dans les
  folksonomies.
\newblock In {\em IC2016: Ing{\'e}nierie des Connaissances}, 2016.

\bibitem{anderson2006long}
Chris Anderson.
\newblock {\em The long tail: Why the future of business is selling less of
  more}.
\newblock Hachette Books, 2006.

\bibitem{ricci2011introduction}
Francesco Ricci, Lior Rokach, and Bracha Shapira.
\newblock Introduction to recommender systems handbook.
\newblock In {\em Recommender systems handbook}, pages 1--35. Springer, 2011.

\bibitem{adomavicius2005toward}
Gediminas Adomavicius and Alexander Tuzhilin.
\newblock Toward the next generation of recommender systems: A survey of the
  state-of-the-art and possible extensions.
\newblock {\em IEEE transactions on knowledge and data engineering},
  17(6):734--749, 2005.

\bibitem{su2009survey}
Xiaoyuan Su and Taghi~M Khoshgoftaar.
\newblock A survey of collaborative filtering techniques.
\newblock {\em Advances in artificial intelligence}, 2009:4, 2009.

\bibitem{rao2008application}
K~Nageswara Rao.
\newblock Application domain and functional classification of recommender
  systems--a survey.
\newblock {\em DESIDOC Journal of Library \& Information Technology}, 28(3):17,
  2008.

\bibitem{celma2009music}
{\`O}scar Celma~Herrada et~al.
\newblock {\em Music recommendation and discovery in the long tail}.
\newblock Universitat Pompeu Fabra, 2009.

\bibitem{naak2009papyres}
Amine Naak.
\newblock Papyres: un système de gestion et de recommandation d’articles de
  recherche.
\newblock 2009.

\bibitem{magnini2001improving}
Bernardo Magnini and Carlo Strapparava.
\newblock Improving user modelling with content-based techniques.
\newblock In {\em International Conference on User Modeling}, pages 74--83.
  Springer, 2001.

\bibitem{degemmis2007content}
Marco Degemmis, Pasquale Lops, and Giovanni Semeraro.
\newblock A content-collaborative recommender that exploits wordnet-based user
  profiles for neighborhood formation.
\newblock {\em User Modeling and User-Adapted Interaction}, 17(3):217--255,
  2007.

\bibitem{middleton2002exploiting}
Stuart~E Middleton, Harith Alani, Nigel~R Shadbolt, and David~C De~Roure.
\newblock Exploiting synergy between ontologies and recommender systems.
\newblock In {\em Proceedings of the 3rd International Conference on Semantic
  Web-Volume 55}, pages 41--50. CEUR-WS. org, 2002.

\bibitem{aciar2007informed}
Silvana Aciar, Debbie Zhang, Simeon Simoff, and John Debenham.
\newblock Informed recommender: Basing recommendations on consumer product
  reviews.
\newblock {\em IEEE Intelligent systems}, 22(3), 2007.

\bibitem{linden2003amazon}
Greg Linden, Brent Smith, and Jeremy York.
\newblock Amazon. com recommendations: Item-to-item collaborative filtering.
\newblock {\em IEEE Internet computing}, 7(1):76--80, 2003.

\bibitem{arnautu2013mures}
Octavian~Rolland Arnautu.
\newblock Mures: Un syst{\`e}me de recommandation de musique.
\newblock 2013.

\bibitem{nguyen2006cocofil2}
An-Te Nguyen.
\newblock {\em COCoFil2: Un nouveau syst{\`e}me de filtrage collaboratif
  bas{\'e} sur le mod{\`e}le des espaces de communaut{\'e}s}.
\newblock PhD thesis, Universit{\'e} Joseph-Fourier-Grenoble I, 2006.

\bibitem{gorgoglione2011effect}
Michele Gorgoglione, Umberto Panniello, and Alexander Tuzhilin.
\newblock The effect of context-aware recommendations on customer purchasing
  behavior and trust.
\newblock In {\em Proceedings of the fifth ACM conference on Recommender
  systems}, pages 85--92. ACM, 2011.

\bibitem{sassi2012situation}
Imen~Ben Sassi, Chiraz Trabelsi, Amel Bouzeghoub, and Sadok~Ben Yahia.
\newblock Situation-aware user’s interests prediction for query enrichment.
\newblock In {\em International Conference on Database and Expert Systems
  Applications}, pages 191--205. Springer, 2012.

\bibitem{sassi2013recherche}
Imen~Ben Sassi, Chiraz Trabelsi, Amel Bouzeghoub, and Sadok~Ben Yahia.
\newblock Recherche d'information contextuelle bas{\'e}e sur la pr{\'e}diction
  des int{\'e}r{\^e}ts des utilisateurs et leurs relations sociales.
\newblock {\em Revue des Sciences et Technologies de l'Information-S{\'e}rie
  ISI: Ing{\'e}nierie des Syst{\`e}mes d'Information}, 18(1):59--84, 2013.

\bibitem{sassi2017fuzzy}
Imen~Ben Sassi, Sadok~Ben Yahia, and Sehl Mellouli.
\newblock Fuzzy classification-based emotional context recognition from online
  social networks messages.
\newblock In {\em Fuzzy Systems (FUZZ-IEEE), 2017 IEEE International Conference
  on}, pages 1--6. IEEE, 2017.

\bibitem{souid2017hypergraph}
Hazem Souid, Chiraz Trabelsi, Gabriella Pasi, and Sadok~Ben Yahia.
\newblock Hypergraph fuzzy minimals transversals mining: A new approach for
  social media recommendation.
\newblock In {\em Fuzzy Systems (FUZZ-IEEE), 2017 IEEE International Conference
  on}, pages 1--7. IEEE, 2017.

\bibitem{abowd1999towards}
Gregory Abowd, Anind Dey, Peter Brown, Nigel Davies, Mark Smith, and Pete
  Steggles.
\newblock Towards a better understanding of context and context-awareness.
\newblock In {\em Handheld and ubiquitous computing}, pages 304--307. Springer,
  1999.

\bibitem{dey2001understanding}
Anind~K Dey.
\newblock Understanding and using context.
\newblock {\em Personal and ubiquitous computing}, 5(1):4--7, 2001.

\bibitem{daoud2007recherche}
Mariam Daoud.
\newblock Recherche contextuelle d'information.
\newblock In {\em CORIA}, pages 467--472, 2007.

\bibitem{adomavicius2011context}
Gediminas Adomavicius and Alexander Tuzhilin.
\newblock Context-aware recommender systems.
\newblock In {\em Recommender systems handbook}, pages 217--253. Springer,
  2011.

\bibitem{koller1996toward}
Daphne Koller and Mehran Sahami.
\newblock Toward optimal feature selection.
\newblock Technical report, Stanford InfoLab, 1996.

\bibitem{liu2012feature}
Huan Liu and Hiroshi Motoda.
\newblock {\em Feature selection for knowledge discovery and data mining},
  volume 454.
\newblock Springer Science \& Business Media, 2012.

\bibitem{chatterjee2015regression}
Samprit Chatterjee and Ali~S Hadi.
\newblock {\em Regression analysis by example}.
\newblock John Wiley \& Sons, 2015.

\bibitem{trabelsi2011folksonomy}
Chiraz Trabelsi, Bilel Moulahi, and Sadok~Ben Yahia.
\newblock Folksonomy query suggestion via users’ search intent prediction.
\newblock In {\em International Conference on Flexible Query Answering
  Systems}, pages 388--399. Springer, 2011.

\bibitem{trabelsi2011auto}
Chiraz Trabelsi, Nader Jelassi, and Sadok~Ben Yahia.
\newblock Auto-compl{\'e}tion de requ{\^e}tes par une base g{\'e}n{\'e}rique de
  r{\`e}gles d'association triadiques.
\newblock In {\em CORIA}, pages 9--24, 2011.

\bibitem{trabelsi2012hmm}
Chiraz Trabelsi, Bilel Moulahi, and Sadok~Ben Yahia.
\newblock Hmm-care: Hidden markov models for context-aware tag recommendation
  in folksonomies.
\newblock In {\em Proceedings of the 27th Annual ACM Symposium on Applied
  Computing}, pages 957--961. ACM, 2012.

\bibitem{trabelsi2012scalable}
Chiraz Trabelsi, Nader Jelassi, and Sadok~Ben Yahia.
\newblock Scalable mining of frequent tri-concepts from folksonomies.
\newblock In {\em Pacific-Asia Conference on Knowledge Discovery and Data
  Mining}, pages 231--242. Springer, 2012.

\bibitem{trabelsi2012bgrt}
Chiraz Trabelsi, Nader Jelassi, and Sadok~Ben Yahia.
\newblock Bgrt: une nouvelle base g{\'e}n{\'e}rique de r{\`e}gles d'association
  triadiques.
\newblock {\em Document num{\'e}rique}, 15(1):101--124, 2012.

\bibitem{trabelsi2013integrated}
Chiraz Trabelsi and Sadok~Ben Yahia.
\newblock An integrated approach for context-aware query recommendation in
  folksonomies.
\newblock In {\em CORIA}, pages 253--268, 2013.

\bibitem{trabelsi2016harnessing}
Chiraz Trabelsi and Sadok~Ben Yahia.
\newblock Harnessing the potential of hmm for movie rating recommendation.
\newblock {\em Procedia Computer Science}, 96:1543--1550, 2016.

\bibitem{ricci2010mobile}
Francesco Ricci.
\newblock Mobile recommender systems.
\newblock {\em Information Technology \& Tourism}, 12(3):205--231, 2010.

\bibitem{baltrunas2009towards}
Linas Baltrunas and Xavier Amatriain.
\newblock Towards time-dependant recommendation based on implicit feedback.
\newblock In {\em Workshop on context-aware recommender systems (CARS’09)},
  2009.

\bibitem{di2012linked}
Tommaso Di~Noia, Roberto Mirizzi, Vito~Claudio Ostuni, Davide Romito, and
  Markus Zanker.
\newblock Linked open data to support content-based recommender systems.
\newblock In {\em Proceedings of the 8th International Conference on Semantic
  Systems}, pages 1--8. ACM, 2012.

\bibitem{baltrunas2014experimental}
Linas Baltrunas and Francesco Ricci.
\newblock Experimental evaluation of context-dependent collaborative filtering
  using item splitting.
\newblock {\em User Modeling and User-Adapted Interaction}, 24(1-2):7--34,
  2014.

\bibitem{baltrunas2009context}
Linas Baltrunas and Francesco Ricci.
\newblock Context-based splitting of item ratings in collaborative filtering.
\newblock In {\em Proceedings of the third ACM conference on Recommender
  systems}, pages 245--248. ACM, 2009.

\bibitem{panniello2009experimental}
Umberto Panniello, Alexander Tuzhilin, Michele Gorgoglione, Cosimo Palmisano,
  and Anto Pedone.
\newblock Experimental comparison of pre-vs. post-filtering approaches in
  context-aware recommender systems.
\newblock In {\em Proceedings of the third ACM conference on Recommender
  systems}, pages 265--268. ACM, 2009.

\bibitem{nasrabadi2007pattern}
Nasser~M Nasrabadi.
\newblock Pattern recognition and machine learning.
\newblock {\em Journal of electronic imaging}, 16(4):049901, 2007.

\bibitem{breiman2001random}
Leo Breiman.
\newblock Random forests.
\newblock {\em Machine learning}, 45(1):5--32, 2001.

\bibitem{herlocker1999algorithmic}
Jonathan~L Herlocker, Joseph~A Konstan, Al~Borchers, and John Riedl.
\newblock An algorithmic framework for performing collaborative filtering.
\newblock In {\em Proceedings of the 22nd annual international ACM SIGIR
  conference on Research and development in information retrieval}, pages
  230--237. ACM, 1999.

\bibitem{koren2010collaborative}
Yehuda Koren.
\newblock Collaborative filtering with temporal dynamics.
\newblock {\em Communications of the ACM}, 53(4):89--97, 2010.

\bibitem{witten2016data}
Ian~H Witten, Eibe Frank, Mark~A Hall, and Christopher~J Pal.
\newblock {\em Data Mining: Practical machine learning tools and techniques}.
\newblock Morgan Kaufmann, 2016.

\bibitem{ling2003auc}
Charles Ling, Jin Huang, and Harry Zhang.
\newblock Auc: a better measure than accuracy in comparing learning algorithms.
\newblock {\em Advances in Artificial Intelligence}, pages 991--991, 2003.

\bibitem{rendle2010pairwise}
Steffen Rendle and Lars Schmidt-Thieme.
\newblock Pairwise interaction tensor factorization for personalized tag
  recommendation.
\newblock In {\em Proceedings of the third ACM international conference on Web
  search and data mining}, pages 81--90. ACM, 2010.

\bibitem{rendle2009factor}
Steffen Rendle and Lars Schmidt-Thieme.
\newblock Factor models for tag recommendation in bibsonomy.
\newblock In {\em ECML/PKDD 2008 Discovery Challenge Workshop, part of the
  European Conference on Machine Learning and Principles and Practice of
  Knowledge Discovery in Databases}, pages 235--243, 2009.

\bibitem{biancalana2011context}
Claudio Biancalana, Fabio Gasparetti, Alessandro Micarelli, Alfonso Miola, and
  Giuseppe Sansonetti.
\newblock Context-aware movie recommendation based on signal processing and
  machine learning.
\newblock In {\em Proceedings of the 2nd Challenge on Context-Aware Movie
  Recommendation}, pages 5--10. ACM, 2011.

\bibitem{lathia2010temporal}
Neal Lathia, Stephen Hailes, Licia Capra, and Xavier Amatriain.
\newblock Temporal diversity in recommender systems.
\newblock In {\em Proceedings of the 33rd international ACM SIGIR conference on
  Research and development in information retrieval}, pages 210--217. ACM,
  2010.

\bibitem{deshpande2004item}
Mukund Deshpande and George Karypis.
\newblock Item-based top-n recommendation algorithms.
\newblock {\em ACM Transactions on Information Systems (TOIS)}, 22(1):143--177,
  2004.

\bibitem{berkovsky2006predicting}
Shlomo Berkovsky, Tsvi Kuflik, Lora Aroyo, Dominik Heckmann, Alexander
  Kr{\"o}ner, Francesco Ricci, and Geert-Jan Houben.
\newblock Predicting user experiences through cross-context reasoning.
\newblock In {\em LWA}, pages 27--31, 2006.

\bibitem{said2013introduction}
Alan Said, Shlomo Berkovsky, and Ernesto~W De~Luca.
\newblock Introduction to special section on camra2010: Movie recommendation in
  context.
\newblock {\em ACM Transactions on Intelligent Systems and Technology (TIST)},
  4(1):13, 2013.

\bibitem{liu2010adapting}
Nathan~N Liu, Bin Cao, Min Zhao, and Qiang Yang.
\newblock Adapting neighborhood and matrix factorization models for context
  aware recommendation.
\newblock In {\em Proceedings of the Workshop on Context-Aware Movie
  Recommendation}, pages 7--13. ACM, 2010.

\bibitem{brenner2010predicting}
Antoine Brenner, Bruno Pradel, Nicolas Usunier, and Patrick Gallinari.
\newblock Predicting most rated items in weekly recommendation with temporal
  regression.
\newblock In {\em Proceedings of the Workshop on Context-Aware Movie
  Recommendation}, pages 24--27. ACM, 2010.

\bibitem{campos2010simple}
Pedro~G Campos, Alejandro Bellog{\'\i}n, Fernando D{\'\i}ez, and J~Enrique
  Chavarriaga.
\newblock Simple time-biased knn-based recommendations.
\newblock In {\em Proceedings of the Workshop on Context-Aware Movie
  Recommendation}, pages 20--23. ACM, 2010.

\bibitem{shi2010mining}
Yue Shi, Martha Larson, and Alan Hanjalic.
\newblock Mining mood-specific movie similarity with matrix factorization for
  context-aware recommendation.
\newblock In {\em Proceedings of the workshop on context-aware movie
  recommendation}, pages 34--40. ACM, 2010.

\bibitem{wu2010novel}
Pei Wu, Weiping Liu, and Cihang Jin.
\newblock A novel recommender system fusing the opinions from experts and
  ordinary people.
\newblock In {\em Proceedings of the Workshop on Context-Aware Movie
  Recommendation}, pages 41--44. ACM, 2010.

\bibitem{diez2010movie}
Fernando D{\'\i}ez, J~Enrique Chavarriaga, Pedro~G Campos, and Alejandro
  Bellog{\'\i}n.
\newblock Movie recommendations based in explicit and implicit features
  extracted from the filmtipset dataset.
\newblock In {\em Proceedings of the Workshop on Context-Aware Movie
  Recommendation}, pages 45--52. ACM, 2010.

\bibitem{liu2010incorporating}
Bin Liu and Zheng Yuan.
\newblock Incorporating social networks and user opinions for collaborative
  recommendation: local trust network based method.
\newblock In {\em Proceedings of the Workshop on Context-Aware Movie
  Recommendation}, pages 53--56. ACM, 2010.

\bibitem{rahmani2010three}
Hossein Rahmani, Beau Piccart, Daan Fierens, and Hendrik Blockeel.
\newblock Three complementary approaches to context aware movie recommendation.
\newblock In {\em Proceedings of the Workshop on Context-Aware Movie
  Recommendation}, pages 57--60. ACM, 2010.

\bibitem{candillier2009state}
Laurent Candillier, Kris Jack, Fran{\c{c}}oise Fessant, and Frank Meyer.
\newblock State-of-the-art recommender systems.
\newblock {\em Collaborative and Social Information Retrieval and
  AccessTechniques for Improved User Modeling}, 2009.

\bibitem{winoto2010role}
Pinata Winoto and Tiffany~Y Tang.
\newblock The role of user mood in movie recommendations.
\newblock {\em Expert Systems with Applications}, 37(8):6086--6092, 2010.

\bibitem{mardani2015multiple}
Abbas Mardani, Ahmad Jusoh, Khalil MD~Nor, Zainab Khalifah, Norhayati Zakwan,
  and Alireza Valipour.
\newblock Multiple criteria decision-making techniques and their
  applications--a review of the literature from 2000 to 2014.
\newblock {\em Economic Research-Ekonomska Istra{\v{z}}ivanja}, 28(1):516--571,
  2015.

\bibitem{manouselis2007analysis}
Nikos Manouselis and Constantina Costopoulou.
\newblock Analysis and classification of multi-criteria recommender systems.
\newblock {\em World Wide Web}, 10(4):415--441, 2007.

\bibitem{saaty2004decision}
Thomas~L Saaty.
\newblock Decision making—the analytic hierarchy and network processes
  (ahp/anp).
\newblock {\em Journal of systems science and systems engineering},
  13(1):1--35, 2004.

\bibitem{bouneffouf2014recommandation}
Djallel Bouneffouf.
\newblock Recommandation mobile, sensible au contexte de
  contenus$\backslash$'evolutifs: Contextuel-e-greedy.
\newblock {\em arXiv preprint arXiv:1402.1986}, 2014.

\end{thebibliography}
\end{document}